%% file: _main_thesis_file.tex
\documentclass[plain]{hvdthesis}

\input{setup}

\input{mycommands}

\author{John H. Madden}
\title{The Color of Habitability}
\department{Cornell Department of Astronomy and Space Sciences}
\subject{Astrophysics}
\month{August}
\year{2020}
\advisor{Lisa Kaltenegger}

\begin{document}



\frontmatter

\makecover
\copyright
\pagenumbering{gobble}
\input{./0a_abstract/abstract.tex}

\pagenumbering{gobble}

\clearpage
\thispagestyle{empty}
\begin{figure*}[p]
\centering
\includegraphics[width=0.8\textwidth]{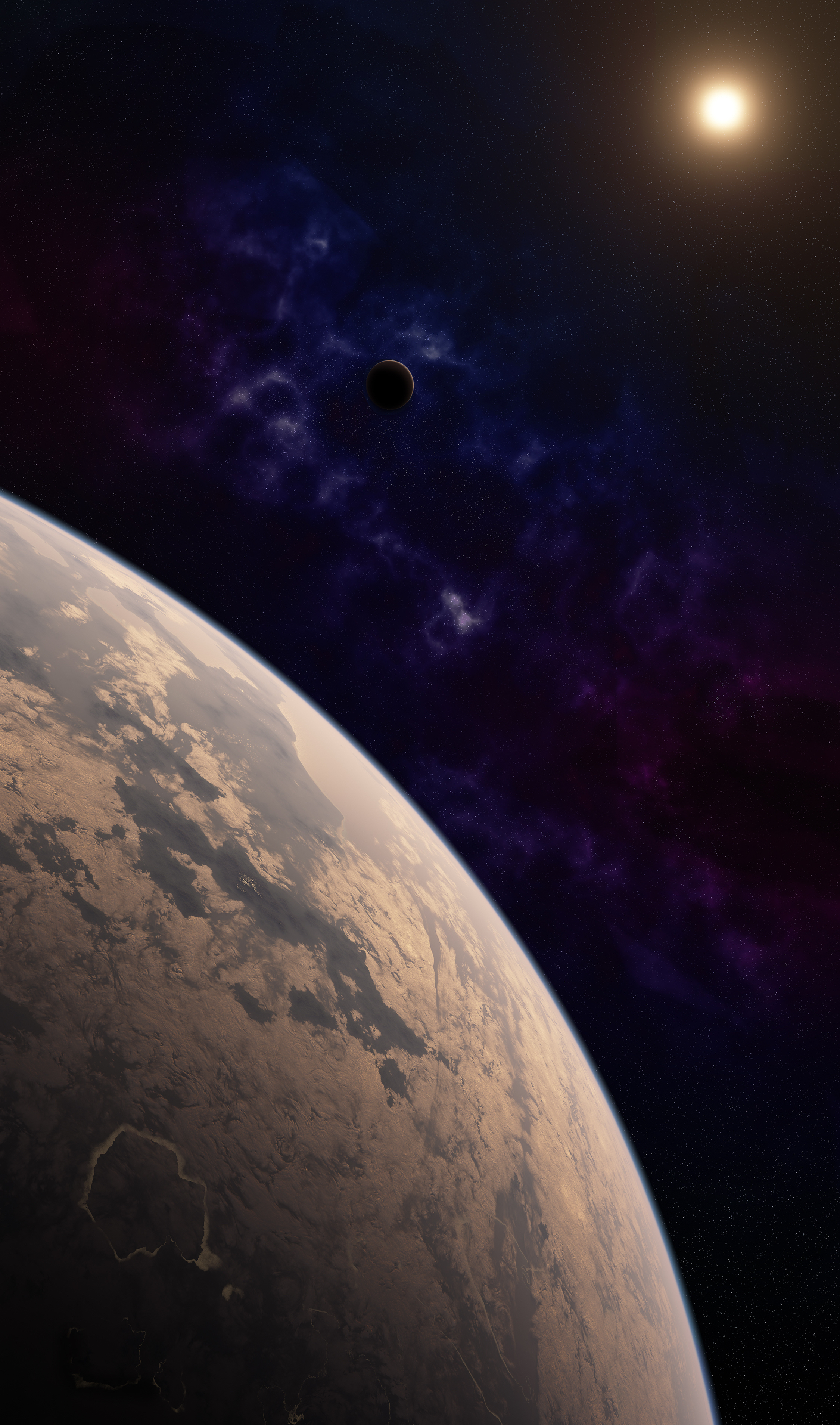}
\end{figure*}

{\singlespace
\tableofcontents
}
\newpage
\clearpage


\pagenumbering{roman}
\setcounter{page}{1}

\thispagestyle{plain}
\addcontentsline{toc}{chapter}{Biographical Sketch}
\chaptermark{Biographical Sketch}
\vskip 0.5cm
{\centerline {\Large \bf Biographical Sketch}}
\vskip 0.5cm
\normalsize
\input{./0b_biographical/biographical.tex}

\clearpage

\thispagestyle{plain}
\addcontentsline{toc}{chapter}{Acknowledgments}
\chaptermark{Acknowledgments}
\vskip 0.5cm
{\centerline {\Large \bf Acknowledgments}}
\vskip 0.5cm
\normalsize
\input{./0c_acknowledge/acknowledge.tex}

\clearpage

\blankpage

\thispagestyle{empty}
\addcontentsline{toc}{chapter}{Dedication}
\vspace*{\fill}
{\centerline {\em To the stars that made me.}} 
\vspace*{\fill}

\clearpage

\mainmatter
\pagestyle{fancy}

\clearpage
\thispagestyle{empty}
\addtocounter{page}{-1}
\begin{figure*}[p]
\centering
\includegraphics[width=\textwidth]{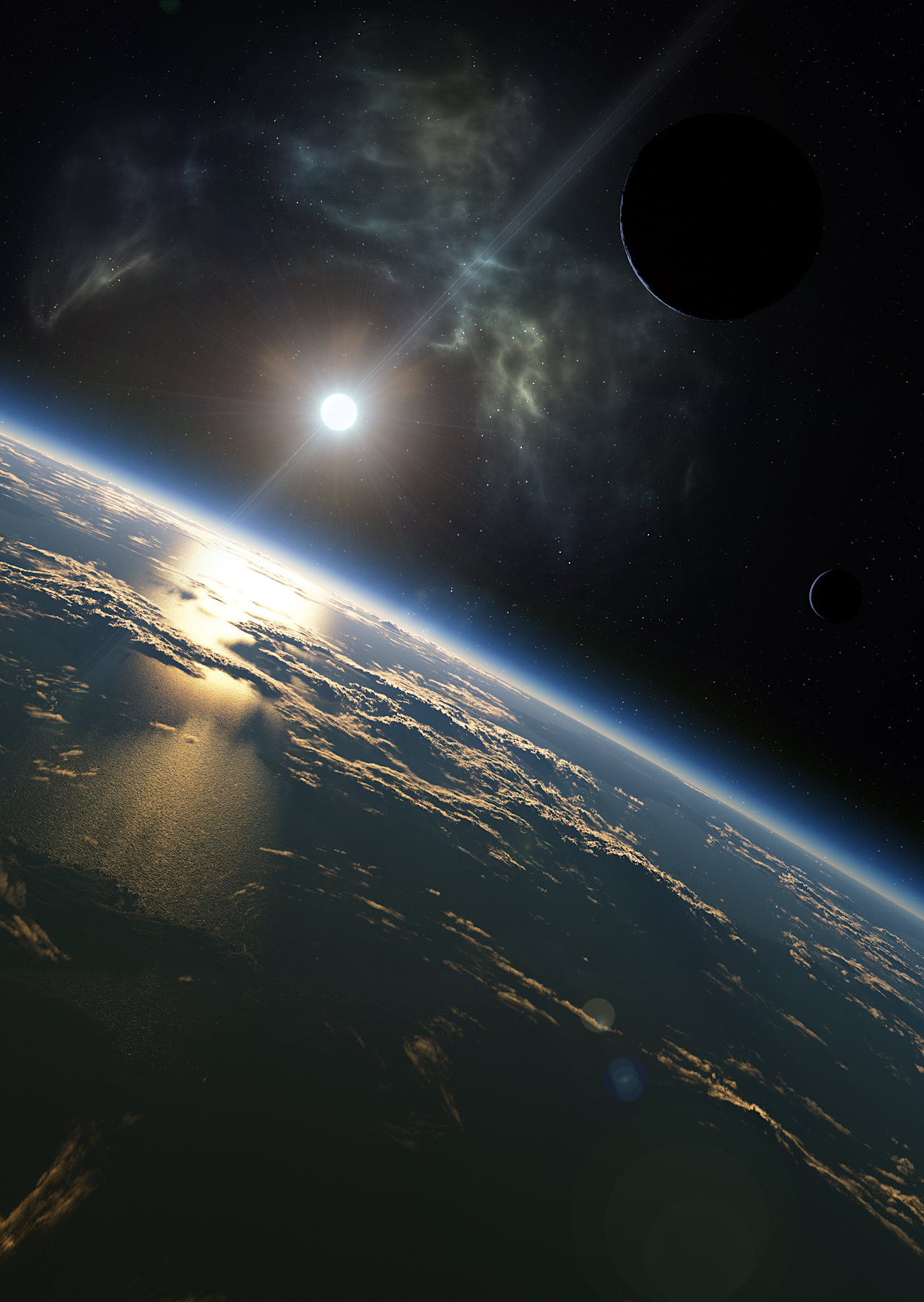}
\end{figure*}

\singlespace
\chapter{Introduction}\label{c:intro}
\chaptermark{Introduction}
\doublespace
\input{./1_introduction/intro}

\clearpage
\thispagestyle{empty}
\begin{figure*}[p]
\centering
\includegraphics[width=\textwidth]{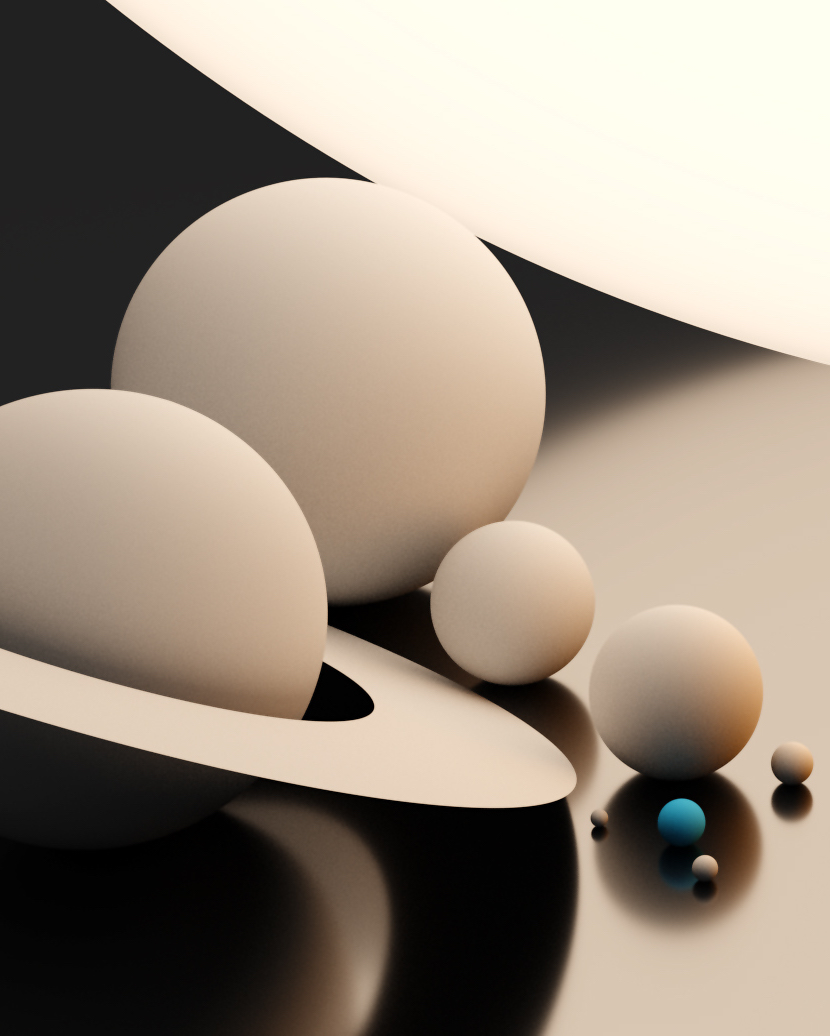}
\end{figure*}

\singlespace
\chapter{A Catalog of Spectra, Albedos, and Colors of Solar System Bodies for Exoplanet Comparison}\label{c:chap2}
\chaptermark{Solar System Spectra and Albedos}
\begin{quote}
{\em ~~~~~~~This thesis chapter originally appeared in the literature as} \\
{J. Madden and L. Kaltenegger,
{\em Astrobiology, Volume 18, Issue 12, pp.1559-1573 (2018)}}
\end{quote}
\doublespace
\input{./2_SolarSystem/ms.tex}

\clearpage
\thispagestyle{empty}
\begin{figure*}[p]
\centering
\includegraphics[width=0.9\textwidth]{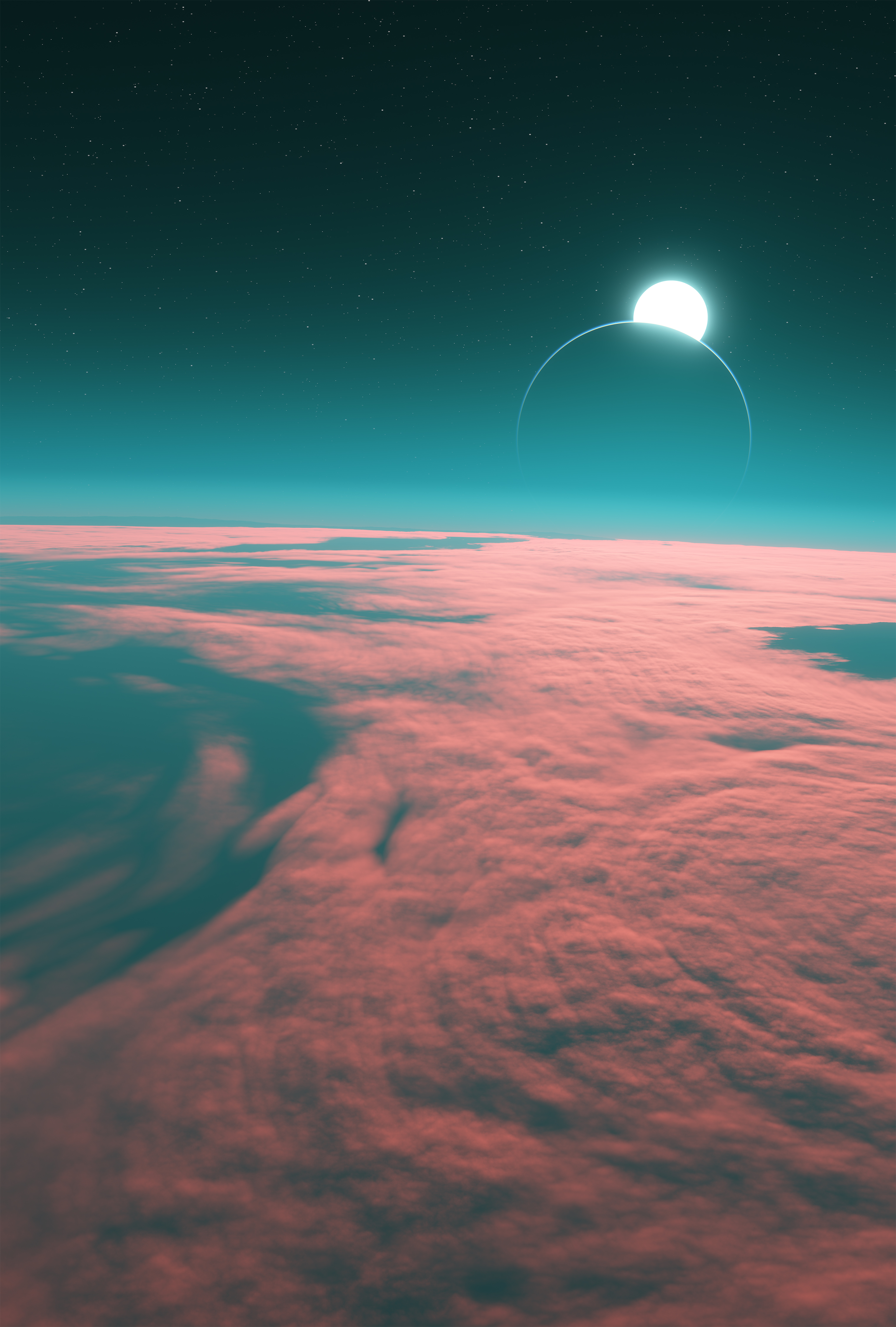}
\end{figure*}

\singlespace
\chapter{How surfaces shape the climate of habitable exoplanets}\label{c:chap3}
\chaptermark{Surfaces of habitable exoplanets}
\begin{quote}
{\em ~~~~~~~This thesis chapter originally appeared in the literature as} \\
{J. Madden and L. Kaltenegger,
{\em Monthly Notices of the Royal Astronomical Society (2020)}}
\end{quote}
\doublespace
\input{./3_surfaces/ms.tex}

\clearpage
\thispagestyle{empty}
\begin{figure*}[p]
\centering
\includegraphics[width=0.8\textwidth]{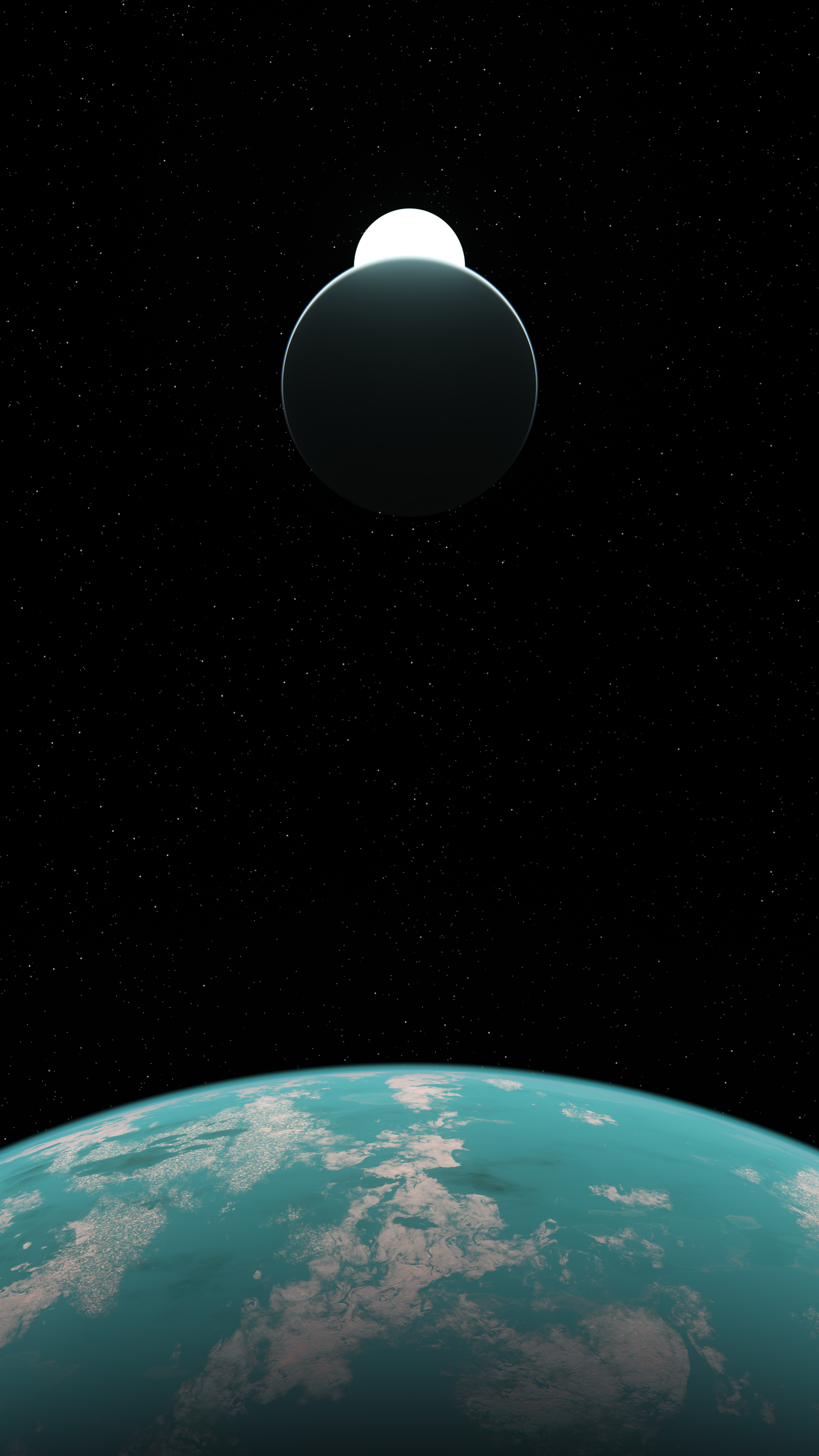}
\end{figure*}

\singlespace
\chapter{Ready Student One: Exploring the predictors of student learning in virtual reality}\label{c:chap4}
\chaptermark{Exploring student learning in VR}
\begin{quote}
{\em ~~~~~~~This thesis chapter originally appeared in the literature as} \\
{J. Madden, S. Pandita, J. P. Schuldt, B. Kim, A. S. Won, and N. G. Holmes,
{\em PLOS ONE 15(3): e0229788 (2020)}}
\end{quote}
\doublespace
\input{./4_VR/ms.tex}

\clearpage
\thispagestyle{empty}
\begin{figure*}[p]
\centering
\includegraphics[width=\textwidth]{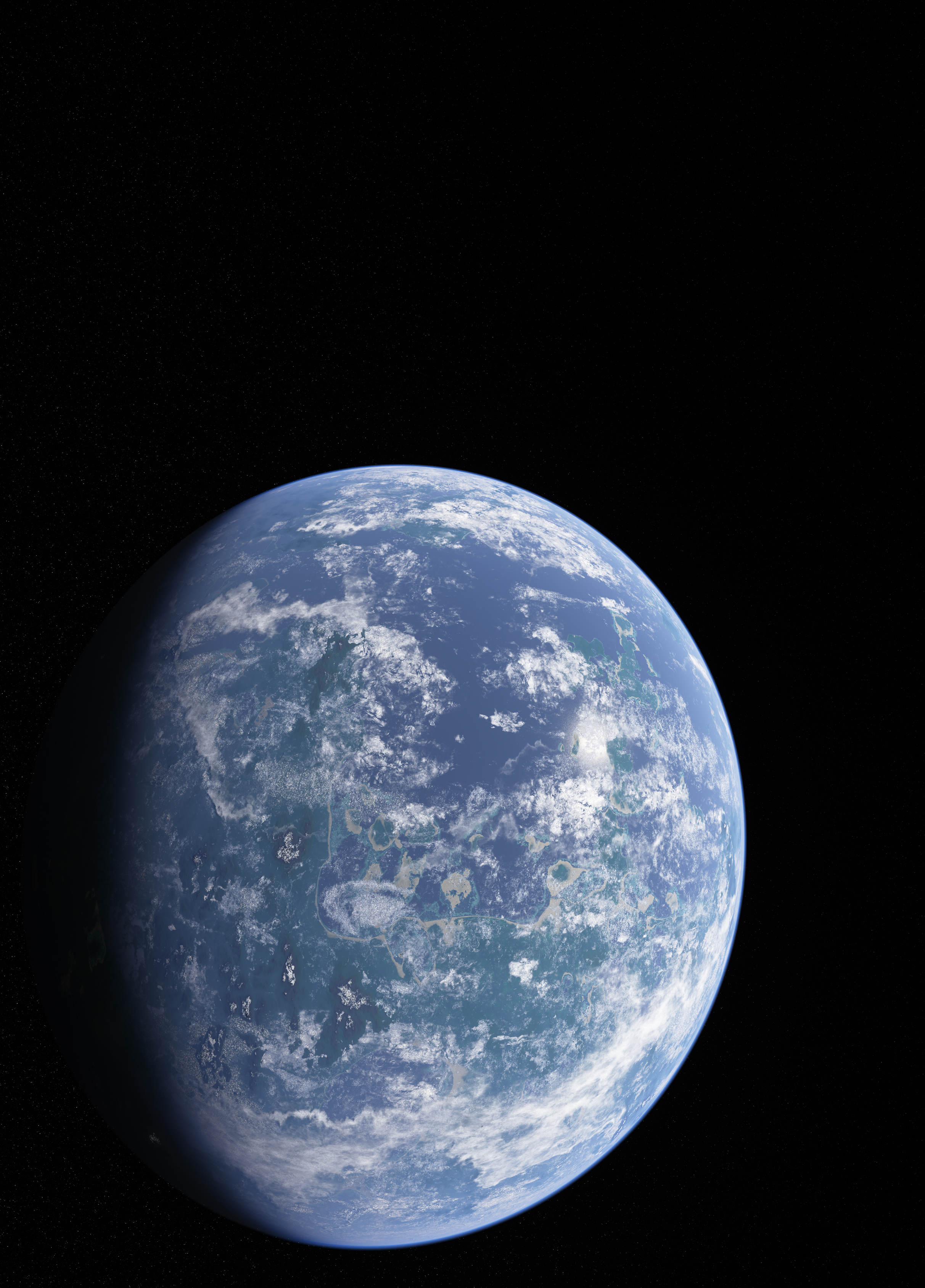}
\end{figure*}

\singlespace
\chapter{High-resolution Spectra for a Wide Range of Habitable Zone Planets around Sun-like stars}\label{c:chap5}
\chaptermark{High-resolution Spectra}
\begin{quote}
{\em ~~~~~~~This thesis chapter originally appeared in the literature as} \\
{J. Madden and L. Kaltenegger,
{\em The Astrophysical Journal Letters (2020)}}
\end{quote}
\doublespace
\input{./5_observe/ms.tex}

\blankpage

\thispagestyle{empty}
\begin{figure*}[p]
\centering
\includegraphics[width=\textwidth]{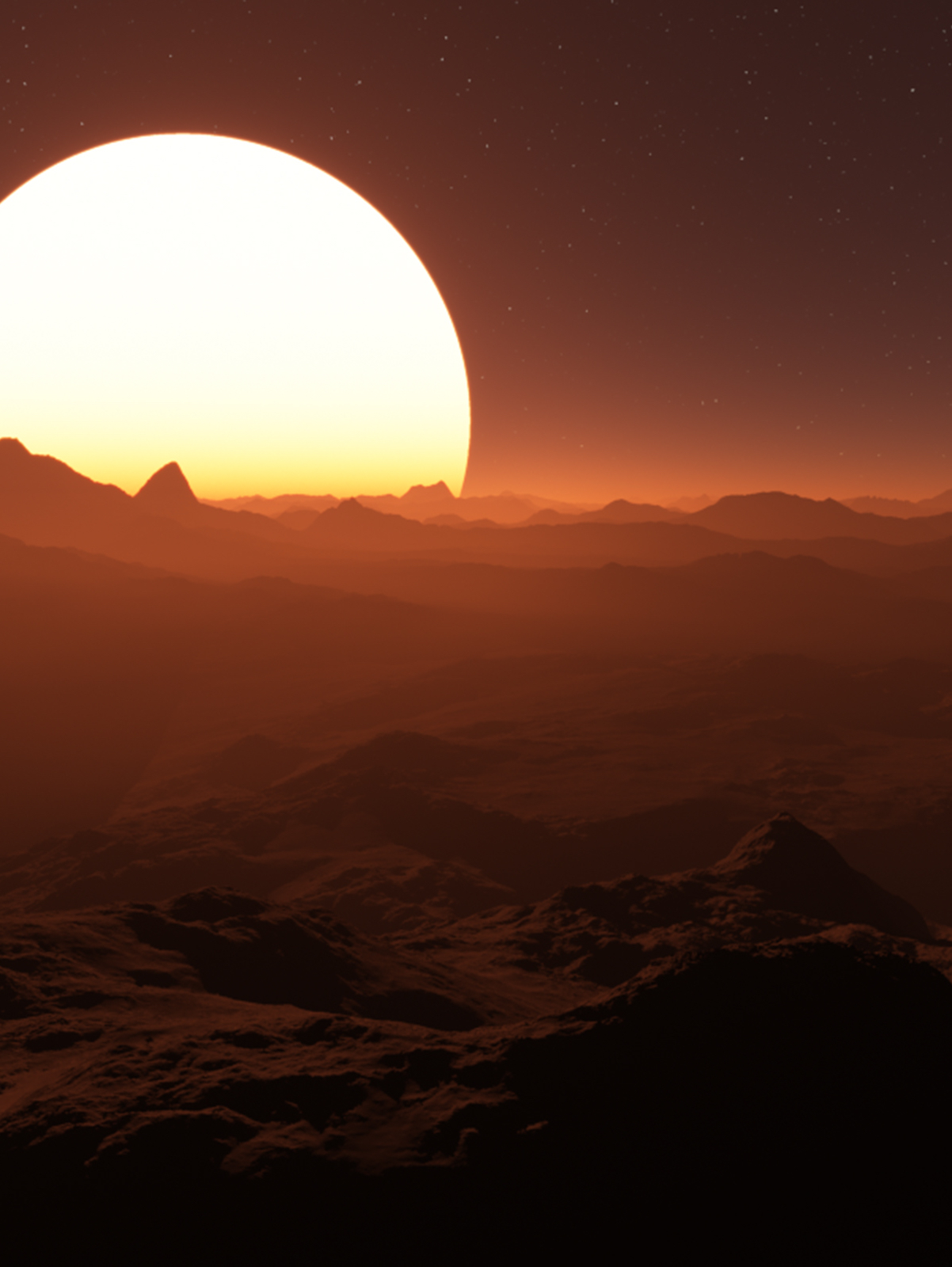}
\end{figure*}

\singlespace
\bibliographystyle{apalike}
\providecommand{\noopsort}[1]{}\providecommand{\singleletter}[1]{#1}%

\blankpage

\thispagestyle{empty}
\begin{figure*}[p]
\centering
\includegraphics[width=\textwidth]{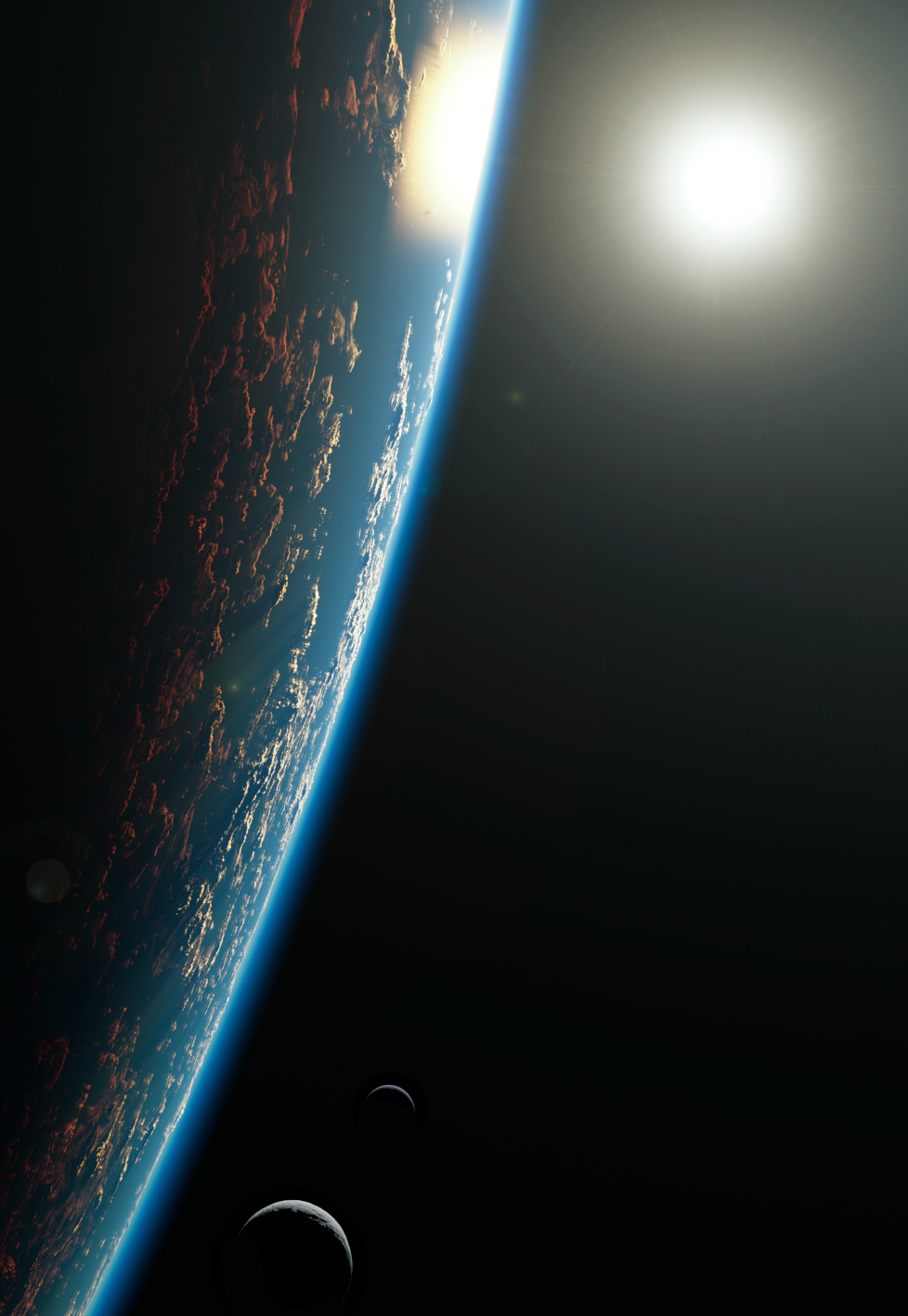}
\end{figure*}

\end{document}

%% file: setup.tex
\bibpunct{(}{)}{;}{a}{}{,}

\usepackage{epstopdf}
\usepackage{epsfig}
\usepackage{aasmod}
\usepackage{amsmath}
\usepackage{amssymb}
\usepackage{rotating}
\usepackage{verbatim}
\usepackage{longtable}
\usepackage{lscape}
\usepackage{float}
\usepackage{color}
\usepackage{fltpage}
\usepackage{url}
\usepackage{rotating}
\usepackage[innercaption]{sidecap}
\usepackage{pdfpages}
\usepackage{xspace}

\usepackage{graphicx}
\usepackage{xcolor}
\usepackage{adjustbox}
\usepackage[utf8]{inputenc}
\usepackage{nameref,hyperref}
\usepackage{changepage}
\usepackage{textcomp,marvosym}
\usepackage{wrapfig}
\usepackage[caption=false]{subfig}
\usepackage{booktabs}
\usepackage{multirow}
\usepackage{footmisc}

\usepackage{fancyhdr}


%% file: mycommands.tex
\definecolor{orange}{rgb}{.8,0.4,0}

\mathchardef\mhyphen="2D
\definecolor{alice}{rgb}{0.0784, 0.5882, 0.7333}
\definecolor{488}{rgb}{0.0, 0.705, 0.896}
\definecolor{green}{rgb}{0.0, 1.0, 0.0}
\definecolor{blue}{rgb}{0.0, 0.0, 1.0}
\definecolor{gas}{RGB}{230, 159, 0}
\definecolor{ice}{RGB}{0, 114, 178}
\definecolor{rock}{RGB}{205, 121, 167}

\let\cleardoublepage\clearpage
\def\blankpage{%
      \clearpage%
      \thispagestyle{empty}%
      \null%
      \clearpage}

%% file: 0a_abstract/abstract.tex
\abstract{\doublespace
From life on other planets to virtual classrooms this thesis spans a wide array of research topics all based on how we see other worlds. Our understanding of everything from moon phases, the planets in our Solar System, and exoplanet atmospheres come from our interpretation of light and one day, our knowledge of light will be used as evidence for the discovery of life on another planet. 

In the time before we scattered rovers, landers, and brave souls across the Solar System we only knew of the planets and moons from the light they reflected from the Sun back to us. We are in much the same situation with exoplanets today. Our telescopes can gather the light from distant worlds but they are too far out of reach to confirm our observations with in-situ measurements. Soon we will be able to gather light from even smaller exoplanets and eventually Earth-sized exoplanets orbiting in their star's habitable zone. As a reference guide to these upcoming observations what better place to compare to than our own Solar System. What we've done is take measurements of planets in our own Solar System and treat them as exoplanets to determine how different surface types can be differentiated. The result is a database of the spectra, geometric albedos, and color of 19 Solar System objects for use as an exoplanetary field guide.

A step beyond the field guide is a way to explore worlds only physics and our imaginations are limited by. By using computer models, we can create thousands of planets to determine the physical and chemical stability of any environment. One parameter domain of interest is the role of surface color on a planet's habitability. Different materials have unique thermal properties that either cool or heat a surface depending on their color and the light that hits them. Dark oceans absorb light well and heat up while white sand is highly reflective and keeps cool. Stars emit different types of light depending on their temperature, cool stars are red while hot stars are blue. Blue starlight on a blue surface will stay cool while blue light on a red surface will heat up and vice-versa for red starlight. This results in a complicated relationship for exoplanets and stars that is important to understand if we are to properly plan observations, analyze data, and make predictions with their results.

The near future is bright for exoplanet science and as new observations dramatically change our understanding of celestial bodies so too do our methods of communicating that science. At the same time the first exoplanets were being discovered, NASA was developing the use of virtual reality to train astronauts and drive rovers on Mars. Virtual reality (VR) has come a long way since then and immersive VR headsets are now used by a fast growing audience for video games, entertainment, and education. As VR education begins to grow, we need a good understanding of its strengths and weaknesses as a teaching tool. So far, very little is know about how employing VR compares to traditional methods of teaching. By designing a learning activity on Moon phases in VR, we were able to compare the learning gains of students who used the VR version with students who used a desktop simulation or a hands-on activity. This information, coupled with the demographics of the participants, allows for a detailed breakdown of who is learning best under which conditions and plan for further study into the use of VR as a teaching tool. 
}

%% file: 0b_biographical/biographical.tex
\singlespace
Jack Madden grew up in Beach Haven, New Jersey before moving to Princeton, New Jersey. After graduating from Princeton High School in 2010 he went to Franklin and Marshall College in Lancaster, Pennsylvania. He worked with Dr. Froney Crawford on a pulsar survey of the Small and Large Magellanic clouds and spent the summer of 2013 at Goddard Spaceflight Center working with Dr. Lynn Carter, and Dr. Catherine Neish on Lunar impact melts. He received his B. A. in astrophysics from F\&M in 2014. He went on to pursue a Ph. D. in astrophysics at Cornell University in Ithaca, New York working with Dr. Lisa Kaltenegger on exoplanet habitability. After successfully defending this thesis in May of 2020, he went on to pursue an MFA at the Rhode Island School of Design. 

%% file: 0c_acknowledge/acknowledge.tex
\textbf{Mom and Dad.} It's so relaxing to know your love is unconditional. Having you here for me as a given means I can calmly focus on what I love to do, without worry. I appreciate every decision you've made because it got me to where I am today. Everything I do, you've made possible.    

\textbf{Will}, I'm so lucky to have such a supportive brother. I know you're always there for me and your music has been perfect to write my thesis to. 

\textbf{Lisa}, thank you. This has been an amazing experience. I don't think I'll ever meet a more enthusiastic and motivated scientist. You have been so generous and supportive. Together we have done some really great work and I'm proud of what we've accomplished.    

\begin{center}
    \noindent\rule[0.5ex]{0.6\linewidth}{0.5pt}
\end{center}

\textbf{First years:} Nic Kutsop, Victoria Calafut, Avani Gowardhan, Georgios Valogianis, Daisy Leung, Michelle Vick, Sam Birch, Eldritch, and Paul Coriles

\textbf{Fellow astronomers:}  Yury Aglyamov, Kassandra Anderson, Crist\'obal Armaza, Catie Ball, Gabe Bonilla, Ligia Fonseca Coelho, Dylan Cromer, Nils Dippe, Trevor Foote, Andrew Foster, Alex Grant,Matt Hankins, Jeremy Hodis, Jason Hofgardner, Dante Iozzo, Jonathan Jackson, Ross Jennings, Abhinav Jindal, Mike Jones, Byungdoo Kim, Thea Kozakis, Michael Lam, J.T. Laune, Marika Leitner, Cody Lemarche, Jiaru Li, Zifan Lin, C.J. Llorente, Tory Lynch, Sean Marshall, Sarah Millholland, Ishan Mishra, Kelsey Moore, Emily Moser, Eamonn O’Shea, Stella Ocker, Chris O'Connor, Swati Pandita, Tyler Pauly, Riccardo Pavesi, Bo Peng, Karen Perez, Bonan Pu, Tyler Robinson, Christopher Rooney, Morgan Saidel, Carly Snell, Tristan Stone, Yubo Su, Yilu Sun, Akshay Suresh, Christian Tate, Alex Teachey, Zoe Todd, Ngoc Truong, Amit Vishwas, Nicole Wallack, Lukas Wenzl, and JJ Zanazzi  

\textbf{Astronomy Department:} Doug, Nena, Jason Jennings, Jessica Jones, Patricia Fernandez de Castro Martinez, Mary Mulvanerton, Cheryl Neville, Dave Pawelczyk, Zoe Ponterio, Tom Shannon, Lynda Sovocool, Jill Tarbell, Bez Thomas, and Melanie Wetzel

\textbf{Cornell EMS:} 
Catherine Appleby, Evy Baer, Hannah Bukzin, Steve Davies, Jessie Dobler, Michael Downey, Jacob Eisner, Anthony Gariolo, Meg Gordon, Ed Jaffe, George Jakubson, Alex Katz, Elise Kleine, Troy Laurence, Chad Lazar, Esteban Lopez, Dan Maas, Lakshmi Mahajan, Sean McCoy, Kyle Otto, Richa Parikh, Zac Pertucci, Dillon Sumanthiran, Galen Weld, Lily Woolf, the 18s, and all the CUPD officers who kept me safe 

\textbf{DBER}
Frank Castelli, Claire Meaders, Katherine Quinn, Emily Smith, Martin Stein, Ryan Tapping, Cole Walsh, and Monica Xu

\textbf{Cornell Faculty:}
 Nic Battaglia, Rachel Bean, Don Campbell, David Chernoff, Jim Cordes, Peter Gierasch, Riccardo Giovanelli, Alex Hayes, Martha Haynes, Terry Herter, Dong Lai, Nikole Lewis, Jamie Lloyd, Richard Lovelace, Phil Nicholson, Mason Peck, Jonathon Schuldt, Michelle Smith, Gordon Stacey, Saul Teukolsky, and Kim Williams

\textbf{Researchers:} 
Shami Chatterjee, Sid Hegde, Paul Helfenstein, Robert Hurt, Jack O'Malley James, Larry Kidder, Illeana Gomez Leal, Thomas Loredo, Ryan MacDonald, Maryame El Moutamid, Valerio Poggiali, Ramses Ramirez, Julie Rathbun, Andrew Ridden-Harper, Marina Romanova, Sarah Rugheimer, Henrik Spoon, and Jake Turner

\textbf{Cornell Staff:}
 Florio Argillas, Janine Brace, Carla DeMello, Blaine Friedlander, Linda Glaser, Sara Xayarath Hern\'andez, and all the printing services at Cornell 

\textbf{Faculty and staff at Franklin and Marshall:} 
Greg Adkins, Andy de Wet, Etienne Gagnon, Roger Godin, Lynn Johnson, Ken Krebs, Christie Larochelle, Andrea Lommen, Amy Lytle, Dan Porterfield, Beth Praton, and Calvin Stubbins

\textbf{Middle and high school teachers: }
Mr. Komada, Mr. Floor, Mr. Carson, Mr. Skalka, Mr. Kosa, Mr. Hoffman, Ms. Agrusti-Taha, Ms. Nickman, Mrs. Cody, and Mr. Mckenna  

\textbf{Friends:} 
Emily Christie, Heather Croy, Zoe Fuhrman, Hilary Gorgol, Emma Handzo, Matt Klimuszka, Adina Klingman, Dan Levin, Matt Momjian, Caitlin Rose, Charlotte Roth, Jennie Anne Simson, Betsy Sweemer, Brenna Vaughn, and Klariobaldo Zavala 

\textbf{Family:} 
Grandmom, Aunt Kathryn, Nanny, Pop-pop, Katie, Patrick, Thomas, Aunt Barbara, Uncle Beard, Uncle Will, Sue, Matt, Melissa, Uncle Walt, Aunt Jan, Charlie, Alicia, Walter, Uncle Robert, Aunt Nancy, and Claire 

\begin{center}
    \noindent\rule[0.5ex]{0.6\linewidth}{1pt}
\end{center}

\textbf{Jacob Peacock} has been the most inspirational example to me of what a scientist should be. Disguised as adventure, my love of exploring the natural world grew from the footprints we left on LBI.    

\textbf{Mr. Anderson} was more than just that one science teacher we all had who was amazing. He helped me tap into my affinity for science and direct it purposefully to projects and independent studies. Mr. Anderson reminded all of us that science could be fun, even for high school seniors, and that science has a massive impact on our lives and future.

\textbf{Froney Crawford} was my first research advisor and I couldn't have asked for a better mentor. His confidence in me, generous support, and rigorous yet lighthearted approach to science drove me to enjoy research more than I ever expected. I am constantly using what I've learned from Froney in my science and my life.  

\textbf{Jonathan Lunine} has shown me that there's always time for respect. He provides all of us with a great example of how scientists have no excuse when it comes to respecting others and valuing professionalism, diversity and inclusion, communication, and mentorship. 

\textbf{Andrea Stevenson Won's} fascination, expertise, and encouragement enabled me to produce work I am really proud of and didn't think I was capable of. She inspires dedication, passion, curiosity, and everyone’s best.  

\textbf{Natasha Holmes}, it's hard to put into words how dramatically working with Natasha has impacted me. Her guidance through our project taught me so much and exposed me to whole new ways of thinking about science. In addition, her confidence, trust, patience, enthusiasm, and support made for one of the most rewarding research experiences I've ever had. 

\textbf{Monica Carpenter} always had my back through grad school ever since I missed the first day of TA orientation. Monica sent me my acceptance letter to Cornell and the announcement email for my defense. I could always count on her for a laugh and help with anything that came up.    

\begin{center}
    \noindent\rule[0.5ex]{0.6\linewidth}{1pt}
\end{center}

I'd like to recognize and thank my committee members \textbf{Lisa Kaltenegger}, \textbf{Natasha Holmes}, \textbf{Ira Wasserman}, \textbf{Dmitry Savransky}, and \textbf{Steve Squyres} for helping shape my graduate career and being a part of my degree conferral. 

I'd also like to give special thanks to those that have supported my Ph.D. work. \textbf{Cornell University}, \textbf{The Departmemnt of Astronomy and Space Sciences}, \textbf{The Carl Sagan Institute}, \textbf{The NY Space Grant Consortium}, \textbf{The Simons Foundation}, \textbf{The Brinson Foundation}, and \textbf{Oculus Education}

%% file: 1_introduction/intro.tex
\section{Color}
Astronomy is an inherently visual endeavor. Whether experienced virtually or through a thirty-meter telescope, we gather information about what lies beyond Earth using light, shape, color, and motion from up to 13.8 billion light-years away. We know the universe not through our nose, mouth, ears, or hands but our eyes. If you're seeing \textcolor{488}{this} on paper, you're seeing a wavelength of light around 488nm reflected off a mix of pigments reflecting light at just the right levels across the whole visible spectrum. If you're looking at a computer screen, you're seeing a combination of \textcolor{green}{green} (530nm) light dimmed to 70.5\%  and \textcolor{blue}{blue} (465nm) light dimmed to 89.6\% that is being processed by your brain into a single color. The qualia we experience as color is a complex system of quantum mechanics and perception. But, as hard as it would be to describe a color with words to match its visual representation, its relatively easy to break down color into its components and distinguish its source. Breaking down light into spectra has become indispensable in science and perhaps the most essential tool in modern astronomy. 

The color of an object is an aggregate of light at wavelengths determined by the properties of the source and the medium through which the light passes. For the most part, the resulting array of wavelengths is unique to the certain source and medium combination. Breaking light apart in a way to see the wavelengths is called spectroscopy and allows us to determine the composition of the source and medium from a distance using only light. This simple principle has allowed astronomy to reach beyond what we can touch and understand what the universe is made of from light-years away. Spectroscopy is how we will find life beyond our Solar System. 

\section{Habitability}
\begin{figure}[ht!]
\centering
\includegraphics[width=\textwidth]{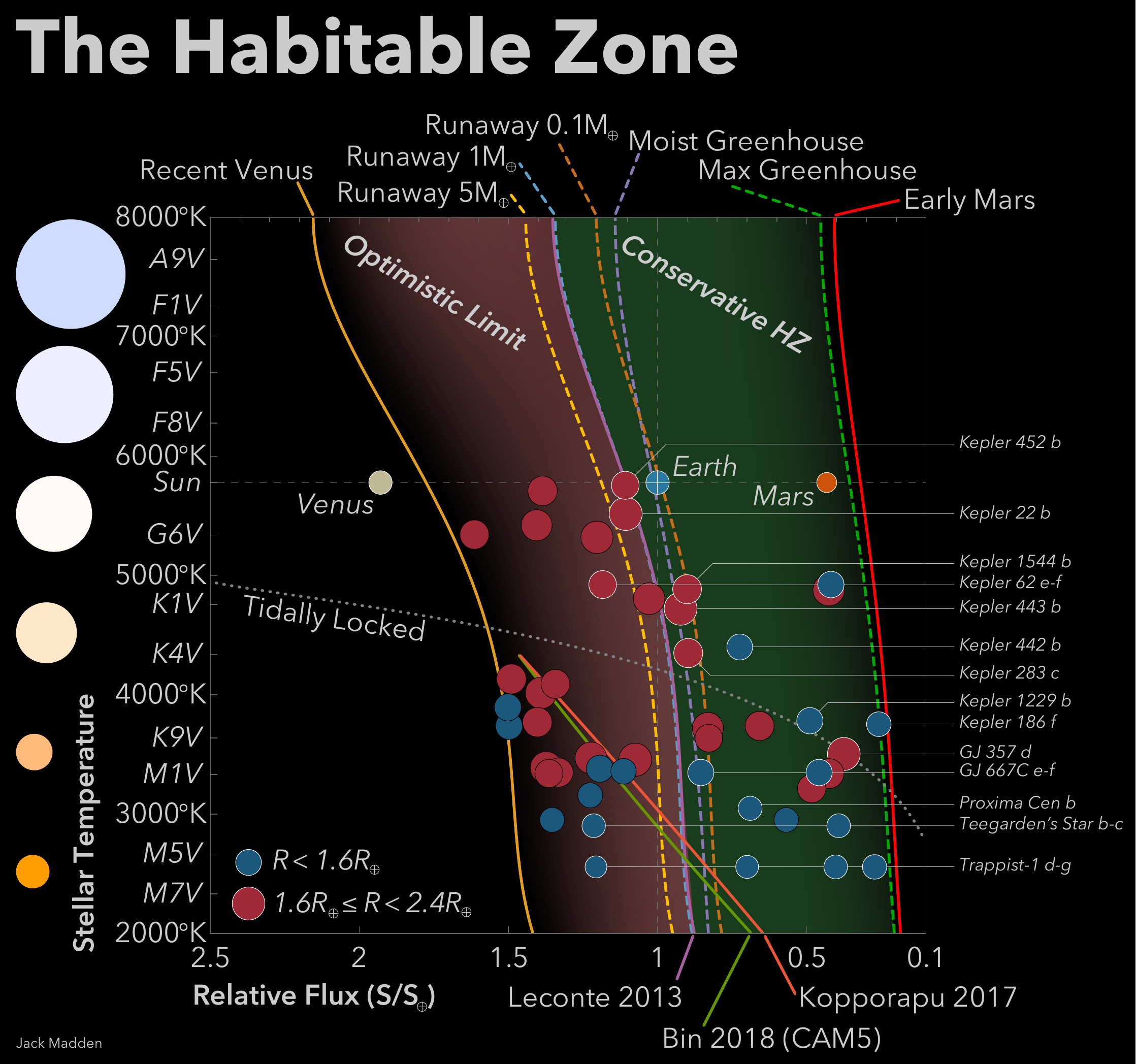}
\caption{ \singlespace The Habitable Zone.  The lines from \cite{Kopparapu2017}, \cite{Bin2018}, and \cite{Leconte2013nat} are 3D models. The other lines are from 1D models in \cite{Kopparapu2013} and \cite{Kopparapu2014}. \cite{Kasting1993} supplied the tidal locking line.  
\label{fig:HZ}}
\end{figure}

Our search for life beyond the Solar System is focused on the habitable zone. This zone is the area around a star in which a terrestrial planet could sustain liquid water on its surface. The boundaries of the habitable zone have gone through many recalculations since the 1960s \citep{Kasting1993,Kopparapu2013,Leconte2013nat,Ramirez2017,Bin2018,Kopparapu2017,GOMEZLEAL2019}, but roughly speaking at relative fluxes between Earth and Mars we have a good understanding of how a planet here could maintain its surface water. Venus shows evidence for past liquid water so at fluxes between Venus and Earth there is the possibility of liquid surface water but it may not be part of a stable climate \citep{Ingersoll1969,Kasting1984}. As shown in Figure \ref{fig:HZ}, the habitable zone is significantly modulated by star type. The hotter the star the bluer the light hitting the planet. Bluer light has more of its energy in the visible portion of the spectrum and little in infrared allowing planets to receive more total flux from blue stars but remain the same temperature. Redder, cooler stars are the opposite. Red stars emit most of their energy in the infrared so less total flux is needed to achieve a warm surface temperature. 

The light from the host star of a planet also drives the photochemistry in the atmosphere. The amount and wavelengths of ultraviolet (UV) hitting an atmosphere control the production and destruction of molecules in an atmosphere. This process controls the composition and structure of an atmosphere and greatly affects what we can see when observing an atmosphere from a telescope \cite{Rugheimer2015Spectra}. 

If the conditions are just right, we may be able to see signs of life through the observation of biosignatures. This was curiously tested by \cite{Sagan1993} in which a strong case was built for the remote detection of life on Earth using a range of measurements. The biosignatures I'll mainly be talking about in this thesis are the spectroscopic measurements of molecules in an atmosphere or on the surface that when seen together provide evidence for a biologic origin. We focus on searching for spectroscopic biosignatures because of the limitations of our current and near-future telescopes only being able to measure the bulk compositions of the atmospheres and surfaces of distant planets. 

Commonly mentioned biosignatures include the disequilibrium chemistry of oxidizing and reducing gases in an atmosphere suggesting a constant biotic source of one or more of the gases. Oxygen (O2), when present with methane (CH4) will eventually break down into carbon dioxide (CO2) and water (H2O). Methane can occur in significant quantities abiotically but for oxygen, large quantities can be produced quickly through mainly biologic sources. Seeing oxygen and methane together tells us that there is a constant supply of each molecule into the atmosphere. Seeing enough oxygen in the presence of methane indicates a large continuous source of oxygen which suggests a biological origin \citep{Lederberg1965,Lippincott1967}.

A more direct biosignature would be detecting the reflected light from the surface of the planet and matching its color to known reflectance spectra of life \citep{Hegde2013}. The vegetation red-edge is an example of such a biosignature \citep{OMalleyJames2018}. Many photosynthesizing organisms absorb visible light but reflect infrared light. The spectral boundary of this transition is often very distinct and is a strong indicator of life.

\section{The Color of Habitability}
The dark blue of an ocean, the verdant green of a forest, the faint reds in oxygen, and the heat-absorbing properties of carbon dioxide and methane. Life as we know it comes with a variety of unique colors and effects on light that, when seen in combination, provide strong evidence for biological activity. However, the challenge of making these spectroscopic observations is immense. With a vast quantity of exoplanets to choose from and more being discovered daily, the focus is now on preparing observational tools and models with the express purpose of finding life \citep{ASTRO2010}. 

Like any scientific endeavor, there is a theoretical approach and experimental approach that work in tandem to advance our understanding. The experimental approach to finding life involves building larger telescopes with more sensitive instruments and attempting observations of biosignatures. The theoretical approach includes building an understanding of the types of observations to make, potential biosignatures, and which places we should be looking. The testing of this theoretical work is done through forward and inverse modeling. Inverse modeling takes observations and attempts to reconstruct the source using a library of references and our understanding of climate. We have an accurate enough idea of what different molecules look like to see an atmosphere with a complex soup of elements floating around and understand its composition through inverse modeling. Inverse modeling requires observations to be made in the first place, which we do not yet have for habitable zone planets. In lieu of observational data, forward modeling takes our understanding of climate and chemistry to create novel simulations of planets allowing us to imagine which types of environments could exist.   

\section{Forward modeling}
Forward modeling gives us insight into the types of planets that could exist within our understanding of physics. Through the process of exploring large parameter spaces, we can discover what makes worlds suitable for life and which biosignatures are most likely to be present. When it comes to modeling entire planets there are a lot of parameters available to us. Models tend to make deliberate assumptions to assure simulations run quickly while remaining accurate. The goal is to simulate climate, not weather, as many small scale atmospheric features can be ignored leaving general processes that keep an atmosphere stable. 

One major simplification that can be taken advantage of is reducing the dimensionality of the model. Instead of modeling the entire atmosphere of a planet only a single vertical dimension of the atmosphere is simulated. For these 1D models, the parameters and calculations are constructed to provide an approximate look at the atmosphere as if the planet was modeled in 3D then averaged. Reducing the simulation to one dimension dramatically increases the speed of the model, but sacrifices a detailed understanding of the planet and limits where the model can be applied. One-dimensional models are best used to explore massive parameter spaces for trends in climate because of their ability to quickly model the average climate of a planet using a wide range of free parameters. While each individual simulation may not be applicable to a specific situation, the result of hundreds of models with slightly altered parameters reveals the strength and direction of an effect on climate.

In the search for habitable planets, the parameters we are interested in exploring are centered around maintaining temperatures and pressures suitable for liquid surface water. This may include altering planet-star distance, surface pressure, greenhouse gas abundance, surface albedo, or star type, among many others \citep{Kozakis2018,Kozakis2019,OMalleyJames2018,Rugheimer2015Surface}. By exploring how these parameters affect the climate of a planet in the habitable zone of its host star, we can determine where best to look for habitable planets, what their atmospheres will look like, and which exoplanets we have found may be the best targets for intensive study. Once we have atmospheric data from a planet, we can also use our models to learn more about a planet's climate than can be directly observed. 

The primary code used in this thesis for atmosphere modeling is based on updated versions of a 1D climate  ~\citep{KastingAckerman1986,Pavlov2000,HaqqMisra2008}, and a 1D photochemistry model ~\citep{Pavlov2002,Segura2005,Segura2007}. This code has a long history of being used for exoplanet modeling and adaptations have been used in the most cited constructions of the habitable zone \citep{Kasting1993,Kopparapu2013}.

\section{Thesis work summary}
Beginning with \cite{Madden2018Abio} in chapter \ref{c:chap2} this thesis will start by looking at how we have been able to use the bodies in our Solar System as a reference guide for exploring exoplanets. In chapter \ref{c:chap3}, \cite{Madden2020mnras} will show how surface albedo influences the climate of potentially habitable planets. We will take a break from exoplanet science with \cite{Madden2020vr} in chapter \ref{c:chap4} by examining how virtual reality can be used in physics and astronomy education before getting back to \cite{Madden2020apjl} in chapter \ref{c:chap5} to look at the reflection and emission spectra of exoplanets with different surfaces.

\subsection{The (exo)Solar System}
The Solar System contains a wide variety of bodies from dusty red Mars and icy white moons to orange gas giants and a very special green planet with a blue sky. We have a much more complete understanding of the bodies in our Solar System compared to any exoplanet. By treating the spectra and albedos of objects in our Solar System as exoplanets we can create a reference guide for comparing new observations of exoplanets to bodies we know well. 

In \cite{Madden2018Abio}, we used past observations to calculate the geometric albedos and spectra for 19 different Solar System bodies as if they orbited F, G, K, and M-stars. This allowed us to create a large database of absolute spectra for use as references in the planning and follow-up observations of large ground-based and space-based telescopes. We also examined the photometric colors of each object as we changed its host star. As a result, we determined the optimal filters in the visible and near-IR to use in order to distinguish the bodies by surface type (rocky, icy, or gaseous) depending on the host star type. 

We gathered data from many sources \citep{Lundock2009,Spencer1987,Filacchione2012,Fanale1974,Mallama2017,McCordWestphal1971,Lorenzi2016,Protopapa2008,Meadows2006,Pollack1978} in the range from 0.45-2.5 microns for our calculations. Much of the data was relatively calibrated and required additional work to calculate the absolute values for the spectra and geometric albedos. During this process, we noted several studies that contained data that were not reduced properly or contained observation contamination previously unnoticed. After calculating a reliable set of spectra, geometric albedos, and colors for all 19 objects in our study we could then simulate their observations as exoplanets around star types other than our Sun.     

Along with producing a catalog of simulated observations of the Solar System we found interesting results by looking at the data holistically. We noted that the color combination of J-K and R-J was the optimal combination in this wavelength range using standard colors to distinguish the surface types between rocky, icy, and gaseous \cite[see also][]{Krissansen2016,Cahoy2010,Lundock2009,Traub2003}. This was also true independent of star type and resolution. We also observed that the CO2 atmosphere of Venus positioned it within the icy looking objects using this color combination, and positioned the Earth within the rocky objects despite the majority of its surface being covered with water. 

A catalog of these spectra and albedos allows for observation planning and comparative planetology between exoplanets and objects in our Solar System \citep{Krissansen2016,Traub2003}. Simulated observations can be made with this data for different telescopes to gather statistics for planning and retrieval. Once observations of exoplanets are made, this catalog can be used in comparison to determine its closest analog in our Solar System. We have also shown that a simple determination of surface type can be achieved using broad filters and low resolution, opening up the possibility of smaller telescopes being able to contribute to the classification of exoplanets. 

\subsection{Climate and Photochemistry Modeling}
After looking at examples in our Solar System, the next step in understanding terrestrial exoplanets is to model their potential climates. A 1D code has the benefit of rapid processing over 3D models, meaning a large parameter space can be explored to first order. Since the true climates of exoplanets remain unknown, the first-order exploration with 1D models is a process that will provide breadth without detailing our results beyond detection capabilities.

The 1D model we have been using and updating is a coupled climate and photochemistry model. This allows for the simulation of atmospheres around stars unlike our Sun where the UV environment is significantly different. The code iterates between balancing the pressure, temperature, and distribution of molecules with the photochemistry induced by an input stellar spectrum. 

We have made substantial updates to this code's usability and physics. It is now possible to parallelize the code across all available CPUs and automate the input file generation allowing for rapid development and deployment of parameter searches. The data handling of the code has also been streamlined to convert the data for input into a code used for observational analysis. In addition to improvements that reduce project run-time, new physics has been added to the model. With the added functionality to treat the surface as a wavelength-dependent albedo instead of a constant value, we have opened a new parameter space and begun exploring it. 

Several new stellar spectra have also been added. Stellar activity and UV can have a large impact on a planet's atmosphere even when the flux differences seem small \citep{Rugheimer2015Surface}. More accurate models of the stars used in the code provide needed confidence for analyzing the output of photochemically altered species. This issue is particularly in need of attention for M-stars and well-studied systems such as Proxima Centauri and Trappist-1 due to the relatively high levels of activity of cooler stars.   

\subsection{Surfaces of Habitable Exoplanets}
With an updated 1D code, there have been several avenues opened to understand habitability in more detail. In \cite{Madden2020mnras} we take advantage of these updates by looking at the effect of surface color on habitability across star type. 

Surface reflection plays a critical role in a planet's climate due to the varying reflection of incoming starlight on the surface depending on the surface composition. 1D models commonly used to simulate terrestrial exoplanet atmospheres, such as ExoPrime and ATMOS \citep{Segura2010,Segura2005,Segura2003,Arney2016,Rugheimer2018,Rugheimer2013}, use a single, wavelength-independent albedo value for the planet, which has been calibrated to reproduce present-day Earth conditions for present-day Sun-like irradiance. 

Even though this average value for Earth's albedo is well calibrated to reproduce a similar climate to the wavelength-dependent albedo for present-day Earth orbiting the Sun, the particular average value will vary strongly if the incident stellar SED changes from a Sun-analog to a cool M-star or hotter F-star. For example, the cooling ice-albedo feedback is smaller for M-stars than for G-stars because ice reflects strongest in the visible shortwave region, whereas cool M-stars emit most of their energy in the red part of the spectrum \citep{Shields2013}. The climate conditions driving the snowball-deglaciation loop in models show a dependence on the stellar type, which has an impact on long-term sustained surface habitability \citep{Shields2014}. This match or mismatch of the surface albedo to the SED of a star can lead to a strong difference in heating or cooling of the planet. 

No study has explored the feedback of a range of different surfaces on the climate, photochemistry, surface habitability, and the resulting spectra of terrestrial planets in the HZ orbiting a wide range of stellar host stars. The modeled climate of a planet with a flat surface albedo only responds to differences in the total incident flux received, whereas a planet with a non-flat surface albedo responds to the wavelength dependence of that flux. Thus a wavelength-dependent surface albedo is critical to assess the differing efficiencies of incoming stellar SED to heat or cool the planet. Depending on the surface, the effectiveness changes, and can not be captured with one single value for the albedo at all wavelengths for stars with different SEDs.

\subsection{Eyes on atmosphere}
A major component of forward modeling is linking the simulations to real observations. After parameters are decided and a planet is constructed within the model we can take the output and calculate what the planet's spectra would look like from Earth. By taking advantage of the speed 1D models offer we can calculate spectra for hundreds of planets to build a database of references for use in other aspects of exoplanet science \citep{Lin2020,Kaltenegger2020,Kozakis2020}. Simulated spectra can be used in planning and optimizing observations, training atmospheric retrieval algorithms, and provides a field guide for comparison to real observations.

In \cite{Madden2020apjl}, we follow from the previous modeling work for different surfaces \citep{Madden2020mnras} and use a radiative transfer code to generate reflectance and emission spectra for each case. This allowed us to cross-compare the spectra to show which cases show the most distinctive features and provide the best opportunity to observe biosignatures. The database of spectra we have created is made from planet models with 30 different surfaces around 12 types of host stars. Our high resolution is applicable to the observations that will be made in the near future by ground-based telescopes like the Giant Magellan Telescope (GMT), Thirty Meter Telescope (TMT), and Extremely Large Telescope (ELT), and concept telescopes like HabEx, LUVOIR, and Origins \citep{Arney2018,Snellen2017,Snellen2015,Rodler2014,Brogi2014,Fischer2016,LopezMorales2019}. 

As we make our first attempts at finding evidence for life in the atmospheres of potentially habitable planets we must have strong theoretical backing to support the data. Building an understanding of our Solar System, exploring the effects of surface albedo, and calculating new spectra for future reference contribute to the search for life by building up a stronger theoretical framework around the observations we plan to make.

\subsection{Physics Education Research}
Along with exoplanet science, this thesis contains research on education in astronomy using virtual reality. Generally regarded as the oldest natural science, astronomy education has come a long way. Recently, virtual reality (VR) has entered the scene as a potential way to immerse students in a learning experience like never before. Before schools and universities make a large investment in VR technology research needs to be done to determine the circumstances in which VR is beneficial, to what degree it is beneficial and simply if it is beneficial at all for students. 

VR allows users to enter a simulated environment where perspectives, physics, time, and space can be altered to a point only limited by the imagination of the creators. This sandbox allows for educators to build an interactive classroom capable of showing concepts and phenomena in ways never before accessible in a traditional classroom. Just like how desktop computers have been able to show students simulations and play with physics in a virtual program, VR is equally capable of this with the additional feature of 3D immersion in the program. But does that make for a better learning experience?

To test this, we built an immersive 3D environment to recreate the classic Moon phase demonstration involving a ball and a light source to simulate the Moon phases. The environment was build by our team using the Unity game engine for the Oculus Rift. A detailed Earth-Moon-Sun system was created with accurate orbits, inclination, surface textures, and background stars. The simulation allowed users to alter time by dragging the Moon through its phases from three different vantage points in the system. 

After months of program development, we recruited participants to take a test on their Moon phase knowledge before experiencing our VR simulation and afterward to compare with results to students who experienced the traditional analog demonstration. This work provides one of the most comprehensive looks into VR as a teaching tool in physics and provides detailed guidelines for conducting future research in this field and recommendations for using VR in the classroom.

%% file: 2_SolarSystem/ms.tex

\section*{Abstract}
We present a catalog of spectra and geometric albedos, representative of the different types of Solar System bodies, from 0.45 to 2.5 microns. We analyzed published calibrated, un-calibrated spectra, and albedos for Solar System objects and derived a set of reference spectra and reference albedo for 19 objects that are representative of the diversity of bodies in our Solar System. We also identified previously published data that appears contaminated. Our catalog provides a baseline for comparison of exoplanet observations to 19 bodies in our own Solar System, which can assist in the prioritization of exoplanets for time intensive follow-up with next generation Extremely Large Telescopes (ELTs) and space based direct observation missions. Using high and low-resolution spectra of these Solar System objects, we also derive colors for these bodies and explore how a color-color diagram could be used to initially distinguish between rocky, icy, and gaseous exoplanets. We explore how the colors of Solar System analog bodies would change when orbiting different host stars. This catalog of Solar System reference spectra and albedos is available for download through the Carl Sagan Institute. 

\section{Introduction}
\begin{figure}
\centering
\includegraphics[width=0.95\textwidth]{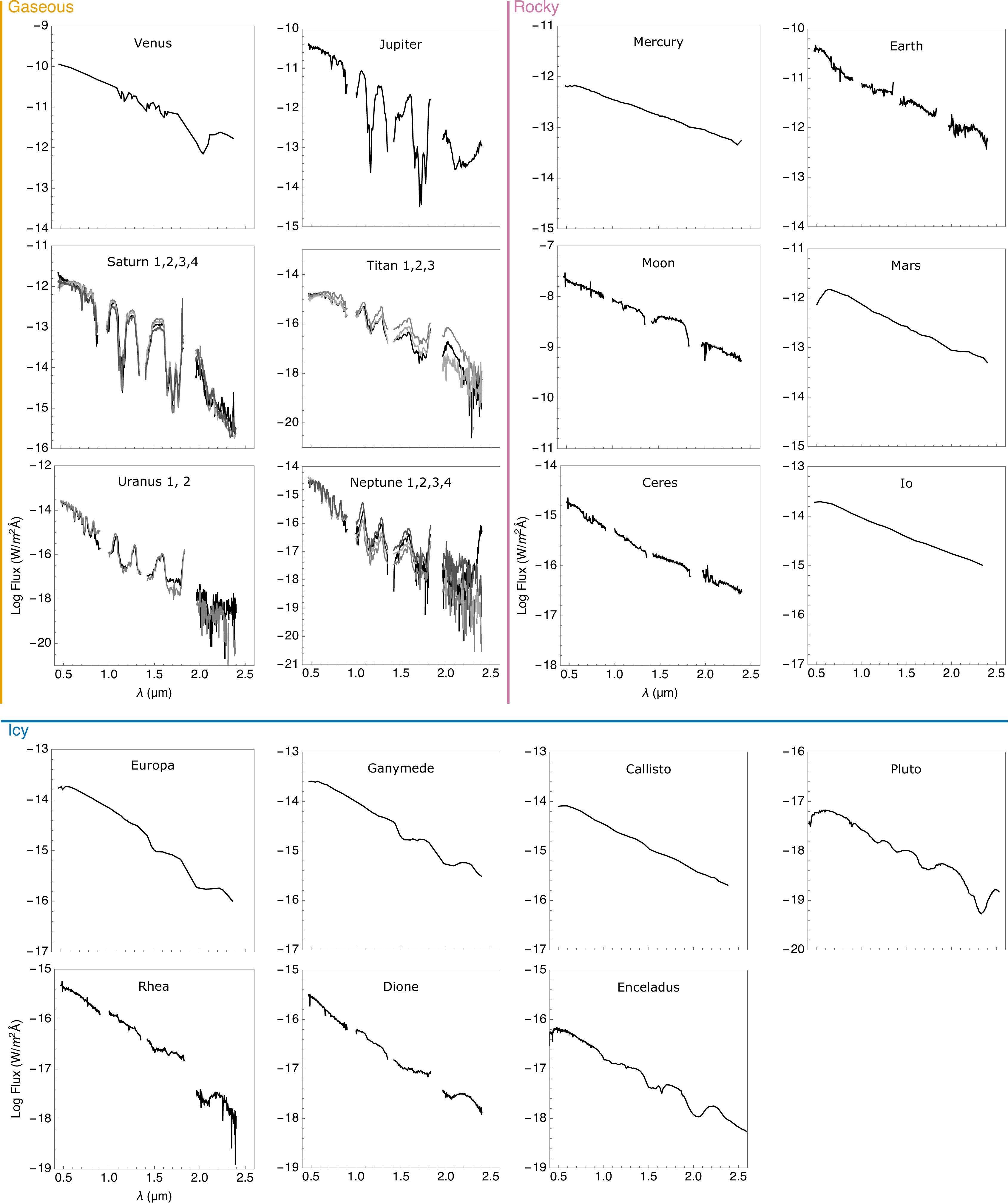}
\caption{ \singlespace Spectra for 19 Solar System bodies for Ceres, Dione, Earth, Jupiter, Moon, Neptune, Rhea, Saturn, Titan, Uranus (albedos calculated in this paper based on un-calibrated data by \cite{Lundock2009}), Callisto \citep{Spencer1987}, Enceladus \citep{Filacchione2012}, Europa \citep{Spencer1987}, Ganymede \citep{Spencer1987}, Io \citep{Fanale1974}, Mars \citep{McCordWestphal1971}, Mercury \citep{Mallama2017}, Pluto \citep{Lorenzi2016,Protopapa2008}, and Venus \citep{Meadows2006,Pollack1978}. Items are arranged by body type then by distance from the Sun.
\label{fig:1}}
\end{figure}

\begin{figure}
\centering
\includegraphics[width=0.92\textwidth]{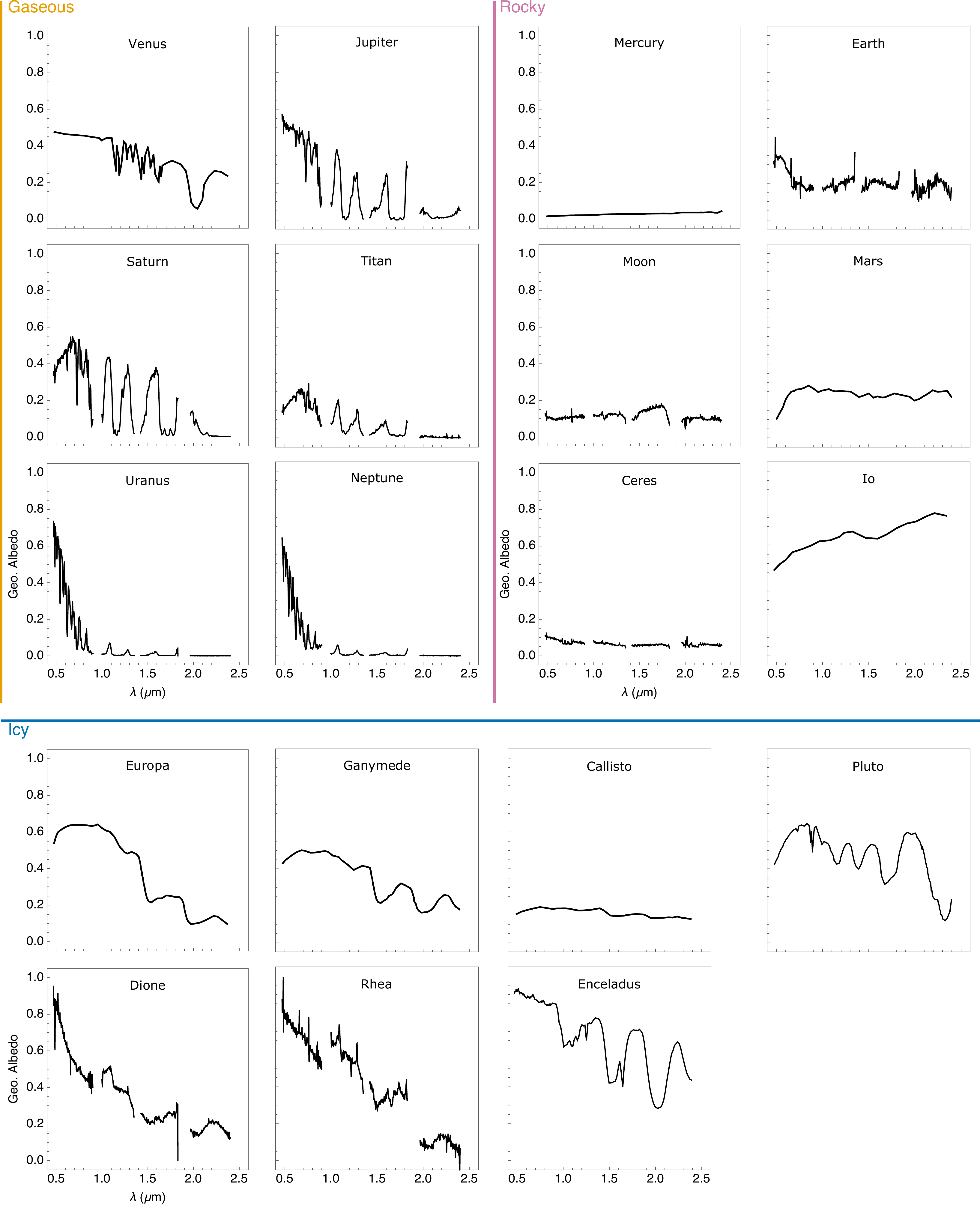}
\caption{ \singlespace Geometric albedos for 19 Solar System bodies for Ceres, Dione, Earth, Jupiter, Moon, Neptune, Rhea, Saturn, Titan, Uranus (albedos calculated in this paper based on un-calibrated data by \cite{Lundock2009}), Callisto \citep{Spencer1987}, Enceladus \citep{Filacchione2012}, Europa \citep{Spencer1987}, Ganymede \citep{Spencer1987}, Io \citep{Fanale1974}, Mars \citep{McCordWestphal1971}, Mercury \citep{Mallama2017}, Pluto \citep{Lorenzi2016,Protopapa2008}, and Venus \citep{Meadows2006,Pollack1978}. Items are arranged by body type then by distance from the Sun.  
\label{fig:2}}
\end{figure}

The first spectra of extrasolar planets have already been observed for gaseous bodies ~\cite[e.g.][]{Dyudina2016,Kreidberg2014,Mesa2016,Sing2016,Snellen2010}. To aid in comparative planetology exoplanet observations will require an accurate set of disk-integrated reference spectra, and albedos of Solar System objects. To establish this catalog for the solar system we use disk-integrated spectra from several sources. We use un-calibrated and calibrated spectra as well as albedos when available from the literature to compile our reference catalog. About half of the spectra and albedos we derive in this paper are based on un-calibrated observations obtained from the Tohoku-Hiroshima-Nagoya Planetary Spectral Library (THN-PSL)\citep{Lundock2009}, which provides a large coherent dataset of un-calibrated data taken with the same telescope. Our analysis shows contamination of part of that dataset, as discussed in section 2.1 and 2.2, therefore we only include a subset of their data in our catalog (see discussion 4.3). 
This paper provides the first catalog of calibrated spectra (Fig. \ref{fig:1}) and geometric albedos (Fig. \ref{fig:2}) of 19 bodies in our Solar System, representative of a wide range of object types: all 8 planets, 9 moons (representing, icy, rocky, and gaseous moons), and 2 dwarf planets (Ceres in the Asteroid belt and Pluto in the Kuiper belt). This catalog is available through the Carl Sagan Institute\footnote{www.carlsaganinstitute.org/data/}  to enable comparative planetology beyond our Solar System.  
Several teams have shown that photometric colors of planetary bodies can be used to initially distinguish between icy, rocky, and gaseous surface types \cite{Krissansen2016,Cahoy2010,Lundock2009,Traub2003} and that models of habitable worlds lie in a certain color space \citep{Krissansen2016,Hegde2013,Traub2003}. We expand these earlier analyses from a smaller sample of Solar System objects to 19 Solar System bodies in our catalog, which represent the diversity of bodies in our Solar System. In addition, we explore the influences of spectral resolution on characterization of planets in a color-color diagram by creating low resolution versions of our data. Using the derived albedos, we also explore how colors of analog planets would change if they were orbiting other host stars. Section 2 of this paper describes our methods to identify contamination in the THN-PSL data, derive calibrated spectra, albedos, and colors from the un-calibrated THN-PSL data and how we model the colors of the objects around the Sun and other host stars. Section 3 presents our results, Section 4 discusses our catalog, and Section 5 summarizes our findings. 

\section{Method}
We first discuss our analysis of the THN-PSL data and how we identified contaminated data in detail, then discuss how we derived spectra and albedo from the uncontaminated data. Finally we discuss spectra and albedo from other data sources for our catalog.
\subsection{Calibrating the Spectra of Solar System Bodies from the THN-PSL}
\begin{figure}
\centering
\includegraphics[width=0.79\textwidth]{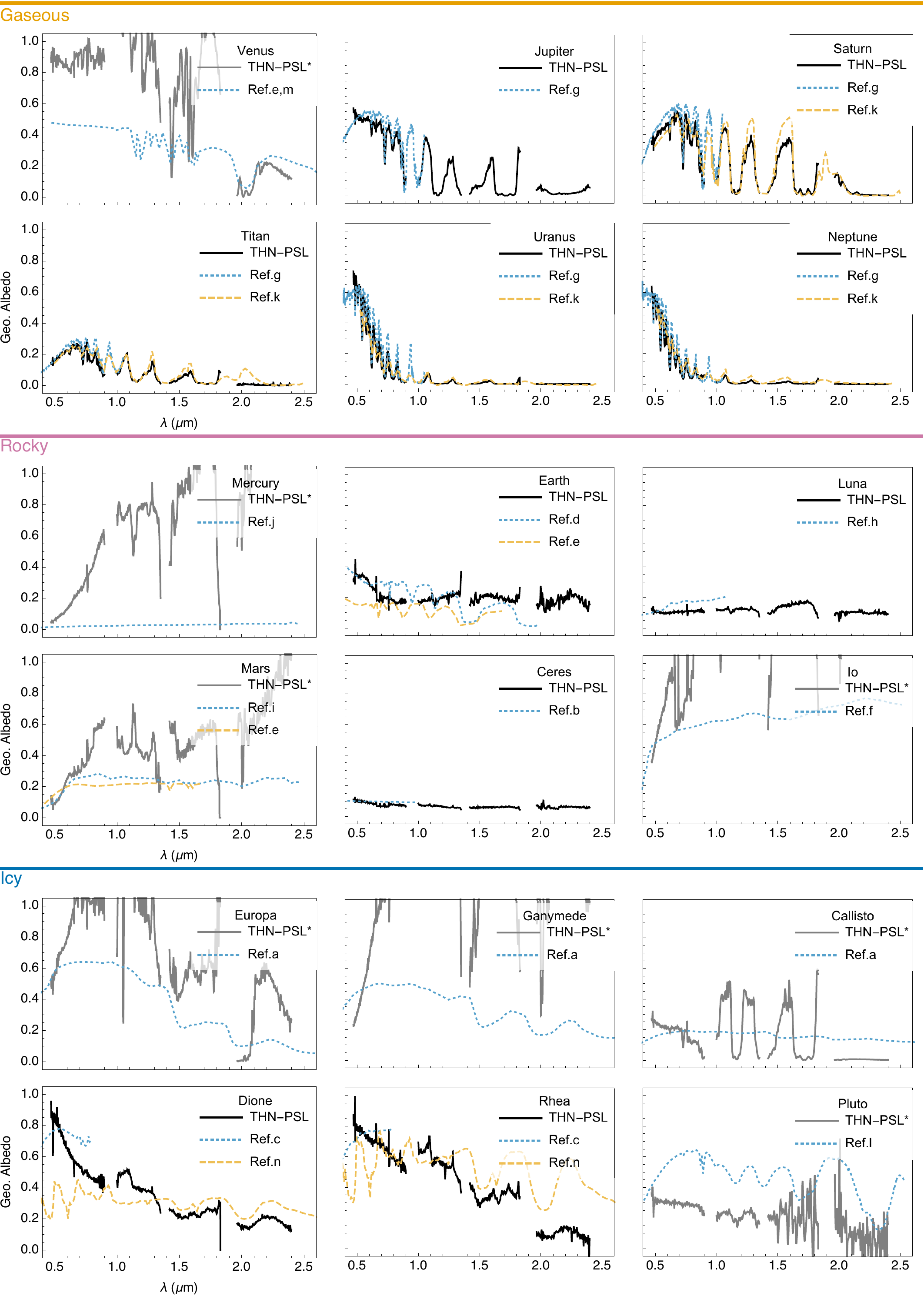}
\vspace*{-5mm}
\caption{ \singlespace A comparison of geometric albedos for the Solar System bodies in our catalog between published values and the albedo calculated from the THN-PSL data. THN-PSL data based albedos are denoted with solid lines for uncontaminated data and with an asterisk and gray line if contaminated. References for comparison albedos: \textsuperscript{a}\citep{Spencer1987}, \textsuperscript{b}\citep{Reddy2015}, \textsuperscript{c}\citep{Noll1997}, \textsuperscript{d}\citep{Kaltenegger2010}, \textsuperscript{e}\citep{Meadows2006}, \textsuperscript{f}\citep{Fanale1974}, \textsuperscript{g}\citep{Karkoschka1998}, \textsuperscript{h}\citep{Lane1973}, \textsuperscript{i}\citep{McCordWestphal1971}, \textsuperscript{j}\citep{Mallama2017}, \textsuperscript{k}\citep{FinkLarson1979}, \textsuperscript{l}\citep{Lorenzi2016,Protopapa2008}, \textsuperscript{m}\citep{Pollack1978}, \textsuperscript{n}(Cassini VIMS – NASA PDS). Items are arranged by body type then by distance from the Sun.  
\label{fig:3}}
\end{figure}

The THN-PSL is a collection of observations of 38 spectra for 18 Solar System objects observed over the course of several months in 2008. The spectra of one of the objects, Callisto, was contaminated and could not be re-observed, while the spectrum for Pluto in the database is a composite spectrum of both Pluto and Charon. We analyzed the data for the 16 remaining Solar System objects for additional contamination and found 6 apparently contaminated objects among them, leaving 10 objects in the database that do not appear contaminated. Their albedos are similar to published values in the literature for the wavelength range such data is available for. We show the derived albedos for both contaminated and uncontaminated data from the database in Fig. \ref{fig:3}, compared to available values from the literature for these bodies.

The THN-PSL data were taken in 2008 using the TRISPEC instrument while on the Kanata Telescope at the Higashi-Hiroshima observatory. TRISPEC \citep{Watanabe2005} splits light into one visible channel and two near-infrared channels giving a wavelength range of 0.45-2.5$\mu m$ . The optical band covered 0.45-0.9$\mu m$  and had a resolution of $R=\lambda/\Delta \lambda=138$. The first IR channel has a coverage from 0.9-1.85$\mu m$  and had a resolution of $R=142$. The second IR channel has a coverage from 1.85-2.5$\mu m$ and had a resolution of $R=360$. Note that the slit subtends 4.5 arcseconds by 7 arcmin meaning that spectra for larger bodies such as Saturn and Jupiter were not disk integrated (see discussion). 
As discussed in the original paper, all spectra are unreliable below 0.47$\mu m$  and between 0.9-1.0$\mu m$  from a dichroic coating problem with the beam splitters. Near 1.4$\mu m$ and 1.8$\mu m$ the Earth’s water absorption degrades the quality and beyond 2.4$\mu m$  thermal contamination is an issue. These wavelength regions are grayed out in all relevant figures in our paper but do not influence our color analysis, due to the choice of filters. The raw data available for download includes all data points. The THN-PSL paper discusses several initial observations of moons that were contaminated with light from their host planet rendering their spectra inaccurate (080505 Callisto, 081125 Dione, 080506 Io, and 080506 Rhea). These objects (with the exception of Callisto) were observed again and the extra light was removed in a different fashion to more accurately correct the spectra \citep{Lundock2009}. Callisto was not re-observed and therefore the THN-PSL Calisto data remained contaminated (Fig. \ref{fig:3}).
The fluxes of the published THN-PSL observations were not calibrated but arbitrary normalized to the value of 1 at 0.7$\mu m$. This makes the dataset generally useful to compare the colors of the uncontaminated objects, as shown in the original paper, but limits the data’s usefulness as reference for extrasolar planet observations because geometric albedos can only be derived from calibrated spectra. The conversion factors used in the original publication were not available (Ramsey Lundock, private communication). 
However, in addition to the V magnitude, the THN-PSL gives the color differences: V-R, R-I, R-J, J-K, and H-Ks for each observation, providing the R, I, and J magnitudes. Therefore, we used the published V, R, I, and J magnitudes to derive the conversion factor for each spectrum to match the published color magnitudes and to calibrate the THN-PSL observations. 
\vspace{-5pt}
\begin{wrapfigure}{r}{.5\textwidth}
\vspace{0pt} 
\centering
\includegraphics[width=0.48\textwidth]{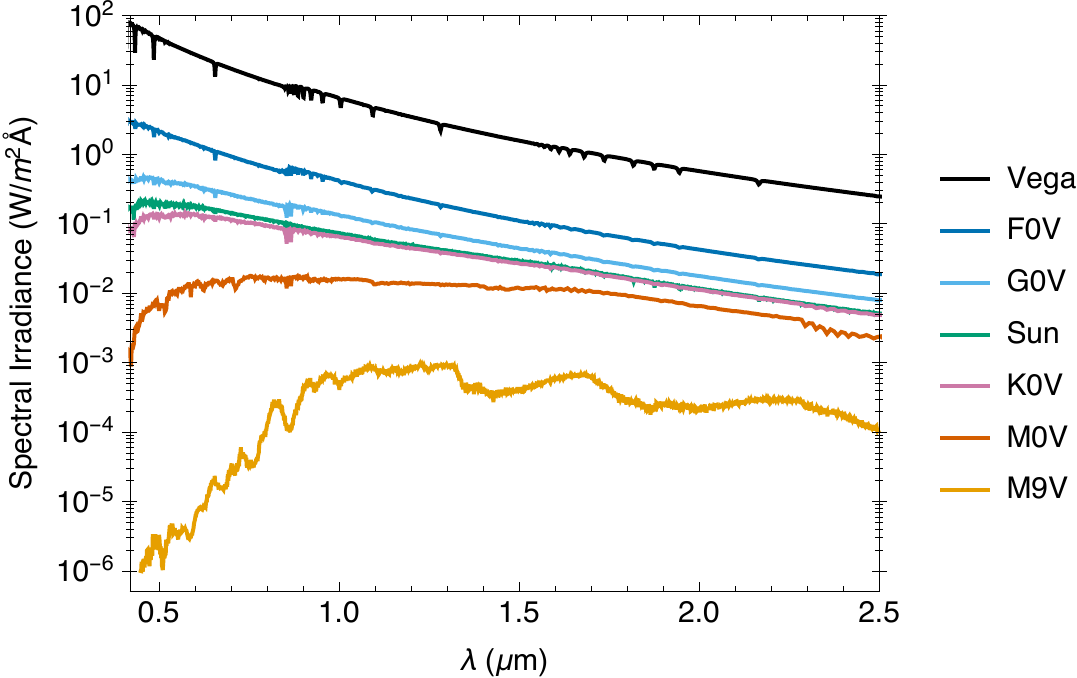}
\caption{ \singlespace Reference spectra used for calibration (Sun and Vega) and model spectra used for host stars at 1 AU (F0V, G0V, K0V, M0V, M9V). Vega was multiplied by $10^{13}$ to fit on the same plot.
\label{fig:4}}
\centering
\vspace{10pt} 
\includegraphics[width=0.48\textwidth]{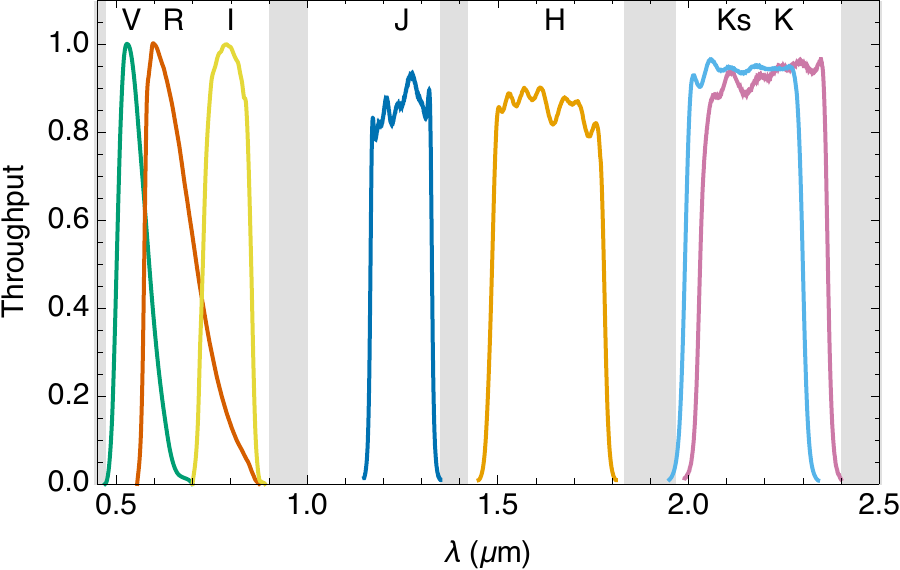}
\caption{ \singlespace Standard filters used for flux calibration and color calculations. Gray bands show the wavelength range where the observed fluxes from the THN-PSL are not reliable.   
\label{fig:5}}
\end{wrapfigure}
\vspace{-20pt} 

We define the conversion factor $k$ such that $kf_{norm}= f$   where $f_{norm}$  and $f$ are the normalized and absolute spectra respectively. Adapting the method outlined in \cite{Fukugita1995} the magnitude in a single band using the filter response, V, and the spectrum of Vega, $f_{vega}$, is given by equation (1). The spectrum of Vega, (Bohlin, 2014)\footnote{www.stsci.edu/hst/observatory/crds/calspec.html (alpha\_lyr\_stis\_008.fits) }  as well as the filter responses, are the same as in the THN-PSL publication and shown in Fig. \ref{fig:4} and Fig. \ref{fig:5} respectively. The filters we used are V \citep{JohnsonMorgan1953J}; R and I \citep{Bessell1979,Cousins1976}; J, H, K, and Ks \citep{Tokunaga2002}. Since the THN-PSL paper recorded the V, R, I, and J color magnitudes for each object we derive the conversion factor to obtain each magnitude and average them to obtain k. For example, we substitute $k_V f_{norm}= f$ in equation (2) and isolate $k_V$ as shown in equation (3) to calculate the conversion factors for the V band. The conversion factors for each band for a single object were averaged and used to calibrate the normalized spectra. 
\vspace{0pt} 
\begin{wrapfigure}{R}{.5\textwidth}
\vspace{-20pt} 
\begin{align}
m_{V} &=-2.5{\rm log}_{10} \Bigg( \frac{\int V f d\lambda}{\int V f_{vega} d\lambda} \Bigg) \\
m_{V} &=-2.5{\rm log}_{10} \Bigg( \frac{\int V k_{V} f_{norm} d\lambda}{\int V f_{vega} d\lambda} \Bigg) \\
k_{V} &= \Bigg( \frac{\int V f_{vega} d\lambda}{\int V f_{norm} d\lambda} \Bigg) 10^{-m_{v}/2.5} \\
k &= (k_V + k_R + k_I + k_J)/4 
\end{align} 
\end{wrapfigure}

We used this method to calibrate the THN-PSL data for each object. When comparing the coefficient of variation ($CV$) of the conversion factors for each body we found that the data showed two distinct groups, one with a $CV$ greater than 14\% and another with a $CV$ smaller than 6\%. We use that distinction to set the level of the conversion factor for uncontaminated spectra to $(CV>6\%)$ over the different filter bins. If the $CV$ value was in the second group $(CV>14\%)$, the data is flagged as contaminated and not used in our catalog. The nature of this contamination is unclear, it could be photometric error during the observation, excess light from the host planet or other effects that influenced the observations. The values calculated for the $k_V$, $k_R$, $k_I$, $k_J$, $k$, and the $CV$ for each observation is given in Table \ref{tab:3}.

\begin{table}[p]
\centering
\caption{\singlespace Calculated k values for each band and their average for each observation in the THN-PSL. The CV and albedo (Fig. 3) was used to determine level of reliability of the observation.\vspace{0.5cm}}
\label{tab:3}
\resizebox{\textwidth}{!}{
\singlespace
\begin{tabular}{ccccccccc}
\hline
Name      & Obs. Date & $k_V$    & $k_R$    & $k_I$    & $k_J$    & $k$      & $StDev$  & $CV$    \\ \hline
\multicolumn{9}{l}{\textit{Uncontaminated ($CV$ \textless 6\% and albedo \textless 1)}}           \\
Ceres     & 11/25/08  & 8.27E-16 & 8.27E-16 & 8.45E-16 & 8.36E-16 & 8.34E-16 & 8.44E-18 & 1.01\%  \\
Dione     & 5/5/08    & 1.34E-16 & 1.34E-16 & 1.35E-16 & 1.34E-16 & 1.34E-16 & 6.75E-19 & 0.50\%  \\
Earth     & 11/21/08  & 1.48E-11 & 1.48E-11 & 1.50E-11 & 1.50E-11 & 1.49E-11 & 1.14E-13 & 0.77\%  \\
Jupiter   & 5/7/08    & 2.17E-11 & 2.16E-11 & 2.08E-11 & 2.19E-11 & 2.15E-11 & 5.03E-13 & 2.34\%  \\
Moon      & 11/21/08  & 1.49E-08 & 1.49E-08 & 1.54E-08 & 1.51E-08 & 1.51E-08 & 2.27E-10 & 1.50\%  \\
Neptune 1 & 5/7/08    & 5.28E-16 & 5.38E-16 & 4.89E-16 & 5.25E-16 & 5.20E-16 & 2.14E-17 & 4.11\%  \\
Neptune 2 & 11/20/08  & 4.44E-16 & 4.53E-16 & 4.33E-16 & 4.42E-16 & 4.43E-16 & 8.44E-18 & 1.90\%  \\
Neptune 3 & 11/25/08  & 2.96E-16 & 3.02E-16 & 2.86E-16 & 2.97E-16 & 2.95E-16 & 6.51E-18 & 2.20\%  \\
Neptune 4 & 11/26/08  & 7.80E-16 & 7.94E-16 & 7.46E-16 & 7.76E-16 & 7.74E-16 & 2.00E-17 & 2.59\%  \\
Pluto     & 5/11/08   & 2.67E-18 & 2.67E-18 & 2.70E-18 & 2.71E-18 & 2.69E-18 & 2.10E-20 & 0.78\%  \\
Rhea      & 11/25/08  & 2.72E-16 & 2.71E-16 & 2.73E-16 & 2.73E-16 & 2.72E-16 & 1.14E-18 & 0.42\%  \\
Saturn 1  & 5/5/08    & 8.29E-13 & 8.31E-13 & 7.86E-13 & 8.49E-13 & 8.24E-13 & 2.70E-14 & 3.28\%  \\
Saturn 2  & 11/19/08  & 1.11E-12 & 1.10E-12 & 1.05E-12 & 1.13E-12 & 1.10E-12 & 3.44E-14 & 3.14\%  \\
Saturn 3  & 11/19/08  & 1.00E-12 & 9.97E-13 & 9.57E-13 & 1.02E-12 & 9.94E-13 & 2.68E-14 & 2.70\%  \\
Saturn 4  & 11/22/08  & 7.57E-13 & 7.52E-13 & 7.21E-13 & 7.62E-13 & 7.48E-13 & 1.81E-14 & 2.42\%  \\
Titan 1   & 5/5/08    & 1.12E-15 & 1.13E-15 & 1.09E-15 & 1.13E-15 & 1.12E-15 & 1.79E-17 & 1.60\%  \\
Titan 2   & 5/6/08    & 1.74E-15 & 1.72E-15 & 1.70E-15 & 1.76E-15 & 1.73E-15 & 2.47E-17 & 1.43\%  \\
Titan 3   & 11/24/08  & 1.17E-15 & 1.16E-15 & 1.12E-15 & 1.18E-15 & 1.16E-15 & 2.45E-17 & 2.11\%  \\
Uranus 1  & 5/11/08   & 3.04E-15 & 3.12E-15 & 2.76E-15 & 3.00E-15 & 2.98E-15 & 1.53E-16 & 5.12\%  \\
Uranus 2  & 11/20/08  & 3.30E-15 & 3.37E-15 & 3.19E-15 & 3.26E-15 & 3.28E-15 & 7.48E-17 & 2.28\%  \\ \hline
\multicolumn{9}{l}{\textit{Contaminated (albedo \textgreater 1)}}                                 \\
Callisto  & 5/5/08    & 7.38E-15 & 7.36E-15 & 6.90E-15 & 7.55E-15 & 7.30E-15 & 2.78E-16 & 3.81\%  \\
Europa    & 5/7/08    & 2.47E-14 & 2.46E-14 & 2.58E-14 & 2.48E-14 & 2.49E-14 & 5.57E-16 & 2.24\%  \\
Ganymede  & 11/26/08  & 4.20E-14 & 4.17E-14 & 4.52E-14 & 4.26E-14 & 4.29E-14 & 1.62E-15 & 3.79\%  \\
Io        & 11/26/08  & 1.26E-14 & 1.25E-14 & 1.41E-14 & 1.29E-14 & 1.31E-14 & 7.45E-16 & 5.70\%  \\
Mars      & 5/12/08   & 1.59E-12 & 1.57E-12 & 1.73E-12 & 1.60E-12 & 1.62E-12 & 7.56E-14 & 4.67\%  \\
Mercury   & 5/11/08   & 7.18E-12 & 7.12E-12 & 8.02E-12 & 7.31E-12 & 7.41E-12 & 4.16E-13 & 5.61\%  \\
Venus     & 11/20/08  & 1.40E-10 & 1.39E-10 & 1.43E-10 & 1.39E-10 & 1.40E-10 & 2.08E-12 & 1.49\%  \\ \hline
\multicolumn{9}{l}{\textit{Contaminated (CV \textgreater{}14\%)}}                                 \\
Earth     & 5/11/08   & 1.84E-11 & 1.79E-11 & 1.62E-11 & 2.26E-11 & 1.88E-11 & 2.71E-12 & 14.43\% \\
Moon      & 11/21/08  & 2.08E-08 & 1.84E-08 & 2.10E-08 & 3.04E-08 & 2.26E-08 & 5.33E-09 & 23.55\% \\
Uranus    & 5/7/08    & 3.28E-15 & 3.22E-15 & 2.60E-15 & 7.30E-16 & 2.46E-15 & 1.19E-15 & 48.52\% \\ \hline
\end{tabular}
}
\end{table}
\clearpage
\subsection{Albedos of Solar System Bodies}
We then derive the geometric albedo from the calibrated spectra as a second part of our analysis (see Table \ref{tab:1} and Table \ref{tab:2} for references) by dividing the observed flux of the Solar System bodies by the solar flux and accounting for the observation geometry as given in equation (5) \citep{deVancouleurs1964}. 
\begin{wrapfigure}{r}{.4\textwidth}
\begin{equation}
p=\frac{d^2 a_{b}^{2}f}{\phi(\alpha) R_{b}^{2} a_{\oplus}^{2} f_{sun}} 
\end{equation}
\end{wrapfigure}
Where d is the separation between Earth and the body, ab the distance between the Sun and the body at the time of observation, and $a_{\oplus}$ the semi major axis of Earth. $f_{sun}$ and $f$ are the fluxes from the Sun seen from Earth and the body seen from Earth respectively, $R_b$ is the radius of the body being observed, and $\phi(\alpha)$ is the value of the phase function at the point in time the observation was taken. For $f_{sun}$, we used the standard STIS Sun spectrum (Bohlin et al., 2001)\footnote{www.stsci.edu/hst/observatory/crds/calspec.html
(sun\_reference\_stis\_002.fits)}  shown in Fig. \ref{fig:4}. If the geometric albedo exceeds 1, the data is flagged as contaminated and not used in our catalog. 
Note that we also compared the spectra that were flagged as contaminated in this 2-step analysis with the available data and models from other groups (Fig. \ref{fig:3}). All flagged spectra show a strong difference in albedo for these bodies observed by other teams, supporting our analysis method (see Fig. \ref{fig:3}). 
\clearpage
\begin{table}[h!]
\centering
\caption{\singlespace Parameters for the 10 Solar System bodies from the THN-PSL we used to calculate the calibrated flux and albedos. References used for phase function and albedo. $\dagger$To obtain the proper geometric albedo for this Earthshine observation a factor of 2.38E5 is needed. *The Pluto-Charon spectrum is added for completeness. See Table \ref{tab:2} for references. \vspace{0.5cm}}
\label{tab:1}
\resizebox{\textwidth}{!}{
\singlespace
\begin{tabular}{cccccccccc}
\hline
Name      & Obs. Date & V Mag. & $d$     & $a_b$  & $R_b$ & $\alpha$ & $\phi (\alpha)$ & Phase & Albedo \\
          &           &        & (AU)    & (AU)   & (km)  & (deg.)   &                 & Ref.  & Ref.   \\ \hline
Ceres     & 11/25/08  & 8.40   & 2.405   & 2.558  & 470   & 23       & 0.34(5)         & a     & a      \\
Dione     & 5/5/08    & 10.40  & 8.97    & 9.298  & 560   & 6        & 0.88(3)         & b     & l,v    \\
Earth     & 11/21/08  & -2.50  & 0.0026† & 0.988† & 6378  & 70       & 0.5             & c     & m,n    \\
Jupiter   & 5/7/08    & -2.40  & 4.68    & 5.199  & 71492 & 10       & 0.91(8)         & d     & o      \\
Moon      & 11/21/08  & -9.30  & 0.0026  & 0.988  & 1738  & 108      & 0.05(1)         & e     & e      \\
Neptune 1 & 5/7/08    & 7.90   & 30.14   & 30.04  & 24766 & 2        & 1               & f     & o,p    \\
Neptune 2 & 11/20/08  & 7.90   & 30.145  & 30.03  & 24766 & 2        & 1               & f     & o,p    \\
Neptune 3 & 11/25/08  & 7.90   & 30.23   & 30.03  & 24766 & 2        & 1               & f     & o,p    \\
Neptune 4 & 11/26/08  & 7.90   & 30.247  & 30.03  & 24766 & 2        & 1               & f     & o,p    \\
Pluto*    & 5/11/08   & 15.00  & 30.72   & 31.455 & 1150  & 1        & 1               &       & q      \\
Rhea      & 11/25/08  & 9.90   & 9.591   & 9.36   & 764   & 6        & 0.87(2)         & b     & l,v    \\
Saturn 1  & 5/5/08    & 1.00   & 8.97    & 9.296  & 60268 & 6        & 0.76(3)         & d     & o,p    \\
Saturn 2  & 11/19/08  & 1.20   & 9.685   & 9.356  & 60268 & 6        & 1               & d     & o,p    \\
Saturn 3  & 11/19/08  & 1.20   & 9.685   & 9.356  & 60268 & 6        & 0.76(3)         & d     & o,p    \\
Saturn 4  & 11/22/08  & 1.20   & 9.6385  & 9.359  & 60268 & 6        & 0.76(3)         & d     & o,p    \\
Titan 1   & 5/5/08    & 8.40   & 8.97    & 9.3    & 2575  & 6        & 0.98(3)         & g     & o,p    \\
Titan 2   & 5/6/08    & 8.40   & 8.97    & 9.3    & 2575  & 6        & 0.98(3)         & g     & o,p    \\
Titan 3   & 11/24/08  & 8.60   & 9.5947  & 9.364  & 2575  & 6        & 0.98(3)         & g     & o,p    \\
Uranus 1  & 5/11/08   & 5.90   & 20.66   & 20.097 & 25559 & 2        & 1               & f     & o,p    \\
Uranus 2  & 11/20/08  & 5.80   & 19.742  & 20.097 & 25559 & 3        & 1               & f     & o,p    \\ \hline
\end{tabular}
}
\end{table}

\begin{table}[ht!]
\centering
\caption{\singlespace Data for the Solar System bodies from the THN-PSL dataset that were contaminated based on the shape of their calculated geometric albedo. *Note that the authors state that the Callisto data is contaminated. \vspace{0.5cm}}
\label{tab:2}
\resizebox{\textwidth}{!}{
\singlespace
\begin{tabular}{cccccccccc}
\hline
Name      & Obs. Date & V Mag. & $d$     & $a_b$  & $R_b$ & $\alpha$ & $\phi (\alpha)$ & Phase & Albedo \\
          &           &        & (AU)    & (AU)   & (km)  & (deg.)   &                 & Ref.  & Ref.   \\ \hline
Callisto* & 5/5/08   & 6.30  & 4.73   & 5.214   & 2410 & 10 & 0.60(2) & h   & r,s   \\
Europa    & 5/7/08   & 5.60  & 4.69   & 5.195   & 1565 & 10 & 0.88(5) & i,b & i,r,s \\
Ganymede  & 11/26/08 & 5.40  & 5.7518 & 5.12066 & 2634 & 8  & 0.80(5) & h   & r,s   \\
Io        & 11/26/08 & 5.80  & 5.7556 & 5.12348 & 1821 & 8  & 0.87(5) & j   & s,t   \\
Mars      & 5/12/08  & 1.30  & 1.68   & 1.6676  & 3397 & 35 & 0.58(5) & d   & u,n   \\
Mercury   & 5/11/08  & 0.00  & 0.99   & 0.37547 & 2440 & 96 & 0.11(5) & d,k & k,x   \\
Venus     & 11/20/08 & -4.20 & 1.077  & 0.72556 & 6052 & 63 & 0.4(1)  & d   & n,m   \\ \hline
\end{tabular}
}
\singlespace{References for Table \ref{tab:1} and \ref{tab:2}: \textsuperscript{a}\citep{Reddy2015}, \textsuperscript{b}\citep{BurattiVeverka1983}, \textsuperscript{c}\citep{Goode2001}, \textsuperscript{d}\citep{Irvine1968}, \textsuperscript{e}\citep{Lane1973}, \textsuperscript{f}\citep{Pollack1986}, \textsuperscript{g}\citep{Tomasko1982}, \textsuperscript{h}\citep{Squyres1981}, \textsuperscript{i}\citep{BurattiVeverka1983}, \textsuperscript{j}\citep{Simonelli1984}, \textsuperscript{k}\citep{Mallama2002}, \textsuperscript{l}\citep{Noll1997}, \textsuperscript{m}\citep{Kaltenegger2010}, \textsuperscript{n}\citep{Meadows2006}, \textsuperscript{o}\citep{Karkoschka1998}, \textsuperscript{p}\citep{FinkLarson1979}, \textsuperscript{q}\citep{Lorenzi2016,Protopapa2008} \textsuperscript{r}\citep{Spencer1987}, \textsuperscript{s}\citep{Spencer1995} \textsuperscript{t}\citep{Fanale1974} \textsuperscript{u}\citep{McCordWestphal1971}, \textsuperscript{v}(Cassini VIMS - NASA PDS), \textsuperscript{w}\citep{Pollack1978}, \textsuperscript{x}\citep{Mallama2017}}
\end{table}

\subsection{Using colors to characterize planets}
\begin{wrapfigure}{r}{.5\textwidth}
\centering
\vspace{-20pt}
\includegraphics[width=0.48\textwidth]{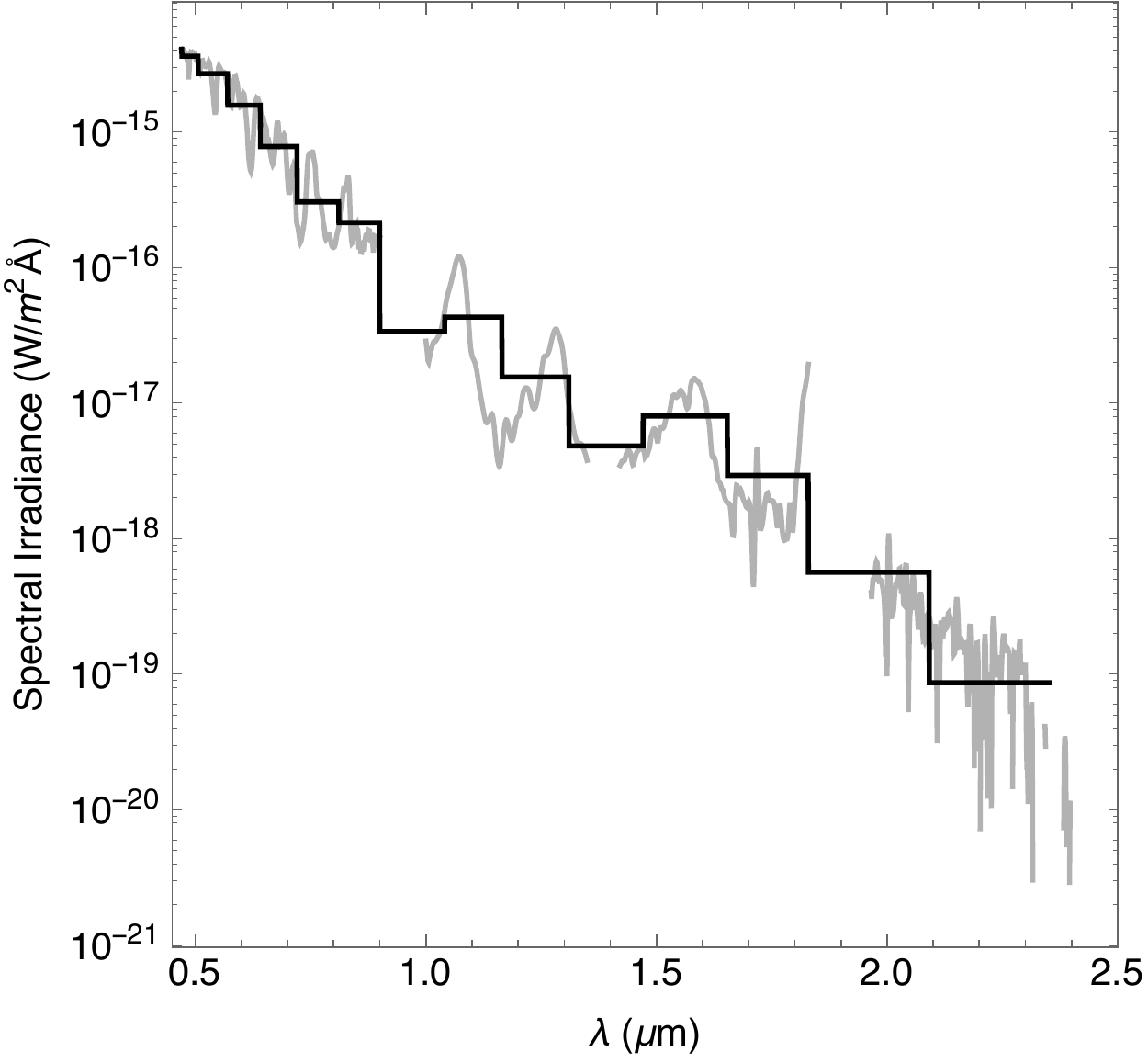}
\caption{ \singlespace An example of a reduced resolution spectrum compared to its high-resolution observations.    
\label{fig:6} }
\end{wrapfigure}

We use a standard astronomy tool, a color-color diagram, to analyze if we can distinguish Solar System bodies based on their colors and what effect resolution and filter choice has on this analysis. Several teams have shown that photometric colors of planetary bodies can be used to initially distinguish between icy, rocky, and gaseous surface types \citep{Krissansen2016,Cahoy2010,Lundock2009,Traub2003}. We calculated the colors from high and low-resolution spectra to mimicearly results from exoplanet observations as well as explored the effect of spectral resolution on the colors and their interpretation.The error for colors derived from the THN-PSL data was calculated by adding the errors used by \cite{Lundock2009} and the error accumulated through the conversion process of 6\% in the $k$ value. This gives $\Delta(J-K)=\pm0.34$ and $\Delta(R-J)=\pm0.28$ for the error values. We reduce the high-resolution data of $R=138-360$ to $R=8$ in order to mimic colors that are generated from low-resolution spectra as shown in Fig. \ref{fig:6}. The colors at high resolutions were used to determine the best color-color combination for surface and atmospheric characterization, a process that was repeated for colors derived from low resolution spectra.
\pagebreak

We also explored how to characterize Solar System analog planets around other host stars using their colors by placing the bodies at an equivalent 
orbital distance around different host stars (F0V, G0V, M0V, and M9V). We used stellar spectra for the host stars from the Castelli and Kurucz Atlas \citep{CastelliKurucz2004}\footnote{www.stsci.edu/hst/observatory/crds/castelli kurucz atlas.html (F0V, G0V, K0V, M0V)}  and the PHOENIX library \citep{Husser2013}\footnote{ http://phoenix.astro.physik.uni-goettingen.de (M9V)}  (Fig. \ref{fig:4}). As a first order approximation, we have assumed that the albedo of the object would not change under this new incoming stellar flux (See discussion).

\section{Results}
\subsection{A spectra and albedo catalog of a diverse set of Solar System Objects}
We assembled a reference catalog of 19 bodies in our Solar System as a baseline for comparison to upcoming exoplanet observations. To provide a wide range of Solar System bodies in our catalog we compiled and analyzed data from un-calibrated and calibrated spectra of previously published disk-integrated observations.
Our catalog contains spectra and geometric albedo of the 8 planets: Mercury \citep{Mallama2017}, Venus \citep{Meadows2006,Pollack1978}, Earth \citep{Lundock2009}, Mars \citep{McCordWestphal1971}, Jupiter, Saturn, Uranus, and Neptune \citep{Lundock2009}. 9 moons: Io \citep{Fanale1974}, Callisto, Europa, Ganymede \citep{Spencer1987}, Enceladus\footnote{Data available on NASA’s Planetary Data Archive: (v1640517972\_1, v1640518173\_1, v1640518374\_1)} \citep{Filacchione2012}, Dione, Rhea, the Moon, and Titan \citep{Lundock2009},  and 2 dwarf planets: Ceres \citep{Lundock2009} and Pluto \citep{Lorenzi2016,Protopapa2008}.
For the 8 planets of the Solar System, 9 moons (Callisto, Dione, Europa, Ganymede, Io, the Moon, Rhea, Titan), and 2 dwarf planets (Ceres and Pluto) we present the absolute fluxes in Fig. \ref{fig:1} and the geometric albedos in Fig. \ref{fig:2}.
\subsection{Contaminated spectra in the THN-PSL dataset}
When we derived the geometric albedo from the calibrated THN-PSL spectra as the second part of our analysis, we found that 6 objects (Io, Europa, Ganymede, Mercury, Mars, and Venus) display geometric albedos exceeding 1, indicating that the measurements are contaminated (see Table \ref{tab:2}, Fig. \ref{fig:3}). We compared the albedo of these six observations to previously published values in the literature \citep{Mallama2017,Meadows2006,Mallama2002,Spencer1995,Spencer1987,BurattiVeverka1983,Pollack1978,Fanale1974,McCordWestphal1971} and found substantial differences over the wavelength covered by the different teams (Fig. \ref{fig:3}). We list the 7 bodies with contaminated THN-PSL measurements in Table \ref{tab:2}.
\subsection{Spectra not flagged as contaminated in the THN-PSL dataset}
Table \ref{tab:1} lists the spectra of the 10 bodies from the THN-PSL database, which were not flagged as contaminated and are part of our catalog, as well as the Pluto-Charon spectrum. It shows the properties we used to calculate their albedos, once we un-normalized the un-calibrated data as well as references to previously published albedos. Note that we did not use the THN-PSL Pluto-Charon spectrum in our analysis because it is not a Pluto spectrum. Instead we use the spectrum for Pluto published by two teams  \citep{Lorenzi2016,Protopapa2008} that cover the wavelength range requires for our analysis. We show both spectra in Fig.3 for completeness. 
We compared the derived albedo of the 10 bodies from the THN-PSL database, which were not flagged as contaminated, against disk-integrated spectra and albedo from observations or models in the literature for the wavelengths available. Our derived albedos are in qualitative agreement with previously published data (Fig. \ref{fig:3}) for Ceres \citep{Reddy2015}; Dione and Rhea \cite[][Cassini VIMS,]{Noll1997}; Earth \citep{Kaltenegger2010,Meadows2006}; the Moon \citep{Lane1973}; Jupiter, Saturn, Uranus, Neptune, and Titan \citep{Karkoschka1998,FinkLarson1979}. We simulated their absolute fluxes with the same observation geometry as the THN-PSL spectra to be able to compare them (Fig. \ref{fig:3}). Note that small changes are likely due to observation geometry as well as the changes in the atmospheres over the time between observations. Giant planets have daily variations in brightness \citep{Belton1981}. For completeness we include the THN-PSL observation of the combined spectrum of Pluto and Charon and compare it to the albedo of Pluto \citep{Lorenzi2016,Protopapa2008}. We averaged several Cassini VIMS observations together and used them as references for Rhea and Dione\footnote{Data available on NASA’s PDS: 
Rhea - v1498350281\_1, v1579258039\_1, v1579259351\_1
Dione - v1549192526\_1, v1549192731\_1, v1549193961\_1}. 
\subsection{Using Color-color diagrams to initially characterize Solar System bodies}
To qualify the Solar System objects in terms of extrasolar planet observables, we consider whether they are gaseous, icy, or rocky bodies and do not distinguish between moons and planets. Thus, Titan and Venus are both gaseous bodies in our analysis since only their atmosphere is being observed at this wavelength range. Fig. \ref{fig:7} shows the spectra as well as the colors for the three subcategories in our catalog. The top panel shows gaseous bodies: Jupiter, Saturn, Uranus, Neptune, Venus, Titan. The middle panel shows rocky bodies: Mars, Mercury, Io, Ceres, Earth, and the Moon. The bottom panel shows icy bodies: Ganymede, Dione, Rhea, Callisto, Pluto, Europa, and Enceladus. Each surface type occupies its own color space in the diagram. To explore how the resolution of the available spectra and thus the observation time available would influence this classification, we reduced the spectral resolution for all spectra to $R=\lambda/\Delta \lambda$ of 8. 

We find that the derived colors of the Solar System bodies do not shift substantially (Fig. \ref{fig:8}), showing that colors derived from high and low-resolution spectra provide similar capabilities for first order color-characterization of a Solar System object. While a slight shift occurs in the color-color diagram, the three different Solar System surface types (gaseous, rocky and icy) can still be distinguished, showing that colors from low resolution spectra can be used for first order characterization of bodies in our Solar System. 

We chose a lower resolution of $R=8$ since the bin width near the K-band becomes larger than the K-band filter itself at lower resolutions. Bandwidth is directly proportional to the amount of light collected by a telescope and thus the time needed for observation. 
If low-resolution spectra could initially characterize a planet, exoplanets could be prioritized for time-intense high-resolution follow-up observations from their colors. At a lower resolution, a higher signal to noise ratio is required to achieve the same distinguishability as an observation at high resolution.  
The ratio of the integral uncertainties of two spectra at different resolutions, $\Delta I_A$ and $\Delta I_B$, \begin{wrapfigure}{r}{0.4\textwidth}
\begin{equation}
    \frac{\Delta I_A}{\Delta I_B} \propto \sqrt{\frac{N_B}{N_A}}\frac{\delta m_A}{\delta m_B}
\end{equation}
\end{wrapfigure} is proportional to the number of bins being integrated over, N, and the measurement uncertainty of each bin, $\delta m$, as shown in equation 6.
\clearpage
\begin{figure}
\centering
\includegraphics[width=0.85\textwidth]{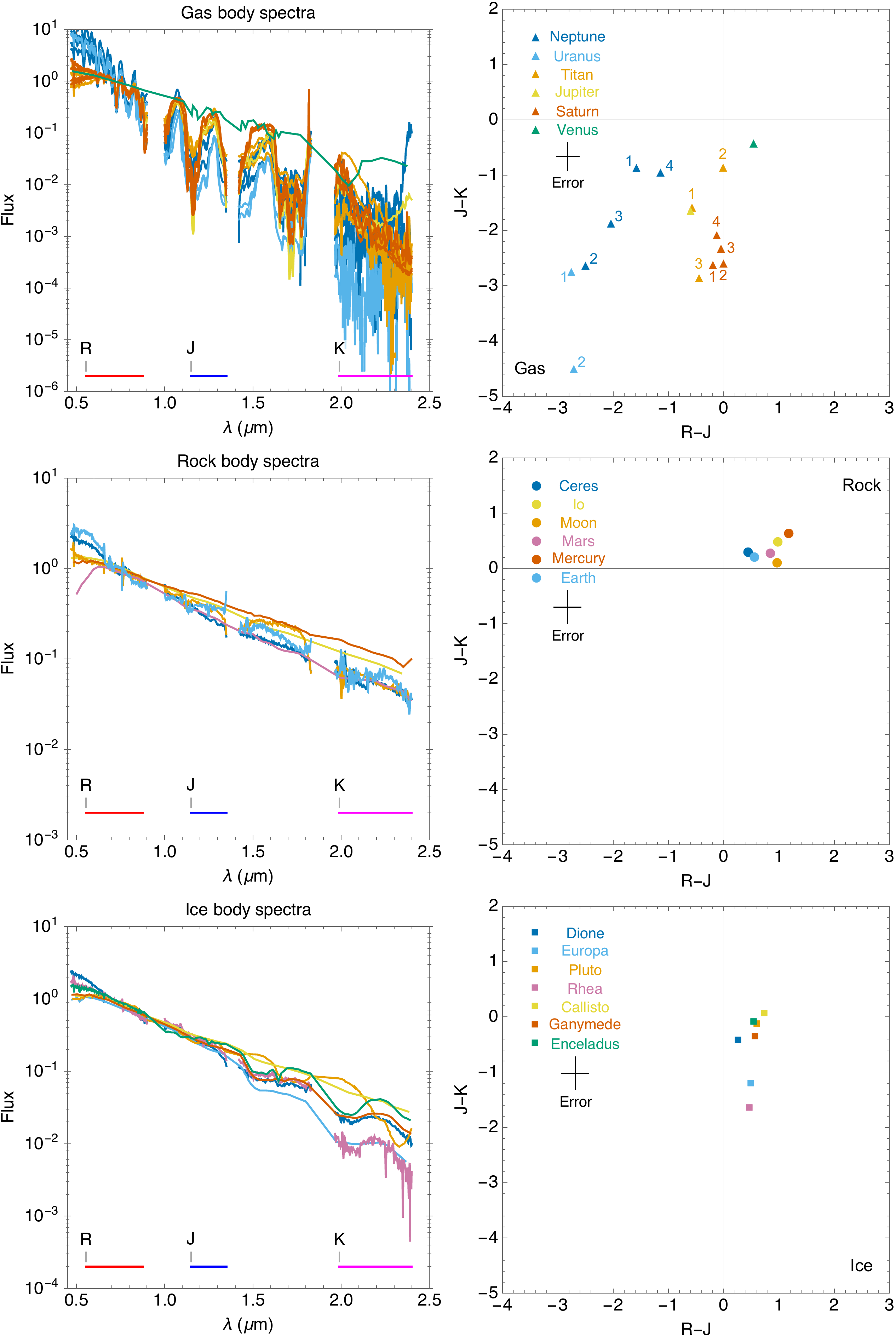}
\caption{ \singlespace Spectra and color-color diagrams for gaseous, rocky and icy bodies of the Solar System. Previously published data was used for bodies that were contaminated in the THN-PSL following the references in Fig. \ref{tab:3}.    
\label{fig:7}}
\end{figure}
\clearpage

\begin{figure}[ht!]
\centering
\includegraphics[width=\textwidth]{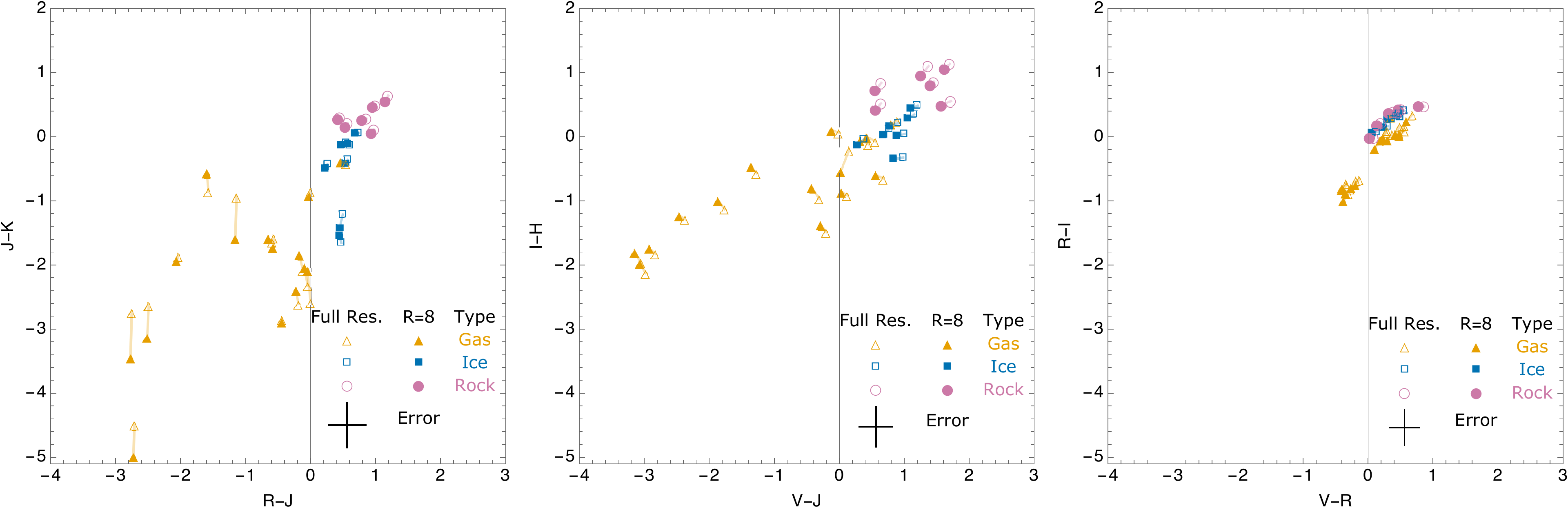}
\caption{ \singlespace Comparison of colors calculated using low (filled symbols, $R=8$) versus high resolution (non-filled symbols, $R = 138$, $142$ and $360$ for THN-PSL data) spectra for 19 Solar System bodies around the Sun. Each panel shows a different filter combination and the symbols represent the three surface types; gaseous (square), icy (triangle) and rocky (circle).
\label{fig:8}}
\end{figure}

We explored different filter combinations to best distinguish between icy, gaseous and rocky bodies. We find that R-J versus J-K colors distinguish the bodies best, \cite[see also][]{Krissansen2016,Cahoy2010,Lundock2009,Traub2003}.
If only a smaller wavelength range is available, such as V through H or V through I, Fig. \ref{fig:8} shows which alternate filter combinations can still separate the surface types. However, Fig. \ref{fig:8} shows that a wider wavelength range improves the characterization of surfaces for Solar System objects substantially. The success of using this method to characterize the Solar System reduces with narrower wavelength coverage. Long wavelengths (J and K band) especially help distinguish different kind of Solar System bodies (Fig. \ref{fig:8}). To characterize all bodies in the Solar System it is important to have wavelength coverage of the visible and near IR at a resolution that distinguishes each band.
\subsection{Colors of Solar System analog bodies orbiting different host stars}
\begin{figure}[ht!]
\centering
\includegraphics[width=\textwidth]{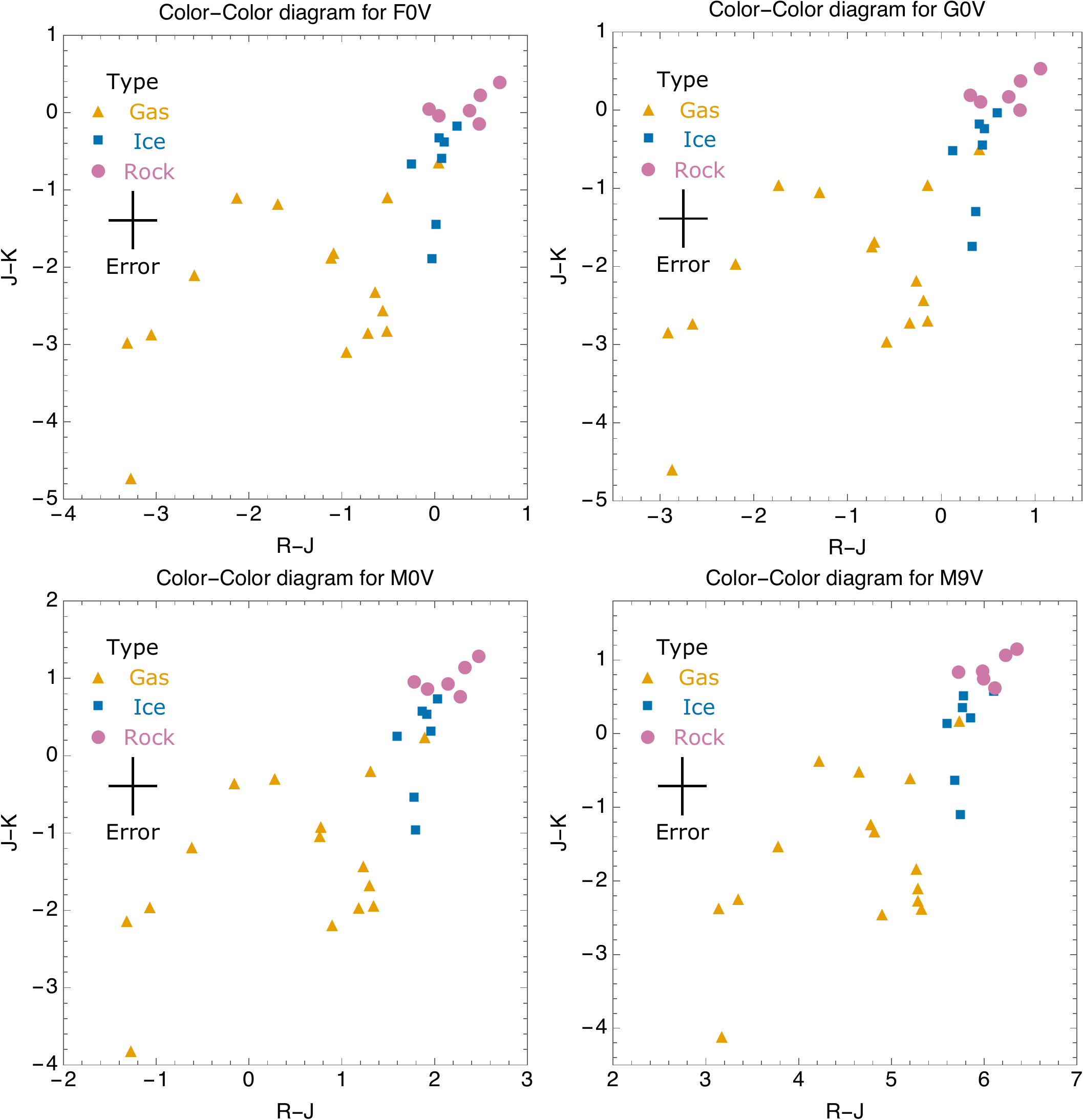}
\caption{ \singlespace Colors of Solar System bodies around different host stars. Here we show the colors of Solar System bodies for an F0V (upper-left), a G0V (upper-right), an M0V (lower-left), and an M9V (lower-right) host star.  
\label{fig:9}}
\end{figure}

To provide observers with the color-space where Solar System analog exoplanets could be found, we use the albedos shown in Fig. \ref{fig:2} to explore the colors of similar bodies orbiting different host stars. For airless bodies the albedo is a direct surface measurement, therefore that assumption should be valid for similar surface composition. For objects with substantial atmospheres that can be influenced by stellar radiation, individual models are needed to assess whether the albedo of a system's bodies would notably change due to the different host star flux. Note that Earth’s albedo would not change significantly from F0V to M9V host stars in the wavelength range 
considered here \cite[see][]{Rugheimer2013,Rugheimer2015FGKM}.
Fig. \ref{fig:9} shows the colors of the Solar System analog bodies orbiting other host stars. Because their albedo is assumed to be constant the shift closely mimics the shift in colors of the host star. For hotter host stars the colors shift to a bluer section of the color-color diagram (F0V). For cooler host stars the colors shift toward a redder portion of the color space (M0V, and M9V). This provides insights for observers into where the divisions in color-space of rocky, icy, and gaseous bodies lie depending on the host star’s spectral class.
\clearpage

\section{Discussion}
\subsection{Change in colors of Gaseous Planets}
Some gaseous bodies in the THN-PSL with multiple spectra (Uranus, Neptune, and Titan) show variations in their colors larger than the error (Fig. \ref{fig:7}). Gaseous bodies are known to vary in brightness over timescales shorter than the time between these observations \citep{Belton1981}, consistent with the THN-PSL data. This indicates that any sub-divide for gaseous bodies would be challenging from their colors alone. However the K-band is also more susceptible to photometric error as discussed in the THN-PSL paper \citep{Lundock2009}, which could add to the observed differences. Multiple uncontaminated observations across the same wavelengths for rocky or icy bodies are not available in the literature, therefore we cannot assess whether a spread in colors also exists for rocky or icy bodies, independent of viewing geometry. 
\subsection{Non disk-integrated spectra of some objects}
Due to the finite field of view of the TRISPEC instrument the observations of the Earth, Moon, Jupiter, and Saturn were not disk integrated. A disk integrated spectrum is preferred because it averages the light from the entire body instead of from a small region of its surface. The spectrograph slit was centered on the planet and aligned longitudinally for Jupiter and Saturn making the spectra as representative of the entire surface as possible. When comparing their spectra to other sources, the spectra shows a good match to disk integrated spectra \citep{Karkoschka1998,FinkLarson1979}.  This could not be done for the Moon and the Earthshine observations leading to variations in their spectra from previously published data. 
\subsection{Spectra derived from the THN-PSL dataset}
We have used several spectra of planets and moons from the THN-PSL dataset that did not appear contaminated in our analysis. The contamination of 6 objects in the THN-PSL database raises questions about the viability of the spectra in this database in general. 
We compared the 10 bodies that we used in our catalog, which were not flagged as contaminated against disk-integrated spectra and albedo from observations or models in the literature. These observations or models were not available for the whole wavelength range, thus we could not compare the full wavelength range, however the range covered shows qualitative agreement with previously published data (Fig. \ref{fig:3}) and thus we have included the spectra and albedos we derived from the un-calibrated THN-PSL data in our analysis. For Earth and the Moon time variability of the spectra can be explained because observations of the Earth and the Moon were not disk integrated, due to the spectrographic slit as discussed in 4.2. Note that for most solar system objects, reliable disc-averaged spectra for different times are lacking, which are observations that would be useful for future exoplanet comparisons.
\subsection{Similarity of the color of water and rock}
The primarily liquid water surface of the Earth is unique in the Solar System however, this is not apparent in the color-color diagrams (Fig \ref{fig:7}, \ref{fig:8}, and \ref{fig:9}). This is because water and rock share a similar, relatively flat, albedo over the 0.5-2.5$\mu m$  wavelength range. This specific color-color degeneracy for rock and water can be broken if shorter wavelength observations are available \cite[see also][]{Krissansen2016}.
\subsection[Color of CO2 atmospheres appear similar to icy surfaces]{Color of $CO_2$ atmospheres appear similar to icy surfaces}
Venus has the interesting position of being a rocky planet that has a gaseous appearance but lies amongst the colors of the icy bodies in the color-color diagrams. This is due to Venus having a primarily CO2 atmosphere which provides a similarly sloped albedo as ice in this wavelength range. This shows the limits of initial characterization through a color-color diagram. It will make habitability assessments from colors alone of terrestrial planets especially on the edges of their habitable zones very difficult since CO2 is likely to be present. Estimates of the effective stellar flux that reaches the planet or moon could help to disentangle the ice/CO2 degeneracy on the inner edge of the Habitable Zone. On the outer edge of the Habitable Zone both surface types should be present, CO2-rich atmospheres as well as icy bodies, therefore higher resolution spectra will be needed to break such degeneracy. 
\subsection{Spotting the absence of methane in a gas planet’s colors}
The absence of methane in Venus’ atmosphere makes it distinguishable from the other gaseous objects in our Solar System in the color-color diagrams. More information about the atmospheric composition of exoplanets and exomoons would be needed before we can assess whether we could derive similar inferences for other planetary systems. 
\subsection{Colors of objects that are made of ‘dirty snow’}
In the color-color diagrams, Ganymede and Callisto fall in the region between rocky and ice bodies due to their high amount of ‘dirty snow’ compared to the other bodies in the icy body category. Given the error bars in their colors, these two bodies could be placed in either the rocky or icy categories. Such rocky-icy bodies are anticipated in other planetary systems as well and should lie in the color space between the icy and rocky bodies like in our Solar System. 

\section{Conclusions}
We present a catalog of spectra, and geometric albedos for 19 Solar System bodies, which are representative of the types of surfaces found throughout the Solar System for wavelengths from 0.45-2.5 microns. This catalog provides a baseline for comparison of exoplanet observations to the most closely studied bodies in our Solar System. The data used and created by this paper is available for download through the Carl Sagan Institute\footnote{www.carlsaganinstitute.org/data/}. 
We show the utility of a color-color diagram to distinguish between rocky, icy, and gaseous bodies in our Solar System for colors derived from high as well as low-resolution spectra (Fig. \ref{fig:7} and \ref{fig:8}) and initially characterize extrasolar planets and moons. The spectra, albedo and colors presented in this catalog can be used to prioritize time-intensive follow up spectral observations of extrasolar planets and moons with current and next generation like the Extremely Large Telescopes (ELTs). Assuming an unchanged albedo, Solar System body analog exoplanets shift their position in a color-color diagram following the color change of the host stars (Fig. \ref{fig:9}). Detailed spectroscopic characterization will be necessary to confirm the provisional categorization from the broadband photometry suggested here, which is only based on planets and moons of our own Solar System.
Planetary science broke new ground in the 70s and 80s with spectral measurements for Solar System bodies. Exoplanet science will see a similar renaissance in the near future, when we will be able to compare spectra of a wide range of exoplanets to the catalog of bodies in our Solar System. 

\clearpage

\section*{Data Access}
DOI for accompanying data: ~\href{https://doi.org/10.5281/zenodo.3930986}{10.5281/zenodo.3930986} 

\section*{Acknowledgments}
We Thank Gianrico Filacchione, Ramsey Lundock, Erich Karkoschka, Siddharth Hegde, Paul Helfenstein, Steve Squyres, and our reviewers for helpful discussions and comments. The authors acknowledge support by the Simons Foundation (SCOL \# 290357, Kaltenegger), the Carl Sagan Institute, and the NASA/New York Space Grant Consortium (NASA Grant \# NNX15AK07H).

%% file: 3_surfaces/ms.tex

\section*{Abstract}
Large ground- and space-based telescopes will be able to observe Earth-like planets in the near future. 
We explore how different planetary surfaces can strongly influence the climate, atmospheric composition, and remotely detectable spectra of terrestrial rocky exoplanets in the habitable zone depending on the host star`s incident irradiation spectrum for a range of Sun-like host stars from F0V to K7V. 
We update a well-tested 1D climate-photochemistry model to explore the changes of a planetary environment for different surfaces for different host stars.
Our results show that using a wavelength-dependent surface albedo is critical for modeling potentially habitable rocky exoplanets.

\section{Introduction}
With dozens of Earth-sized planets already discovered, the next step in the search for life beyond our Solar System will be the characterization of the atmospheres of terrestrial planets and the search for signs of life on planets in the Habitable Zone (HZ). 

Gathering spectra of the atmospheres of potentially habitable exoplanets is one of the aims of several upcoming and proposed telescopes both on the ground and in space like the James Webb Space Telescope (JWST) and the Extremely Large telescopes (ELTs), such as the Giant Magellan Telescope (GMT), Thirty Meter Telescope (TMT), and the Extremely Large Telescope (ELT) and several missions concepts like Arial ~\citep{Tinetti2016}, Origins ~\citep{Battersby2018}, Habex ~\citep{Mennesson2016} and LUOVIR ~\citep{Luvoir2018}. Future ground-based ELTs and JWST are designed to be capable of obtaining first  measurements of the atmospheric composition of Earth-sized planets ~\cite[see e.g.][]{KalteneggerTraub2009,Kaltenegger2010,Stevenson2016,Barstow2016,Hedelt2013,Snellen2013,Rodler2014,Betremieux2014,Misra2014,Munoz2012}.

Surface reflection plays a critical role in a planet's climate due to the varying reflection of incoming starlight on the surface, depending on the surface composition. 1D models commonly used to simulate terrestrial exoplanet atmospheres such as, Chemclim ~\citep{Segura2010,Segura2005,Segura2003}, ATMOS ~\cite[see, e.g.][]{Arney2016}, and earlier versions of ExoPrime ~\citep{Rugheimer2018,Rugheimer2013} use a single, wavelength-independent albedo value for the planet, which has been calibrated to reproduce present-day Earth conditions for present-day Sun-like irradiance. 

Even though this average value for Earth's albedo is well calibrated to reproduce a similar climate to the wavelength-dependent albedo for present-day Earth orbiting the Sun, the particular average value will vary strongly if the incident stellar SED changes from a Sun-analog to a cool M star or hotter F star. For example, the cooling ice-albedo feedback is smaller for M stars than for G stars ~\citep{Shields2013,Abe2011} because ice reflects strongest in the visible shortwave region. In contrast, cool M stars emit most of their energy in the red part of the spectrum. The climate conditions driving the snowball-deglaciation loop in models show a dependence on the stellar type, which has an impact on long-term sustained surface habitability ~\citep{Abe2011,Shields2014,Abbot2018}. The relationship between the surface albedo and the stellar SED can lead to a substantial difference in the heating of a planet. We focus on F, G, and K-stars because M-stars pose a unique challenge for climate modeling due to the potential for planets to be tidally locked in the habitable zone and high stellar activity \citep{Airapetian2017,Johnstone2018}.

Several studies ~\citep{Kaltenegger2007,Cockell2009,Kaltenegger2010,Rugheimer2013,Rugheimer2015Spectra,Schwieterman2015,OMalleyJames2018,Rugheimer2018} have shown that surface albedo, as well as cloud coverage, is an essential factor for atmospheric as well as surface biosignature detection. 

However, no study has explored the feedback for a range of different surfaces on the climate, photochemistry, habitability, and observable spectra of planets in the HZ orbiting a wide range of stars. The modeled climate of a planet with a flat surface albedo only responds to differences in the total incident flux received whereas a planet with a non-flat surface albedo, responds to the wavelength dependence of that flux. Thus, a wavelength-dependent surface albedo is critical to assess the differing efficiencies of incoming stellar SED to heat or cool the planet. Depending on the surface, the effectiveness changes, and cannot be captured with one single value for the albedo at all wavelengths for stars with different SEDs.

Another critical issue that has not been explored is that the average surface albedo value commonly used encompasses the heating as well as cooling of clouds in addition to the reflectivity of a planet's surface. We address this by first exploring the contribution of clouds to the overall albedo for present-day Earth and separating that effect from the surface for present-day Earth. This separation is critical to be able to assess the influence of the surface environment on the planet's climate. Note that there is no self-consistent model that predicts the cloud feedback with different stellar types or stellar irradiance. Thus we keep the cloud component constant in our model comparison for different host stars, to isolate the effect of changing planetary surfaces. 

Our paper demonstrates the importance of including the wavelength-dependent feedback between a planet's surface and a planet's host star for Earth-like planets in the HZ. We focus on the change in a planet`s climate, its surface temperature as well as atmospheric species including biosignatures, which can indicate life on a planet: Ozone and oxygen in combination with a reducing gas like methane or N$_2$O ~\citep{Lovelock1965L,Lederberg1965,Lippincott1967}. Other atmospheric components we highlight are climate indicators like water and CO$_2$, which in addition to estimating the greenhouse gas concentration on an Earth-like planet, can also indicate whether the oxygen production can be explained abiotically  ~\cite[e.g.][]{DesMarais2002,Kaltenegger2017}. 
Section 2 describes our models, section 3 presents our results and section 4 a discussion. 

\section{Methods}
\subsection{Planetary Model}
A star's radiation shifts to longer wavelengths with cooler surface temperatures, which makes the light of a cooler star more efficient in heating an Earth-like planet with a mostly N$_2$-H$_2$O-CO$_2$ atmosphere ~\citep{Kasting1993}. This is partly due to the effectiveness of Rayleigh scattering, which decreases at longer wavelengths. A second effect is an increase in near-IR absorption by H$_2$O and CO$_2$ as the star's spectral peak shifts to these wavelengths. That means that the same integrated stellar flux that hits the top of a planet's atmosphere from a cool red star warms a planet more efficiently than from a hot blue star. Thus the stellar irradiance and the resulting orbital distance where a planet will show a similar surface temperature depends on the stellar host's SED. 

To establish the incident irradiation which produces similar surface temperatures for different host stars, we reduce the incident stellar flux at the moist greenhouse HZ limits for planet models with 1 Earth-mass ~\citep{Kopparapu2013} and fit it to the incident flux of present-day Earth for a G2V star to estimate the stellar irradiation for each stellar type.  The HZ is a concept that is used to guide remote observation strategies to characterize potentially habitable worlds. It is defined as the region around one or multiple stars in which liquid water could be stable on a rocky planet's surface ~\citep{Kasting1993,Kaltenegger2013,Kane2013,Kopparapu2013,Ramirez2016,Ramirez2017}, facilitating the remote detection of possible atmospheric biosignatures ~\cite[see e.g. review][]{Kaltenegger2017}.

This approach provides surface conditions similar to modern Earth of  $288K \pm 2\%$ across all stellar types for a wavelength-independent, fixed surface albedo (284 K for the K7V to 292K for the F0V host star). 

\subsection{Atmospheric Model}
For this study, we update exo-Prime to include a wavelength-dependent surface albedo. Exo-Prime ~\cite[see e.g.][]{Kaltenegger2010}, is a coupled 1D iterative radiative-convective atmosphere code with a line by line radiative transfer code, developed for rocky exoplanets. We update exo-Prime to include i) the updates in ATMOS ~\cite[see][]{Arney2016} in the climate and photochemical model as well as ii) a decoupled cloud and surface albedo and iii) wavelength-dependent albedo in all calculations, instead of a single average value.   

The code is based on iterations of a 1D climate  ~\citep{KastingAckerman1986,Pavlov2000,HaqqMisra2008}, and a 1D photochemistry model ~\citep{Pavlov2002,Segura2005,Segura2007}, which are run to convergence ~\cite[see details in][]{Segura2005}. Visible and near-IR shortwave fluxes are calculated with a two-stream approximation, including atmospheric gas scattering ~\citep{Toon1989}, and longwave fluxes in the IR region are calculated with a rapid radiative transfer model (RRTM). We use a geometrical model in which the average 1D global atmospheric model profile is generated using a plane-parallel atmosphere, treating the planet as a Lambertian sphere, and setting the stellar zenith angle to 60 degrees to represent the average incoming stellar flux on the dayside of the planet ~\cite[see also][]{SchindlerKasting2000}.A reverse-Euler method within the photochemistry code (originally developed by ~\cite{Kasting1985}) contains 220 reactions to solve for 55 chemical species. The radiative transfer code to model reflected planetary spectra in Exo-Prime was originally developed to study Earth spectra ~\citep{TraubStier1976} and later adapted for exoplanet use ~\citep{Kaltenegger2007,KalteneggerTraub2009}. We calculate light transmission at a resolution of $0.01 cm^{-1}$ from 0.4 to 2 microns providing a minimum resolving power of 100,000 at all wavelengths.

We divide the atmosphere into 100 layers for our model up to an altitude of at least 60 km, with smaller spacing towards the ground. We present the spectra at a resolution of 100 for clarity in the figures of this paper.
\vspace{-10pt}
\subsection{Stellar Spectra}
\begin{figure}[ht!]
\centering
\includegraphics[width=0.75\textwidth]{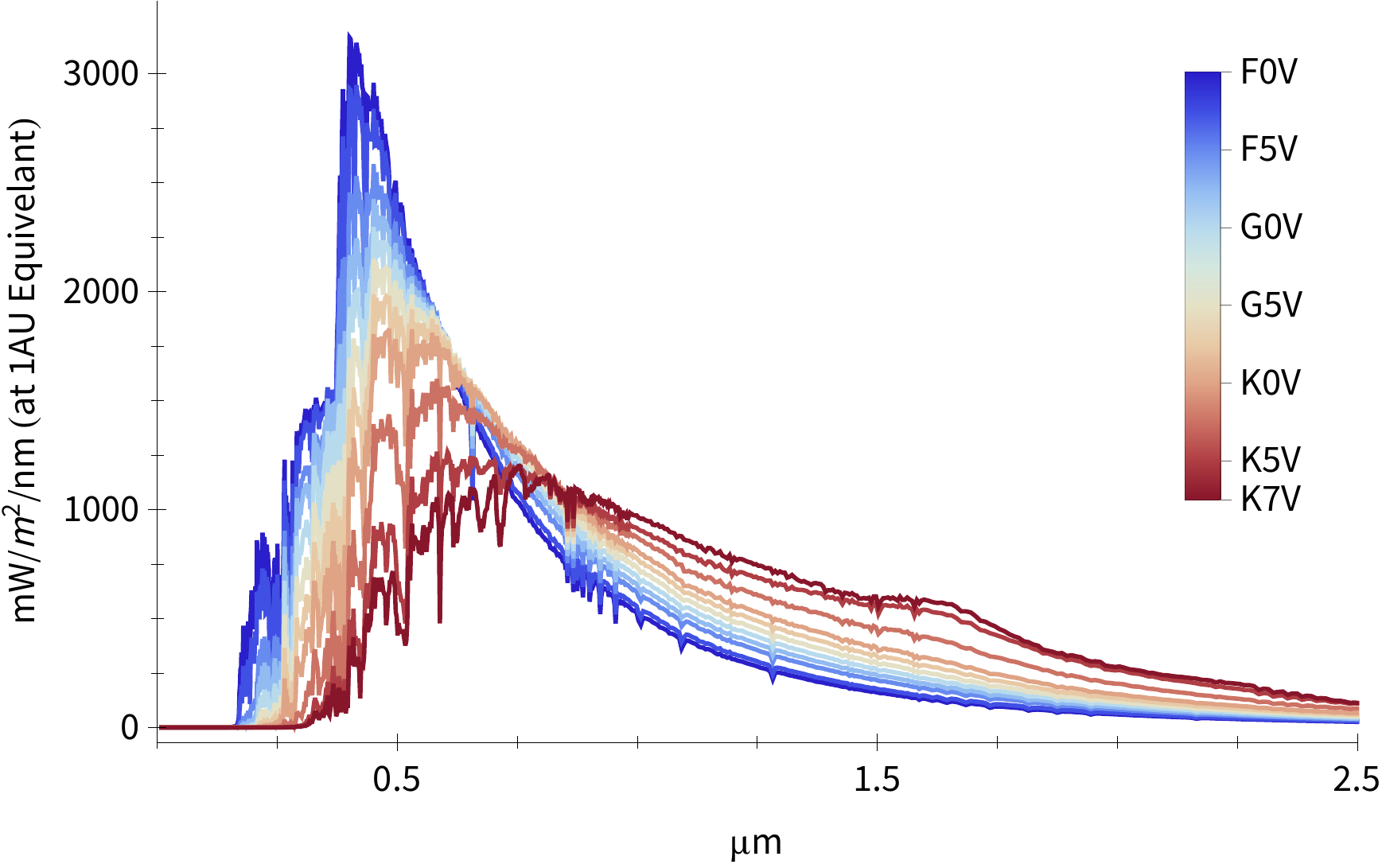}
\caption{\singlespace Incident stellar flux for our model planets for host stars with $T_{eff}$ 7,400K to 3,900K, corresponding to main sequence stars F0V to K7V from ATLAS models. 
\label{fig:stellarspectra}}
\end{figure}

The effects of wavelength-dependent surface and cloud albedo on habitability are most apparent across star type. We used ATLAS models ~\citep{CastelliKurucz2004} for the F, G, and K star input spectra (Fig.  \ref{fig:stellarspectra}). We scale the spectra from these sources to provide stellar irradiation at a model planet's position, which provides similar surface temperatures for the single wavelength-independent albedo setup (as explained in section 2.1).
\vspace{-50pt}

\subsection{Initial Conditions}
In our models, we keep the outgassing rates for H$_2$, CH$_4$, CO, N$_2$O, and CH$_3$Cl constant, and set the mixing ratios of O$_2$ to 0.21 and CO$_2$ to $3.55\times10^{-6}$, with a varying N$_2$ concentration that is used as a fill gas to reach the set surface pressure of the model  ~\cite[see also][]{Segura2005,Segura2003,Rugheimer2013,Rugheimer2015Surface,Rugheimer2018}.Note that by keeping the outgassing rates constant, lower surface pressure atmosphere models initially have slightly higher mixing ratios of chemicals with constant outgassing ratios than higher surface pressure models. The dominant parameters we vary between simulations are the host star type and planetary surface albedo. Other parameters were altered slightly to aid in a more rapid convergence of the model, such as atmosphere height, and height of convection.

\subsection{Albedos}
The surface albedo in the 1D iterative climate-photochemistry code we updated for this study ~\citep{KastingAckerman1986,Kasting1979,Zahnle2006} 
\begin{wrapfigure}{r}{.6\textwidth}
\vspace{-15pt}
\includegraphics[width=0.58\textwidth]{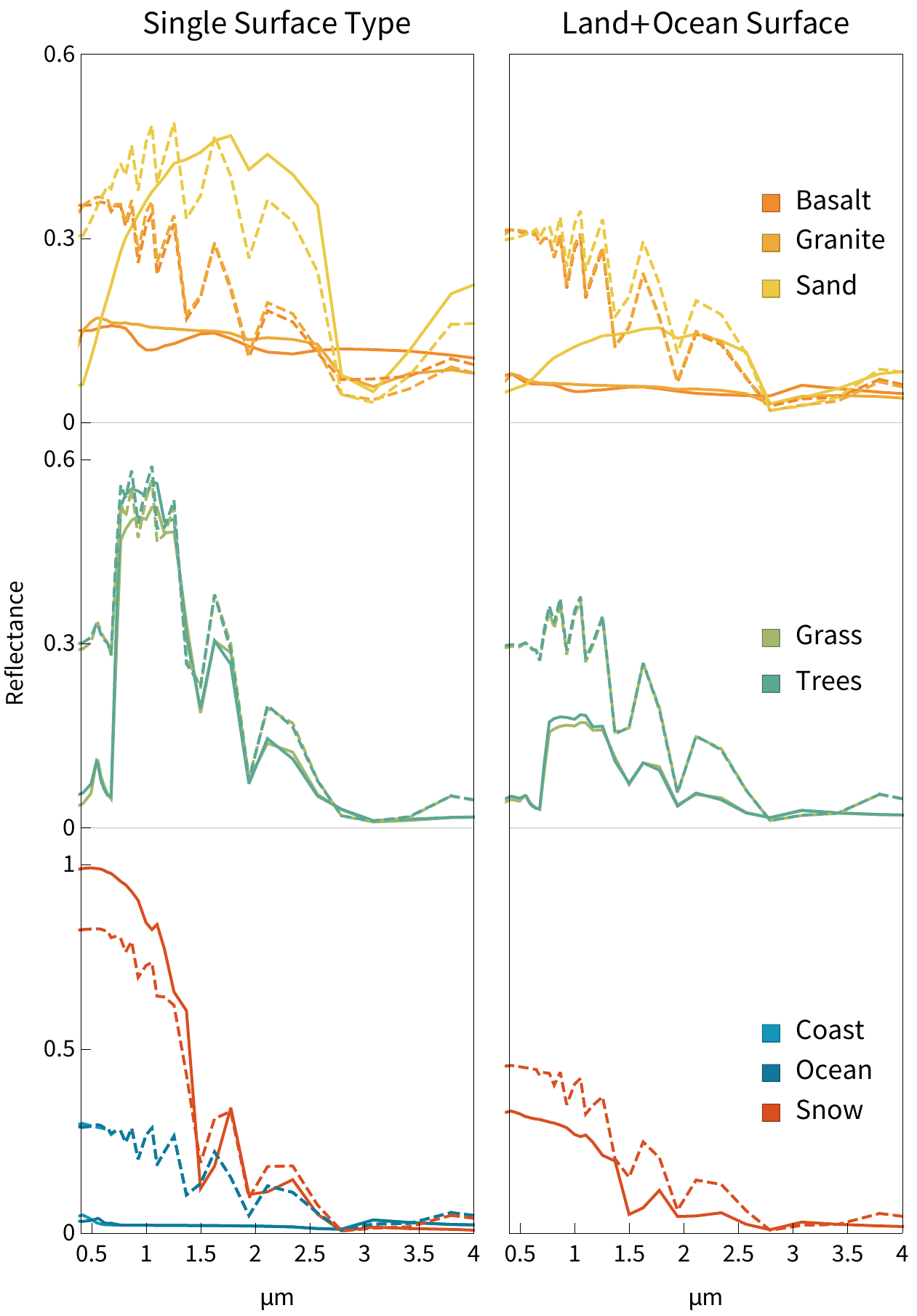}
\caption{\singlespace Surface albedos used in combination to create the modern Earth surface albedo. We also use those surfaces separately to explore the climate effects of (left) a single or (right) mixed ocean-land surface albedo. Albedos are sourced from the ASTER and USGS spectral catalogs. Mixed surfaces in this figure all contain 70\% ocean and 30\% of the specified surface.\label{fig:surfacealbedos}}
\vspace{-15pt}
\end{wrapfigure}
was a single value from 0.237$\mu$m to 4.55$\mu$m. To simulate Earth conditions for solar irradiation at Earth's orbital position with a single albedo, a value of 0.31 is used for all wavelengths ~\cite[e.g.][]{Arney2016}. To examine how different surfaces influence a planet's climate, we updated the code to read in a wavelength-dependent albedo value.

\begin{wrapfigure}{R}{.6\textwidth}
\vspace{-10pt}
\includegraphics[width=0.58\textwidth]{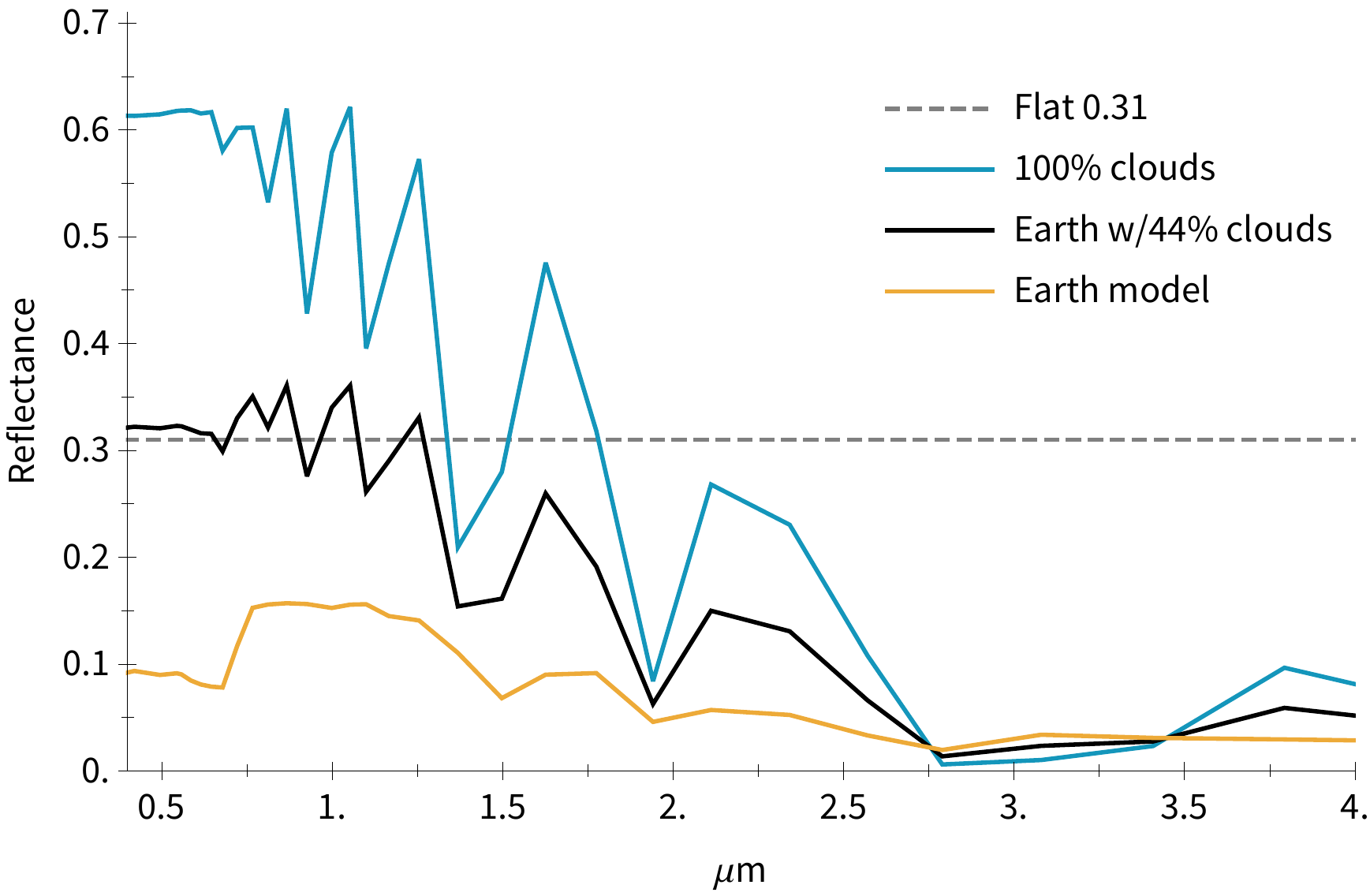}
\caption{\singlespace Difference between individual Earth single-surface albedo, average Earth albedo and flat, 0.31, albedo models used in previous studies. \label{fig:earthalbedo}}
\vspace{-10pt}
\end{wrapfigure}

Following ~\cite{Kaltenegger2007} we gathered surface albedos from the ASTER and USGS spectral libraries ~\citep{Baldridge2009aster,Kokaly2017usgs,Clark2007usgs} to create an average present-day Earth surface albedo from 8 raw albedos of snow, water, coast, sand, trees, grass, basalt, and granite (Fig. \ref{fig:surfacealbedos}). If the data was not complete in the UV, we extended the albedo constantly using the nearest value (these extensions affected regions smaller than 0.1 microns). For Earth clouds, we use the Modis 20$\mu$m cloud albedo model  ~\citep{King1997,RossowSchiffer1999}(Fig. \ref{fig:earthalbedo}), which provides an average for many clouds of different droplet size. We then used this Earth surface albedo to determine what fractional addition of clouds results in the same surface temperature as the model using a flat albedo of 0.31. 
As discussed further in the results section, we find that using 44\% cloud coverage in combination with the wavelength-dependent surface albedo of present-day Earth reproduces the same surface temperature and climate as the original albedo treatment of a single value of 0.31.

Note that the cloud fraction mimics the combined effect of warming and cooling due to clouds, which is why the fraction is lower than modern Earth's actual cloud fraction, which is between 50\% and 70\%  ~\citep{Stubenrauch2013}. Because of unknown cloud feedback for host stars with different SEDs, we then keep the cloud properties, reflectivity, and coverage constant to explore the influence of the surface albedo on the planet's climate and spectra for different host stars.

\section{Results}
\begin{sidewaysfigure*}
      \includegraphics[width = \textheight]{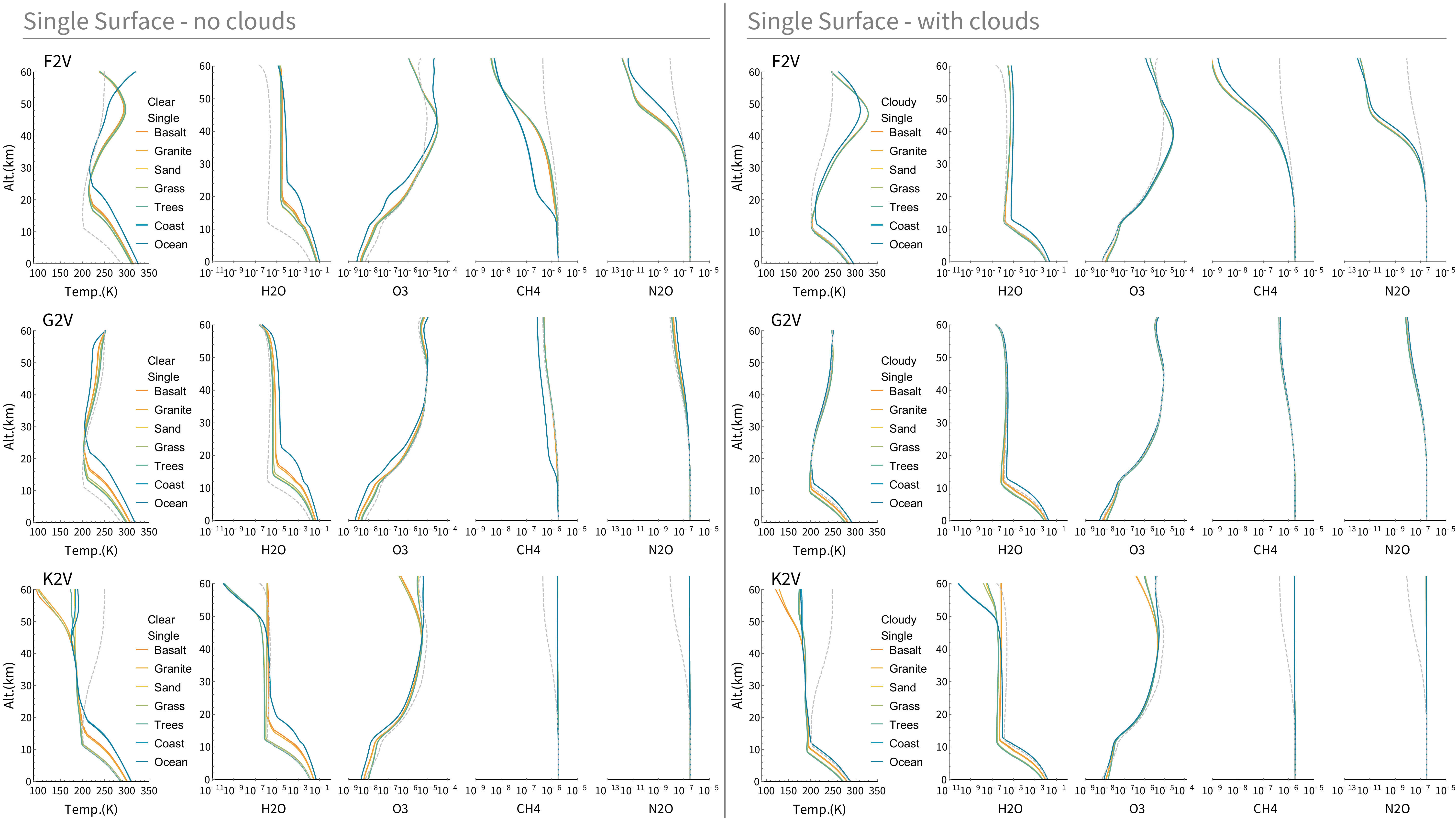}
        \caption{\singlespace Differences in model atmospheres for single-surface rocky planets orbiting in the HZ of FGK host stars with no clouds added (left) and with 44\% cloud coverage (right).\label{fig:profilessingle}}
\end{sidewaysfigure*}

\begin{sidewaysfigure*}
      \includegraphics[width = \textheight]{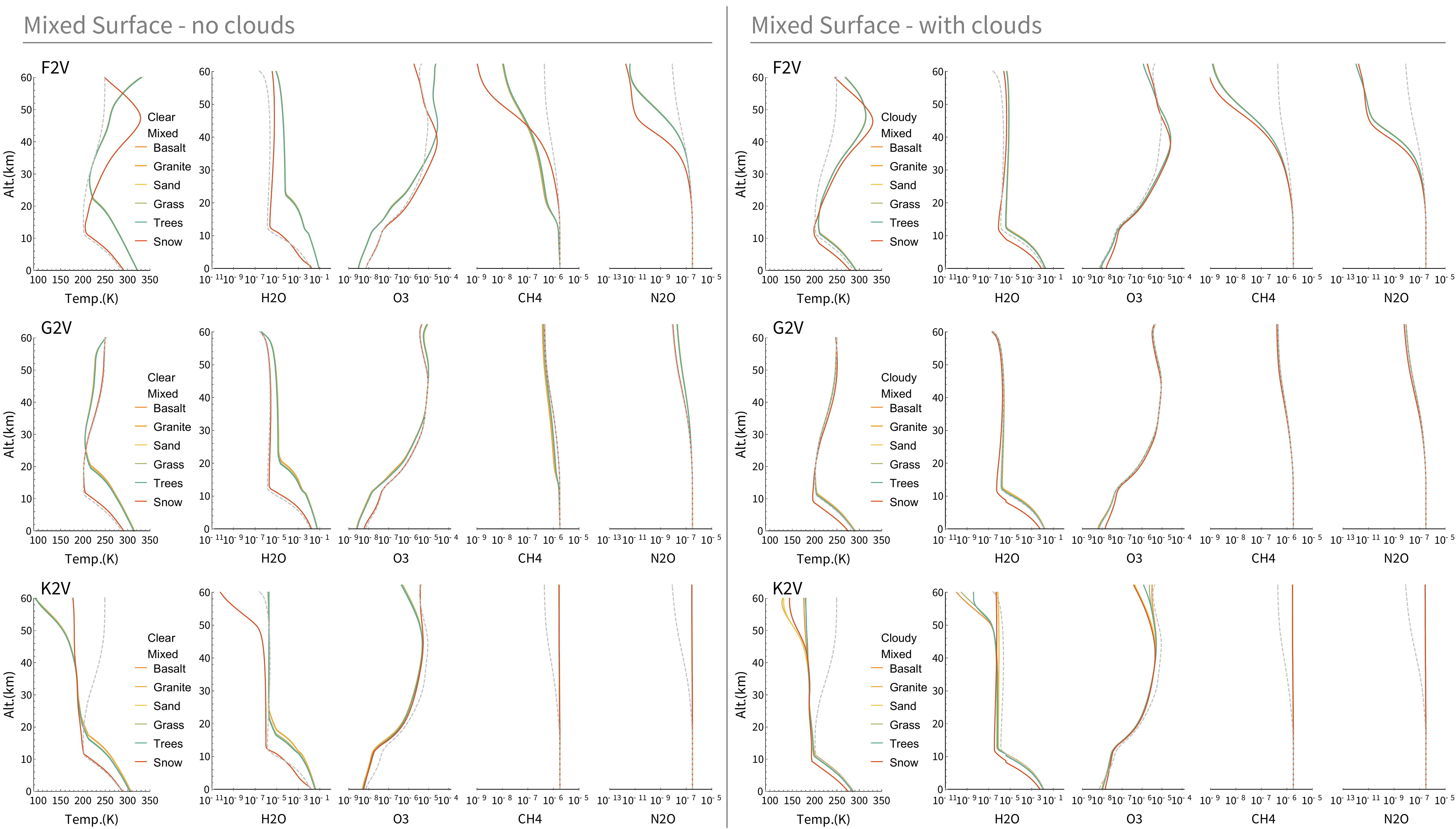}
      \caption{\singlespace Differences in model atmospheres for mixed ocean-land surface rocky planets orbiting in the HZ of FGK host stars with no clouds added (left) and with 44\% cloud coverage (right). \label{fig:profilesmixed}}
\end{sidewaysfigure*}

We model Earth-like planets with different surfaces for a representative grid of 12 host stars from F0 to K7 in approximately 350K effective surface temperature steps.
Figures \ref{fig:profilessingle} and \ref{fig:profilesmixed} show the temperature profiles along with the mixing ratio profiles of the major atmospheric chemicals of interest for characterization and biosignature detection H$_2$O, O$_3$, CH$_4$, and N$_2$O for planets with different surfaces with and without clouds. In this paper, we present the detailed temperature and chemical profiles of a select subset of 3 stars (K2V, G2V, F2V) and the spectra for a subset of 4 stars (K2V, G2V, F2V, F0V) in the figures for clarity. 

\subsection{Surface Temperature for Earth-like Planets}
\begin{figure}[ht!]
\centering
\includegraphics[width=0.75\textwidth]{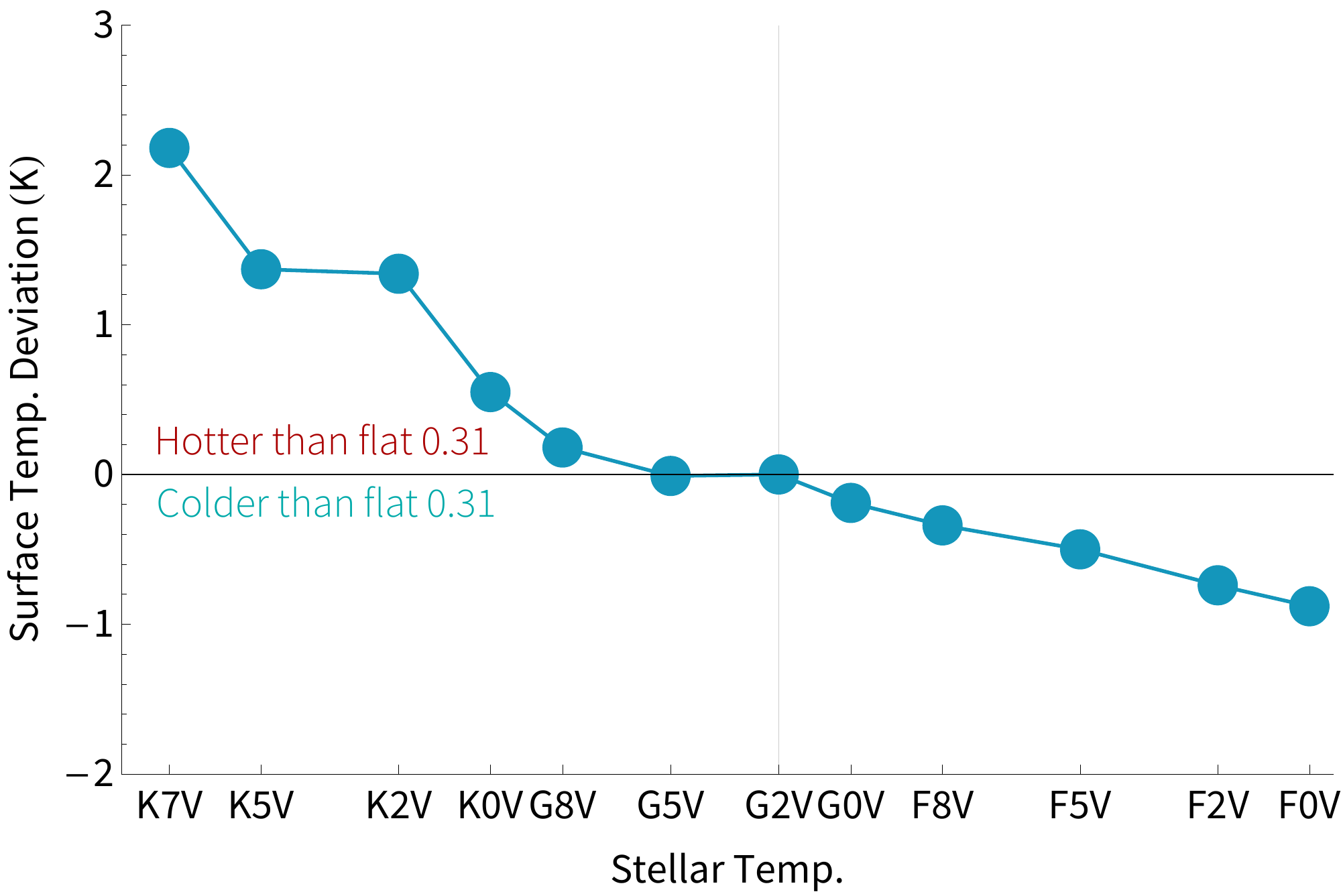}
\caption{\singlespace Surface temperature deviation of model planets across host stars between using a flat 0.31 surface albedo and a wavelength-dependent Earth surface albedo with clouds. The cloud coverage in the Earth model was adjusted to create a deviation of zero for the model orbiting the Sun (G2V).   
\label{fig:surfacetemp}}
\end{figure}

Using our wavelength-dependent Earth surface albedo in combination with a 44\% cloud fraction, we compare the results for our Earth-like planet models for each host star type to the flat albedo models. Fig. \ref{fig:surfacetemp} shows that the deviation of the resulting surface temperature becomes larger as the stellar type becomes more different from the Sun. The wavelength-dependent Earth albedo is less reflective in the near IR than the flat albedo, which causes the planetary surface to become hotter around redder stars and cooler around bluer stars. For an Earth-analog surface, which the 1D flat albedo models were calibrated for, where the surface is dominated by 70\% ocean, our results show that the wavelength dependence of the surface albedo increases the average surface temperature by up to 5K and decreases it by down to 1K (Fig. \ref{fig:surfacetemp}). However, the effects can be much stronger for planets that do not have modern Earth-analog mixed surfaces.

Our models show that planetary surface temperature generally increases with decreasing effective temperature of the host star, and temperature inversions in the upper atmosphere of the planet decrease (Fig.\ref{fig:profilessingle} ~\cite[see also ][]{Segura2007, Rugheimer2015Spectra}. The absolute surface temperature generally is higher for clear atmosphere models as expected due to the high reflectivity of clouds. The surface temperature of ocean-worlds is higher due to the lower reflection of oceans compared to granite or basalt surfaces, which show higher surface temperatures than desert- and jungle-worlds.

While the trend of increasing surface temperature with decreasing host star effective temperature also holds for planets with similar surfaces, different surfaces can reduce the magnitude significantly e.g. an ocean-planet orbiting an F-star shows hotter surface temperature than a rocky planet orbiting a K-star (see Fig.\ref{fig:profilessingle}).

The chemical profiles in Figures \ref{fig:profilessingle} and \ref{fig:profilesmixed} show an increase in methane and N$_2$O concentrations as expected for host stars with lower surface temperatures, especially in the upper atmosphere ~\cite[see also][]{Segura2007, Rugheimer2015Spectra}. The ozone concentration increases for host stars with lower surface temperatures as discussed in several papers  ~\cite[e.g.][]{Segura2005,Segura2007,Rugheimer2015Spectra,Rugheimer2013}. 

\subsection{Surface Temperature of Water-, to Desert-worlds around different host stars}
\begin{figure*}
\centering
\includegraphics[height=0.87\textheight]{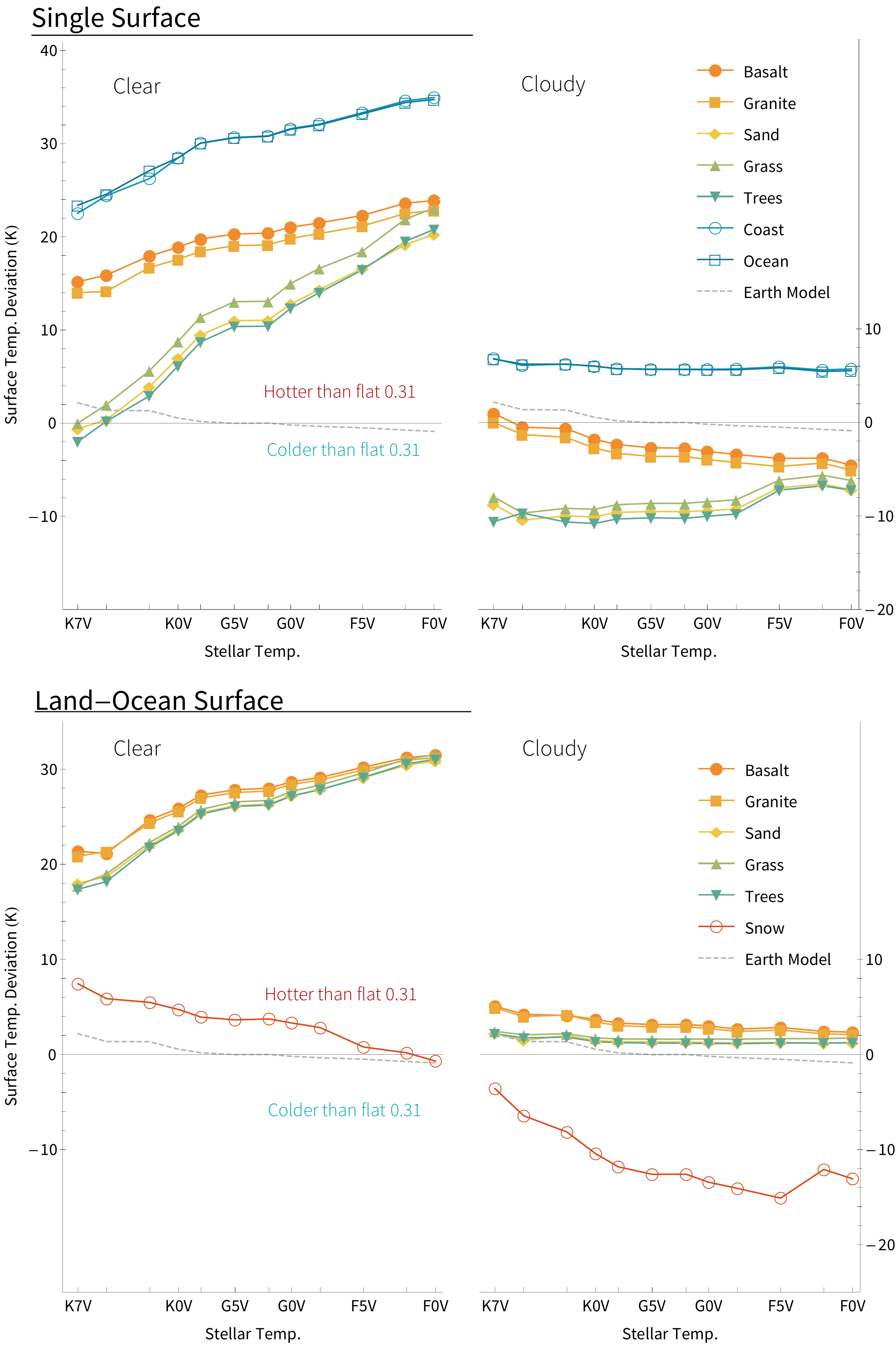} 
\caption{\singlespace Planetary surface temperature deviations between models using a flat 0.31 albedo and wavelength-dependent single-surface-cloud (top) and land-ocean-cloud (bottom) albedo for different star types. Clear denotes atmosphere without and Cloudy atmospheres with 44\% Earth cloud coverage, as discussed in the text.
\label{fig:surfacetempclearcloudy}}
\end{figure*}

To explore how the spectral type of the host star influences different kinds of potentially habitable worlds, we first model a planet covered entirely with a single surface, e.g. ocean covered water worlds (using the surface albedo of oceans), desert worlds (sand), jungle worlds (trees and grass) and rocky worlds (basalt and granite) with and without clouds. Since these single-surface planets have unique, non-flat surface albedos, their climates each respond differently to host stars with different SEDs. As a second step, we created a set of wavelength-dependent ocean-land surface models comprising of the wavelength-dependent albedo of one unique surface combined with 70\% ocean coverage. Fig. \ref{fig:surfacetempclearcloudy} shows the average surface temperature differences between models assuming a flat 0.31 albedo and water-, jungle-, rocky- and desert-planets (top panel) and land-ocean surface coverage for clear and cloudy atmospheres (bottom panel). 
These results show that the surface of a planet can have a significant impact on the surface temperature of an exoplanet, and potentially alter habitability. For our single surface models, using a wavelength-dependent albedo instead of the constant albedo value changes the surface temperature for up to +35 K for an ocean planet orbiting an F0 host star and -10K for a cloudy jungle-planet orbiting a K7 host star (Fig. \ref{fig:surfacetempclearcloudy}).

\subsection{How Surfaces influence Spectra}
We model the reflection spectra of Earth-like planets with different surfaces in the Habitable Zone to explore how different surfaces influence the spectra of potentially habitable worlds seen directly imaged.

\begin{figure}[th!]
\includegraphics[width=\textwidth]{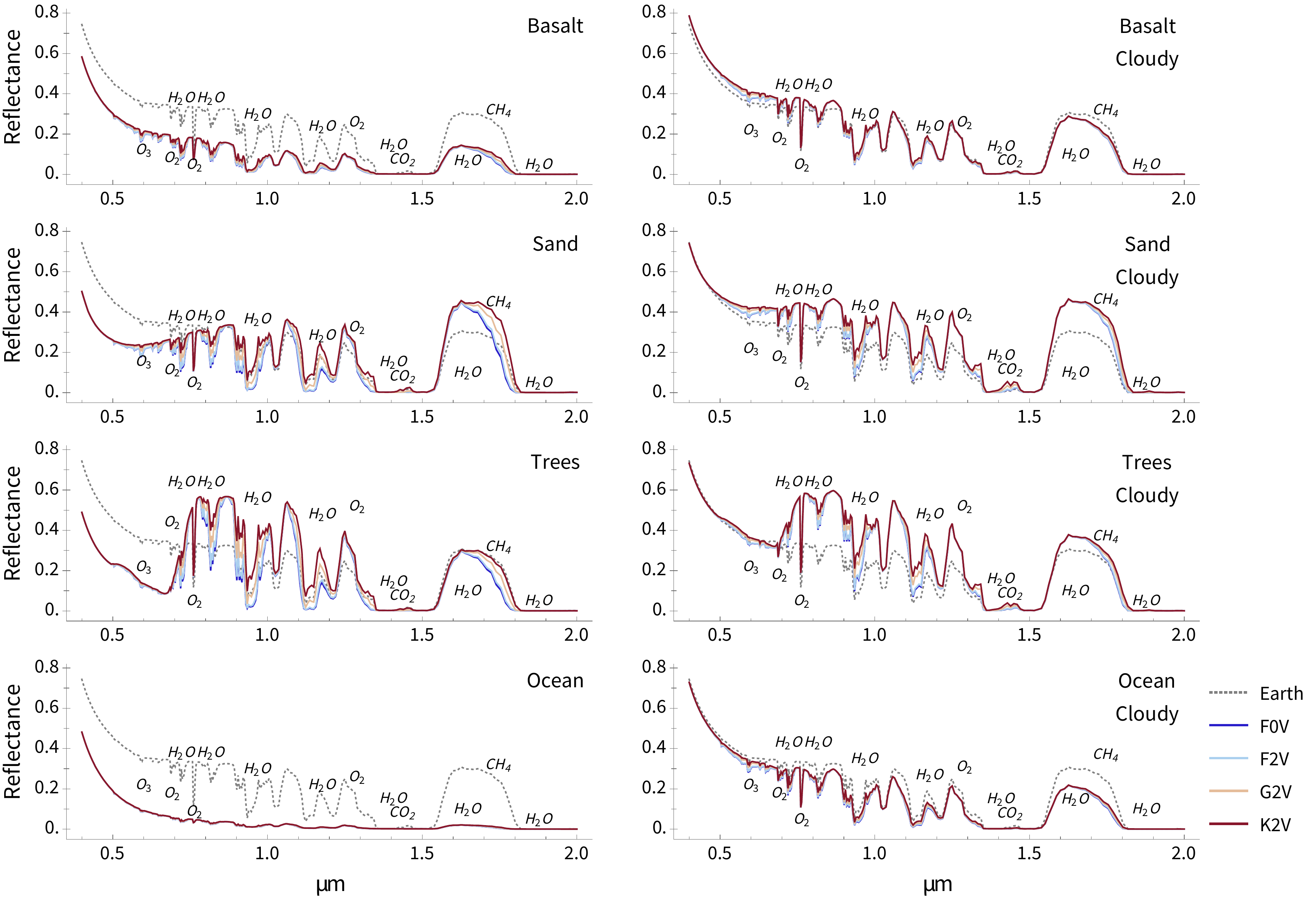}
\caption{\singlespace Reflection spectra for rocky planet models with different surfaces in the HZ of F0V to K2V host stars, assuming a single-surface coverage for (left) clear and (right) cloudy atmospheres.}
\label{fig:spectrasingle}
\vspace{-8pt}
\end{figure}

\begin{figure}[th!]
\includegraphics[width=\textwidth]{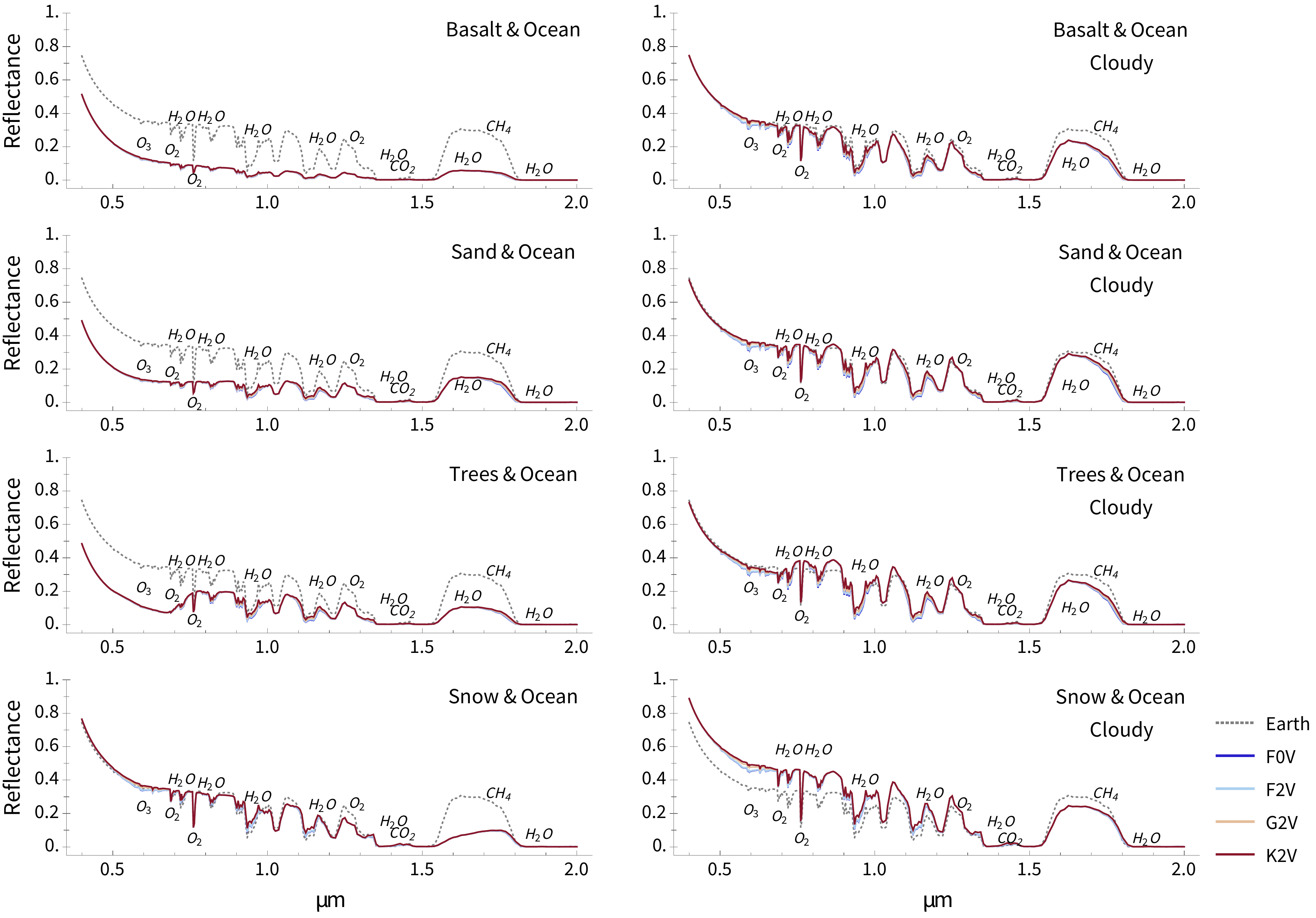}
\caption{\singlespace Reflection spectra for planet models with different surfaces for rocky planets in the HZ of F0V to K2V host stars, assuming 30\% surface and 70\% ocean coverage for (left) clear and (right) cloudy atmospheres.}
\label{fig:spectramixed}
\vspace{-8pt}
\end{figure}

Fig. \ref{fig:spectrasingle} and Fig.\ref{fig:spectramixed} show that some atmospheric species exhibit noticeable spectral features in reflected light (0.4 to 2$\mu$m) as a result directly or indirectly from biological activity: the main ones are O$_2$, O$_3$, CH$_4$, N$_2$O and CH$_3$Cl ~\cite[see e.g.][]{DesMarais2002, Kaltenegger2017}. In the visible wavelength range, the strongest O$_2$ feature is the saturated Frauenhofer A-band at 0.76$\mu$m, with a weaker feature at 0.69$\mu$m. O$_3$ has a broad feature, the Chappuis band, which appears as a shallow triangular dip in the middle of the visible spectrum from about 0.45$\mu$m to 0.74$\mu$m. Methane at present terrestrial abundance (1.65ppm) has no significant visible absorption features, but at high abundance, it shows bands at 0.88$\mu$m, and 1.04$\mu$m, detectable e.g. in early Earth models \citep{Kaltenegger2007, Rugheimer2018}. In addition to biosignatures, H$_2$O shows bands at 0.73$\mu$m, 0.82$\mu$m, 0.95$\mu$m, and 1.14$\mu$m. CO$_2$ has negligible visible features at present abundance, but in a high CO$_2$-atmosphere of 10\% CO$_2$, like in early Earth evolution stages, the weak 1.06$\mu$m band could become detectable. 

As discussed in detail in ~\citep{Rugheimer2013,Rugheimer2015Spectra}, features of Oxygen, water, Methane, and Carbon Dioxide are present in the visible/near IR spectrum. Aside from the dominant water features, the O$_2$ feature near 0.76$\mu$ m is clearly distinguished in the relative reflection around all model stars. Most features appear deeper for the models using the wavelength-dependent Earth albedo compared to the flat albedo. Note that the spectra have not been multiplied by the incident stellar flux, which will reduce the reflected flux in the shorter wavelength range significantly ~\cite[see e.g.][]{Rugheimer2015Spectra}.

\begin{figure}[ht!]
\centering
\includegraphics[width=0.75\textwidth]{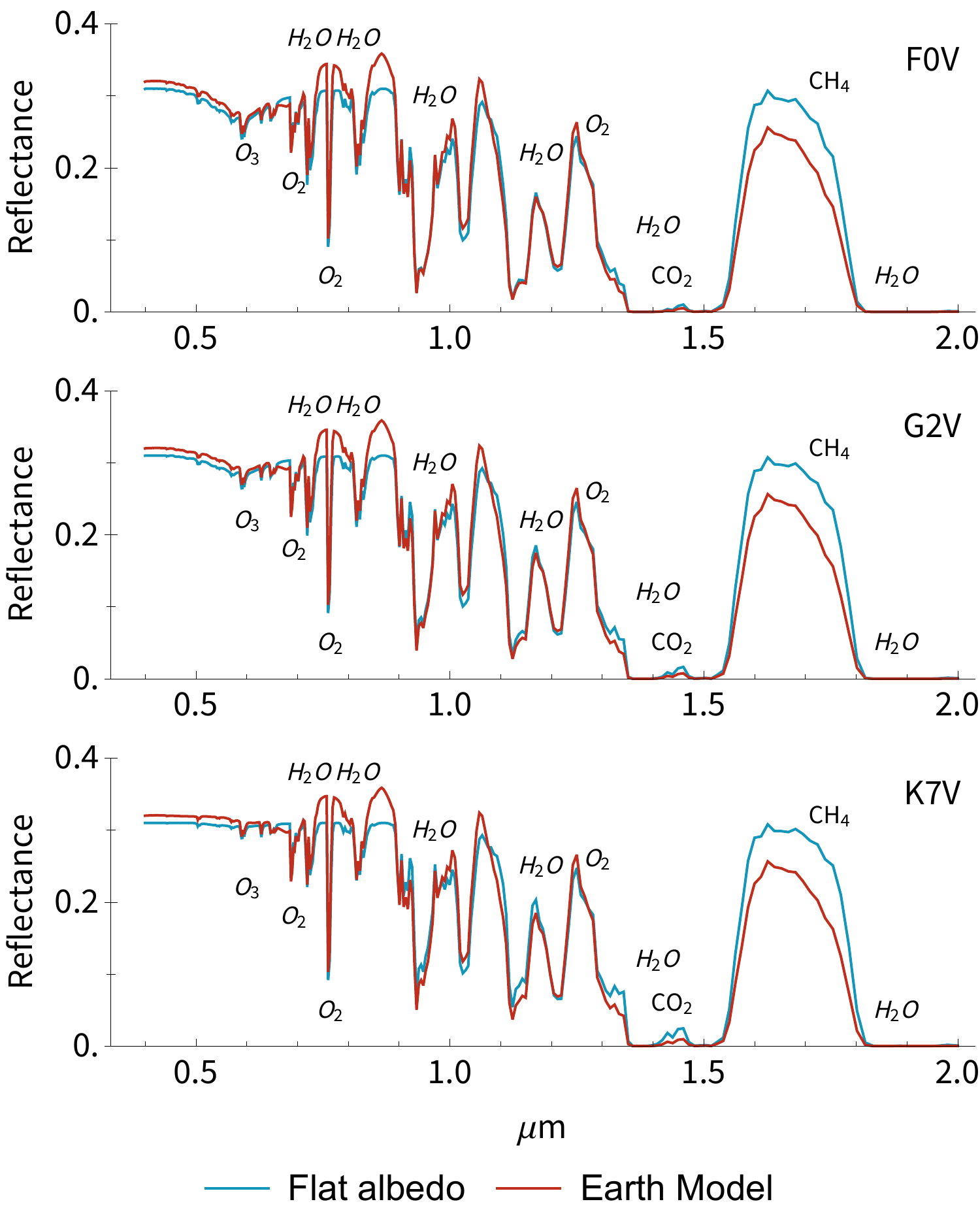}
\caption{\singlespace Reflectance spectra differences for Earth models assuming a flat 0.31 albedo and a wavelength-dependent Earth albedo.
\label{fig:spectraearth}}
\end{figure}

Along with the differences in surface temperature and photochemical differences described previously, the resulting reflectance spectra show changes between a wavelength-dependent Earth-analog surface model versus a flat albedo model (Fig. \ref{fig:spectraearth}). This shows that assuming a flat albedo can both under- or over-estimate the strength of specific chemical atmospheric signatures for Earth-like planets, depending on their host stars. 

Since the wavelength-dependent Earth albedo reflects less than the flat albedo in the near IR, spectral features in the NIR will be overestimated if a flat albedo is used in an Earth spectrum model (Fig. \ref{fig:earthalbedo}).

Fig. \ref{fig:spectrasingle} and Fig. \ref{fig:spectramixed} show relative reflectance for planet models with one surface as well as 30\% land and 70\% ocean coverage, respectively, for different host stars for clear (left) and cloudy (right) atmospheres. Note that the dashed line in both figures is the modern Earth reflection spectra, including clouds. Surfaces with lower albedo (e.g. water) reduce the reflected flux of the planet. In contrast, surfaces like sand and trees increase the overall reflectivity of a planet compared to modern Earth (dashed line). 

The overall effect of the decreased stellar incident flux at shorter wavelengths for cooler stars makes atmospheric features like the 0.76$\mu$m O$_2$ challenging to detect for cooler stars ~\cite[see also e.g.][]{Rugheimer2015Spectra}. However, the shape of the surface albedo (e.g. sand reflection increases with decreasing wavelength compared to modern Earth) can increase and decrease the detectability of specific atmospheric features.

For atmospheric models, including clouds, the overall reflectivity of the planet increases due to the added high cloud albedo. However, the overall results from the single surface planet models discussed above hold, even though the coverage of the surface reduces due to the added 44\% cloud coverage in the planet models.

Earth`s surface consists of about 70\% ocean and 30\% land. Therefore we also model this specific case to provide a second comparison set. Assuming one single surface dominates the remaining landmass (basalt, sand, trees, or snow), we explore the effect on the reflection spectra of such planets. Note that the 100\% ocean covered surface has been shown in Fig. \ref{fig:spectrasingle}. Adding 70\% ocean reduces the overall reflectivity of the planet because oceans only reflect a small amount of incident light (Fig. \ref{fig:surfacealbedos}). While the overall reflected flux reduces for all models (Fig. \ref{fig:spectramixed}), the overall signature changes discussed above hold for the land-ocean models as well under both clear and cloudy conditions. 

The spectra shown in Fig.\ref{fig:spectrasingle} and Fig.\ref{fig:spectramixed} show that increasing cloud coverage and decreasing surface coverage of individual surfaces decrease the detectable differences in reflected flux for planets with different surfaces (as discussed in ~\cite{Kaltenegger2007}.

Several teams has discussed observations of Earth-size planets in the habitable zone with upcoming extremely large telescopes ~\cite[see e.g.][]{KalteneggerTraub2009,Kaltenegger2010,Stevenson2016,Barstow2016,Hedelt2013,Munoz2012,Snellen2013,Rodler2014,Betremieux2014,Misra2014}.
Different surfaces of planets can influence the overall abundance of chemicals in the atmosphere, via their influence on the surface temperature of a planet and the resulting photochemical changes. Note that these changes affect most spectral lines (especially water, oxygen, and methane).

\section{Discussion}

\subsection{Single surface models explore the most extreme effect}
Our exploration of the influence of the surface on an exoplanet`s climate and spectra shows the most extreme effect for planets fully covered with a single surface (Fig. \ref{fig:spectrasingle}). We chose this case to explore whether wavelength-dependent surfaces affected the climate and surface temperature of exoplanet models. In addition to the most extreme case, a single surface coverage, we also show a second comparison set in this paper, analog to Earth, we assume a 70\% ocean coverage of the planet`s surface. While the fraction of different surfaces on exoplanets in unknown, these two cases show i) the maximum effect of a surface on a planet`s climate as well as ii) the reduced effect if only 30\% of a planet`s surface is covered with a specific surface. While many other options are possible, these two cases clearly show how the surface fraction, as well as clouds, influence the climate of Earth-like planets. Any other surface fraction can be interpolated from the cases shown. While we do not know what surfaces exist on Earth-like exoplanets, We chose the 8 dominant surfaces of our own planet here as the test cases in our models (Fig. \ref{fig:earthalbedo}). In an exoplanet context, the interaction of wavelength-dependent surface albedo and stellar spectra has been examined mainly regarding the effect of ice-albedo feedback or special case surfaces ~\citep{Abe2011,Shields2013,Shields2014,Shields2018}. Some of our cases resembled those previously studied with snow/ice surfaces, which we confirmed show increased heating for planets around K-stars compared to G- and F-stars.

\subsection{Cloud feedback is unknown for different host stars}
Cloud feedback is unknown for different host stars. We, therefore, use clouds with similar properties for all our models based on the modern Earth model, as explained in section 2 as a first-order approximation. Cloud properties and coverage depend on many factors, including planetary rotation rate, atmospheric pressure, temperature, aerosol abundance, particle size distribution, and humidity ~\cite[see e.g.][]{Zsom2012}. Cloud coverage on Earth changes seasonally and is thought to have also changed over geologic time  ~\cite[see e.g.][]{OMalleyJames2018,Brierley2009}.

\subsection{3D and 1D model exploration of rocky planets}
3D models are being expanded to explore rocky exoplanets, with a large body of papers advancing a lively discussion in the literature on how to best include the feedback effects of clouds, rotation, surface features, atmospheric dynamics, and full photochemistry in 3D models (see e.g. discussions in \cite{GOMEZLEAL2016,GOMEZLEAL2019,Forget1997,chen2019,Lorenz1997,Williams2002,Joshi2003,Lopez2005,Selsis2007,Edson2011,Zsom2012,Goldblatt2013,Leconte2013,Leconte2013nat,Leconte2015,Vladilo2013,Wordsworth2013,Yang2013,Boschi2013,Ferreira2014,Wolf2015,Linsenmeier2015,Kopparapu2016,Popp2016,Kitzmann2017}.) 3D models and 1D models tend to be in agreement on globally averaged surface temperature for a range of stellar types though some 3D effects can exacerbate this difference due to cloud feedback \cite[][]{Arney2016}, tidal locking \cite[][]{Kopparapu2016}, or at climate extremes such as habitable zone edges \cite[][]{Shields2013,Yang2016,GOMEZLEAL2019}.

Our 1D model includes detailed photochemistry based on Earth's atmosphere and is, therefore, an excellent tool to explore the effect of the host's SED and the wavelength-dependent surfaces on the changes in the atmosphere of Earth-like planets orbiting different host stars. We concentrate on F, G, and K stars in this paper, where rocky planets in the HZ should not be synchronously locked to their host stars, and thus effective heat transfer in an Earth-like atmosphere can be assumed ~\cite[see e.g.][]{Joshi2003}. We use a 1D model for this study to explore a large range of surfaces and star types.

\section{Conclusions}

Our paper demonstrates the importance of including the wavelength-dependent feedback between a planet's surface and a planet's host star for Earth-like planets in the HZ of stars with an effective temperature between 3,900 and 7,400 K, corresponding to K7V to F0V main sequence stars. The heating or cooling effect of a specific surface is due to the interplay between the host star's SED compared to the shape of the wavelength-dependent surface albedo and can substantially change the surface temperature of an Earth-like planet. Our paper demonstrates the importance of including the wavelength-dependent feedback between a planet's surface and a planet's host star for Earth-like planets in the habitable zone. Reflected light from the surface plays a significant role not only on the overall climate but also on the detectable spectra of Earth-like planets.  

\section*{Data Access}
DOI for accompanying data: ~\href{https://doi.org/10.5281/zenodo.3912065}{10.5281/zenodo.3912065}

\section*{Acknowledgements}
This work was supported by the Brinson Foundation and the Carl Sagan Institute. We would like to thank Zifan Lin, Thea Kozakis, and Sarah Rugheimer for comments and insights.

%% file: 4_VR/ms.tex

\section{Abstract}
Immersive virtual reality (VR) has enormous potential for education, but classroom resources are limited. Thus, it is important to identify whether and when VR provides sufficient advantages over other modes of learning to justify its deployment. In a between-subjects experiment, we compared three methods of teaching Moon phases (a hands-on activity, VR, and a desktop simulation) and measured student improvement on existing learning and attitudinal measures.  While a substantial majority of students preferred the VR experience, we found no significant differences in learning between conditions. However, we found differences between conditions based on gender, which was highly correlated with experience with video games. These differences may indicate certain groups have an advantage in the VR setting. 

\section{Introduction}
In twenty-first century education, technology is pervasive in our classrooms~\citep{Maddux2003}. Research has found many ways in which technology benefits student learning and attitudes towards science~\citep{DBER2012}. As new instructional technologies are developed, it is necessary that researchers conduct critical evaluations of their effectiveness. There are many open research questions related to identifying how to most effectively use different kinds of technology for different learning goals or for different points in the learning process. One technology that has received particular attention for its potential in science learning is virtual reality and other immersive media \citep{dede2009immersive, pan2006virtual}.

In a typical college lecture for a science course, an instructor can choose to engage students with a concept using technologies such as specialized equipment for an interactive lecture demonstration~\citep{Sokoloff1997}, a dynamic computer simulation~\citep{Wieman2008}, or classroom polling with personal response systems~\citep{Stains2018}. Science courses also employ technology extensively in instructional labs, where students can use technology to obtain first-hand experiences with the phenomenon~\citep{Hofstein2004}. Instructional labs, however, have failed to provide measurable gains in student learning of those phenomena~\citep{Holmes2017, Etkina2010}, reminding us that what is important for learning is not the technology itself, but how and why it is used~\citep{DBER2012}. For example, using relatively similar equipment, interactive lecture demonstrations have consistently found measurable learning gains when students are actively engaged in making predictions and explaining the observations from the demonstrations~\citep{Sokoloff1997, Crouch2004}. Interactive simulations have also been shown to improve or replicate learning compared to hands-on manipulatives, while reducing the associated resources~\citep{Finkelstein2005, Chini2012}. While there are many potential explanations for these differences~\citep{SmithDemos}, two relate to how learners and technology interact through issues of embodiment and real-world complexities. This study aimed to test the impact of these variables by directly comparing student learning and attitudes from three different instructional technologies (a hands-on activity, desktop simulation, and virtual reality simulation), while taking advantage of their respective affordances along the dimensions of embodiment and real-world complexity.

\subsection{Embodiment}
Theories of learning argue that cognition is inherently embodied: ``the mind must be understood in the context of its relationship to a physical body that interacts with the world.''~\cite[p.625]{Wilson2002}. Research has found that learning benefits from activities that explicitly attend to embodied cognition~\citep{ANDERSON200391, Roth2013,Carbonneau2013}. There are several ways in which embodiment is argued to support learning, generally tied to a hypothesis whereby activities help move cognition from abstract to concrete representations of a phenomenon~\citep{ANDERSON200391}.

Two such notions of embodied cognition focus on how learners off-load cognitive work on to the environment and how off-line cognition is body-based~\citep{Wilson2002, Martin2005}. These notions suggest physical aspects of cognition, whereby learning is supported through engaging perceptuo-motor systems~\citep{Tsang2015}. Indeed, nearly all science education advocates for the use of interactive hands-on activities~\citep{Ruby2001}.  Through hands-on activities and demonstrations, learners connect abstract concepts to their physical environment. For example, researchers in physics education have developed embodied activities for teaching concepts of energy conservation~\citep{Scherr2013}. In these activities, learners assign units of energy to physical objects (either cubes or people)~\citep{Scherr2012}, and then manipulate those objects to represent processes of energy transfer and dynamics. Through these concrete representations of an otherwise abstract phenomenon, learners develop their conceptual and mechanistic understandings of the phenomenon~\citep{Scherr2013}. By manipulating physical objects, students can see deep features of the phenomena, allowing them to effectively integrate the features into their mental models of the phenomena~\citep{Tsang2015, Piaget2013}.

Two other notions of embodied cognition focus on how cognition is situated within the real-world and that the learning environment is part of the cognitive system~\citep{Wilson2002, Martin2005}. Because the knowledge learned is tied to the environment in which it was learned, it is argued that there are limitations to applying the knowledge in new environments~\citep{Brown1989}. In science education, the initial learning environment can often seem too far removed or distinct from the intended environment for applying the knowledge. Context-rich activities, can provide concrete, real-world scenarios for otherwise abstract learning activities. Alternatively, activities may be modified to better represent the real world, such as by providing realistic representations of the phenomena.

These activities also demonstrate that there are degrees of embodiment. These may range from metaphorical groundings of cognition (for example, being able to imagine oneself walking from point A to point B) to physically experiencing the phenomenon (for example, actually walking from A to point B)~\citep{ANDERSON200391}. 
While much research has demonstrated the benefits of physically experiencing the phenomenon through hands-on activities, the results are not universal~\citep{Cunningham1946, Holmes2017, Chini2012, Finkelstein2005,Klahr2007}. One explanation for the lack of clear benefit for hands-on activities is that when the activities and materials are complicated or difficult to manipulate, the learner may experience extraneous cognitive load~\citep{Paas2003} or may be likely to make mistakes~\citep{Sokoloff1997}.
That is, real-world complexity gets in the way of student learning in hands-on activities. 

\subsection{Real-world complexity}
In a real-world experiment, students’ measurements and observations are prone to variability, systematic effects, and measurement mistakes that are not relevant to the theoretical concepts being taught. In general, people struggle to evaluate uncertain events~\citep{Kahneman1972} and use error-prone shortcuts and heuristics in making judgments under uncertainty~\citep{Tversky1974}. Learning an underlying concept, then, becomes difficult if the information being used to develop that understanding is uncertain or probabilistic.

For hands-on activities, the uncertain and probabilistic elements contribute to extraneous cognitive load during the activity. Cognitive load refers to a learner's capacity for processing information in their short-term memory~\citep{Paas2003}. Information can either contribute to extraneous (ineffective) cognitive load or germane (effective) cognitive load~\citep{Paas2003}. Germane cognitive load refers to information that is relevant for learning, such that large amounts of germane cognitive load can improve learning~\citep{Kapur2016}. Extraneous cognitive load refers to information that is irrelevant for learning, such that large amounts of extraneous cognitive load impede learning~\citep{Sweller1991,Paas2003}. In hands-on activities, issues of uncertainty, complicated equipment, and user mistakes contribute to extraneous cognitive load, which may hamper learning. 

To reduce that cognitive load, hands-on activities often use heavily guided instructions that attempt to limit mistakes, tell students how to manipulate the equipment, and reduce uncertainties~\citep{Hofstein2004}. However, constructivist theories of learning argue that students must have the ability to explore and generate their own knowledge~\citep{Bransford1999}. The key is to develop activities with high germane cognitive load, but low extraneous cognitive load~\citep{Kapur2016, Schwartz2011}.

Simulations are one way to remove the extraneous cognitive load of real-world hands-on activities, allowing phenomena to be demonstrated in a consistent and controlled way. With that control, they can still be relatively unstructured, maintaining high germane cognitive load. Students can autonomously and easily change variables, allowing them to learn at their own pace~\citep{Price2018, podolefsky2010}. Several studies have found that computer simulations produce equal~\citep{Darrah2014, Chini2012, Evangelou2018} or better~\citep{Finkelstein2005, Chini2012} learning than hands-on activities. Simulations also provide opportunities for students to see features of a phenomenon that they would be unable to see otherwise, for example abstract concepts such as heat~\citep{Strzys2018} or microscopic objects such as cells~\citep{Chang2016} or electrons~\citep{Finkelstein2005, Kapp2019}.

Simulations, however, provide a more limited embodied cognition experience than hands-on activities, where the learner interacts directly with the phenomenon.
Virtual reality is a potential technology that can employ high-levels of embodiment, while maintaining controlled and simplified representations of the phenomena to be learned.

\subsection{Why virtual reality?}
Immersive virtual reality (VR) may provide the best of both worlds. VR allows embodied simulations and offers a number of other affordances~\citep{Bricken1991, Perone2016} that make it uniquely suited as a teaching tool for basic science. First, students can physically interact with content, providing the engagement of a hands-on activity but with the control and replicability of a simulation.  Second, the simulations provide multiple forms of embodiment, such as changing perspectives to experience phenomena as they would in different circumstances in the real world~\citep{ANDERSON200391}. Third, students can experience these perspectives in ways unavailable in the real world~\citep{Price2018, Strzys2018, Kapp2019}. From a research perspective, the ability to track student movement allows for assessing engagement and learning~\citep{Won2014} to better test the embodiment hypothesis. It also facilitates implementing interventions to increase learning in real time. 

There is thus a need for experimental research that directly tests the effectiveness of VR on science learning, over-and-above that offered by existing hands-on and simulation approaches. Several studies have compared learning between these three modalities and found that, in terms of student attitudes, participants generally prefer learning in VR over other modalities~\citep{Smith2017, Chang2016}. One study even found that students' attitudes towards socio-scientific issues improved more in an augmented reality (AR) simulation over a desktop simulation, even though these aspects were secondary to the activity's primary cognitive goal~\citep{Chang2016}. Related to embodiment, several studies have found that participants became more immersed in VR than in other environments~\citep{Lier2018, Makransky2017, Winn2002}. Generally, this immersion seems to be independent of personal characteristics, such as gender, VR experience, and time spent gaming~\citep{Lier2018}. 

Measures of learning from VR are generally conflicting. Studies have found that participants in VR learn more than~\citep{Winn2002, Strzys2018}, as much as~\citep{Smith2017, Chang2016}, or less than~\citep{Makransky2017} participants in hands-on or desktop conditions. When learning was improved in VR over a hands-on activity, the gains were attributed to 
immediate feedback available through the simulation and visualization of the abstract phenomenon that was otherwise imperceptible in the hands-on activity~\citep{Strzys2018}. 
When learning was hindered in VR over a desktop simulation, the differences were attributed to 
higher cognitive load~\citep{Makransky2017}. In this study, where participants answered multiple-choice questions and performed technical lab procedures in the two types of simulation, researchers found that students in the desktop simulation condition learned more than students in the VR condition on a conceptual test. However, learning was the same between conditions on a transfer test and participants overwhelmingly preferred the VR condition. The researchers also found that participants in the VR condition had significantly higher cognitive load as measured through an electroencephalogram (EEG). They suggested that the physical manipulation of the equipment in VR was more complicated than the desktop condition, which may have increased students' extraneous cognitive load, impacting learning. They recommended experiments that used more natural control systems to manipulate the environment. 
Furthermore, the researchers also argued that the enjoyment associated with VR actually distracted the learners from learning. 

In addition to cognitive load, there are also questions about gender differences and experience with 3D rotations through experience with video games. In one study, men outperformed women during the task itself, but there was no difference on a post-condition recall test~\citep{Leon2018}. In a study that found no overall differences in learning from VR compared with video and static images, the researchers found that men and participants with experience with video games outperformed in the VR condition over other conditions and participants~\citep{Smith2017}. 
Because gender was also correlated with video game experience, they hypothesized that video game experience was a proxy for the gender differences emerging in their study.
This result is somewhat surprising given that neither condition involved much interaction with the simulations (even the VR participants could only look around the simulation). 

Research has found, however, that performance is improved when participants can more fully interact with the simulation: for example, walking around the simulation compared with remaining stationary while looking around the simulation~\citep{Leon2018}. Furthermore, one study found that students' reported sense of presence in a simulation was correlated with their learning from the simulation~\citep{Winn2002}. This study also found that preferential learning from VR was confined to sub-topics involving ``dynamic three-dimensional processes, but not processes that can be represented statically in two dimensions"~\cite[p.1]{Winn2002}. This suggests that VR simulations are more effective when they take advantage of their specific affordances.

The existing research exemplifies the idea that it is not the technology, but how it is used, that promotes learning. In this study, we aimed to develop and test a VR simulation that took advantage of its various affordances, particularly related to embodiment and real-world complexity. We probed these ideas by comparing an interactive VR condition with analogous hands-on and desktop simulation activities for learning about Moon Phases. As per the previous research, we evaluated students' conceptual knowledge, long-term retention, attitudes towards the activity, and socio-scientific beliefs. We also compared differential effects on sub-populations of our participants, including evaluating effects for gender, video game experience, and experience in VR. We found that there were no overall differences in learning on short-term or long-term assessments between conditions and that the immersive VR did not impact students' socio-scientific beliefs. We also replicated previous work in that participants overwhelmingly preferred the VR condition, and that men outperform in the VR condition, which may be attributed to video game experience.

\section{Methods and Materials}
This study had two hypotheses. First, we proposed that there would be a main effect of virtual reality on learning. Second, we proposed that there would be corresponding effects on environmental attitude. We also collected data on other measures in order to explore interactions that might be predicted by the literature. All measures are reported below.

\begin{figure}[ht!]
\centering
\includegraphics[width=0.95\textwidth]{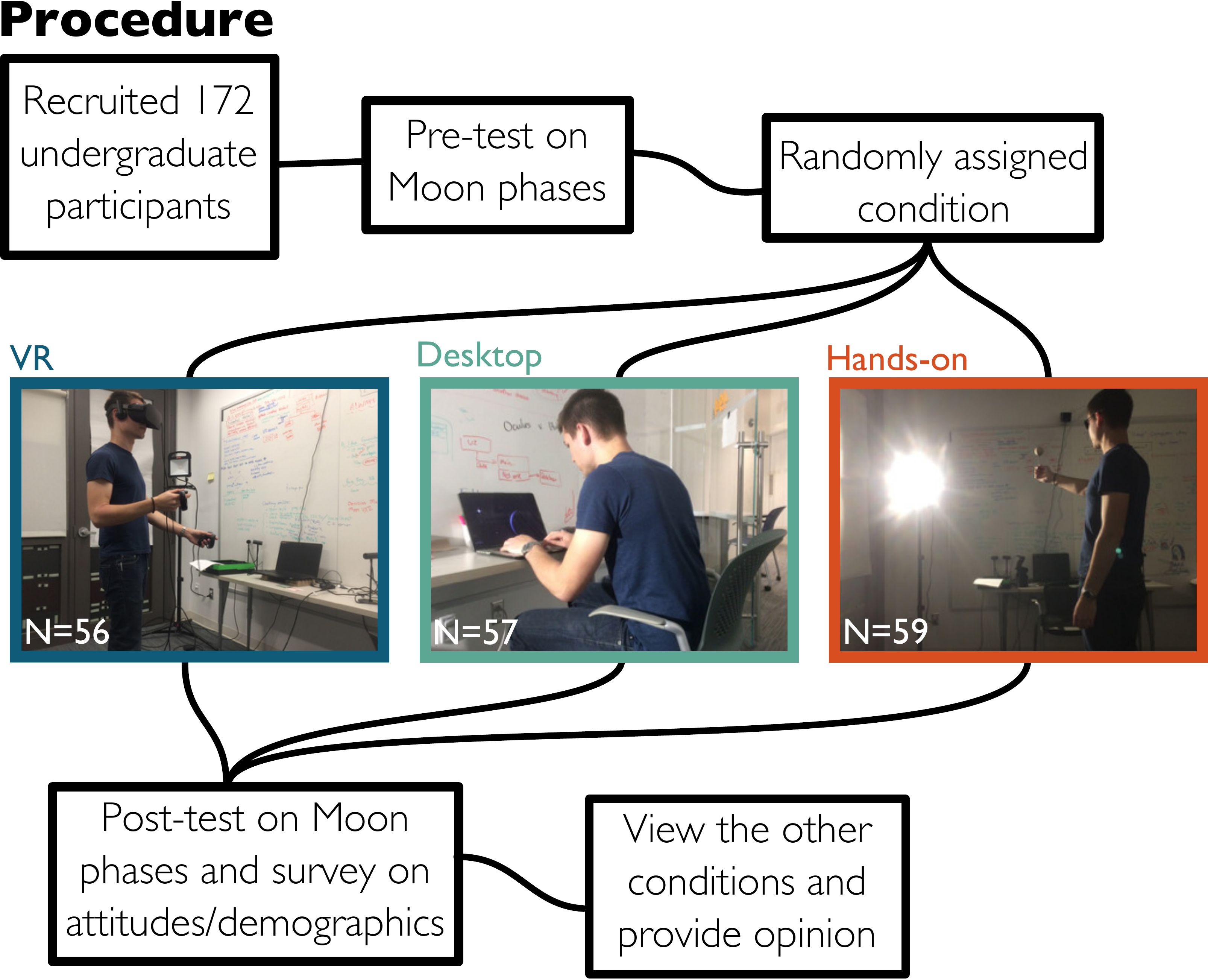}
\caption{\singlespace {\bf Experiment procedure}
Map of the process for our experiment showing the three conditions.\emph{The individual shown has given written consent to have their likeness presented here.}}
\label{fig:flowchart}
\vspace{-20pt}
\end{figure}

In our study, we used a between-subjects pre-post design, with three conditions. The three conditions were designed to express the same content using different educational tools; an immersive simulation using a VR headset, a computer-based interactive desktop simulation, and an analog hands-on activity (Fig. \ref{fig:flowchart}). We chose the concept of Moon phases, as it was expected to benefit from embodiment and reduction of real-world complexities. A review of over 35 years of astronomy education literature found that phases of the Moon was one of the most challenging topics in astronomy education~\citep{Lelliott2010}. A lesson on Moon phases requires the student to place themselves in spatial and temporal perspectives of the Sun-Earth-Moon system that are generally inaccessible, which can be challenging through static images or text~\citep{Galano2018, Turk2015}. Previous work has found that students' spatial reasoning correlates with their understanding of lunar concepts~\citep{Wilhelm2013, Cole2018}. Understanding of Moon phases also requires understanding the dynamic evolution of the phases over time and space~\citep{Cole2018}, likely facilitated through interacting with and moving around a simulation.  Furthermore, the traditional hands-on activity for teaching Moon phases (described below) is susceptible to various real-world inaccuracies, such as creating eclipses every month or rotating or orbiting the wrong way. Research has also found that explanatory features of traditional descriptions and images can interfere with student understanding~\citep{Galano2018, Turk2015}, motivating the need for authentic real-world visualizations over abstracted ones.

\subsection{Participants}
\vspace{-10pt}
Participants were recruited from the undergraduate student population of a medium-sized private university. There were 172 participants, including 138 women, 31 men, and 3 other, all between the ages of 18 and 24. For three conditions to obtain a Cohen's d effect size of 0.25 with 80\% statistical power at 95\% significance level, our power test determined that a sample size of 53 participants per condition (159 total) would be sufficient. We ran a total of 172 participants to buffer against potential exclusions. The study primarily drew from students currently enrolled in an introductory astronomy class (varying in major), and students majoring in Communication. Participants' self-reported race and ethnicity were as follows: 62 Asian/Pacific Islander, 59 Caucasian, 21 African American, 18 bi/multiracial (including Asian/Pacific Islander, Caucasian, African American, Hispanic/Latinx, and Native American), 6 Hispanic/Latinx, 4 preferred not to answer, and 2 specified `other' without elaboration. Participants could select as many categories that applied. All participants signed an informed-consent form before beginning the experiment. All aspects of the experiment were approved by the Cornell Institutional Review Board Protocol \#1708007381. Participants were compensated with course credit or 15 dollars in cash for their participation. The individual pictured in Fig. \ref{fig:flowchart} has provided written informed consent (as outlined in the PLOS consent form) to publish their image alongside the manuscript.

Participants first took a pre-test and then they were randomly assigned to one of three conditions. After the activity (which included consistent self-guided lessons on Moon phases, described below), participants took a post-test that included a demographic and attitudinal survey. Finally, the participants were shown the other two conditions and asked to comment on which they would prefer as their favorite learning method and why. These activities are summarized in Fig. \ref{fig:flowchart}. A large preference for one condition over another may lead to greater retention of knowledge learned in that condition. To explore this potential effect we contacted all participants four months after participating and asked them to complete another learning test (delayed post-test). We received 56 responses to the delayed post-test making a breakdown by demographics difficult.

\subsection{Conditions}
Each condition was designed to give participants a similar learning experience using the three technologies we employed. The overall design was to recreate a Sun-Earth-Moon system that the participant could control in time in order to observe the changes in the Moon's phase and the positions of the Sun, Earth, and Moon during each phase. In each condition the participant could move forward and backward in time and had control over the perspective from which the system was being viewed. Each condition also contained guiding questions for the participant to assist with navigating and learning from the simulation. 

\subsubsection{VR Simulation}
\begin{figure}[!ht]
\centering
\includegraphics[width=0.8\textwidth]{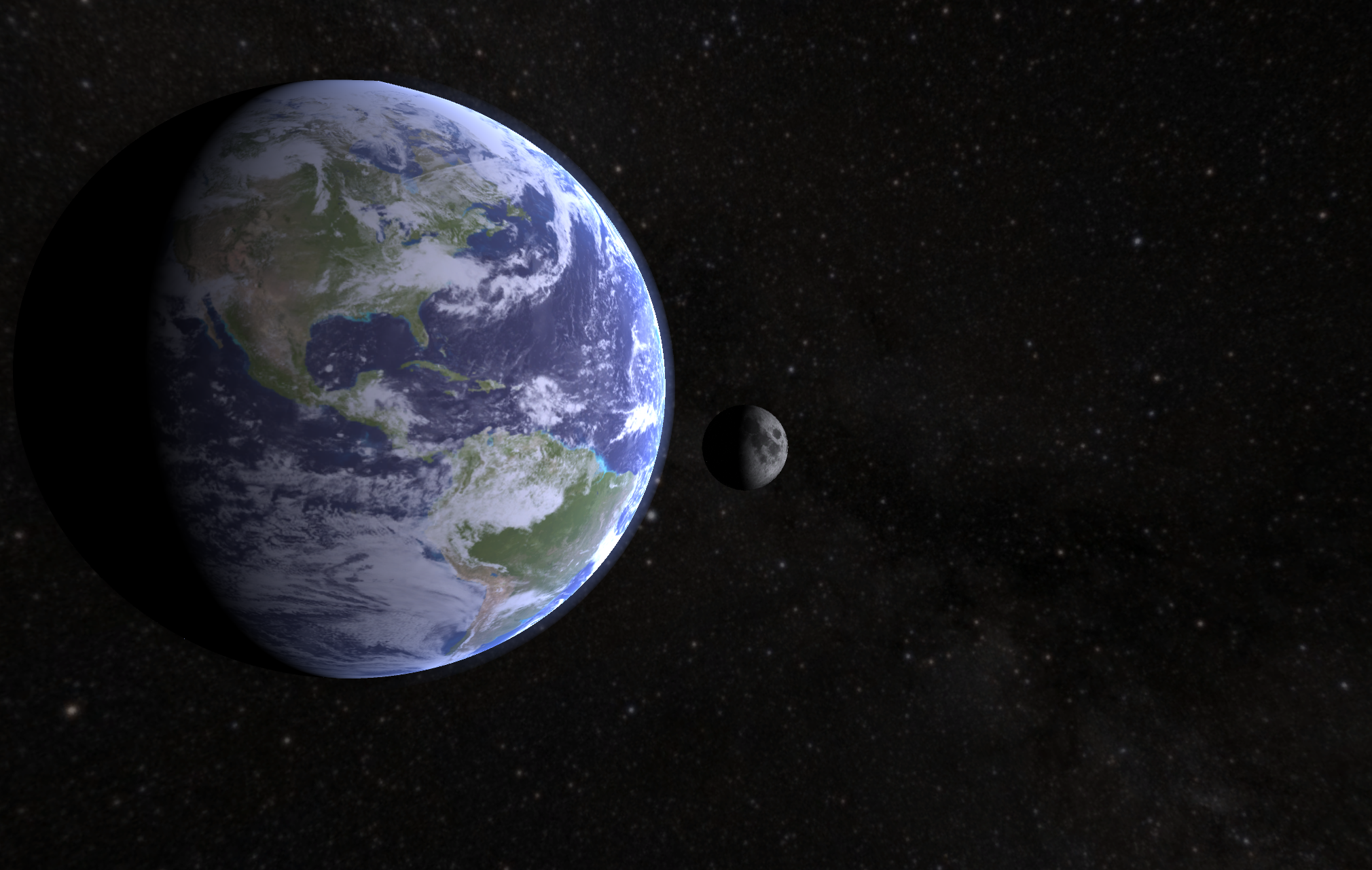}
\caption{\singlespace {\bf Simulated activity}
A screenshot taken from inside the simulated Moon phases activity showing what the VR and Desktop conditions looked like. A video of the VR experience can be found online: \url{https://vimeo.com/310212130} }
\label{pic:earthmoonpic}
\end{figure}

The VR simulation was designed to mimic the hands-on activity as closely as possible, while still taking advantages of the technology's unique affordances. In the VR simulation, participants used a headset and controllers that tracked their motion and rendered the environment, providing an immersive and interactive experience (Fig. \ref{fig:flowchart}, left). The simulation contained a realistic Sun-Earth-Moon system in which the participant had control over time and their viewing location (Fig. \ref{pic:earthmoonpic}). Students could control the Moon phases by moving forwards and backwards in time with simple button presses or grabbing the Moon to move it in its orbit. Upon entering the simulation, participants were initially placed on top of the Earth’s north pole, but they could change their position to be far above the Earth to provide a more diagrammatic view of the system or to be near the surface of the Earth to view a realistic horizon as the Sun or Moon rises and sets. The participants were given the guiding questions throughout the experience via a virtual clipboard attached to their hand to help students interact with and learn from the environment. The names of the Moon phases were displayed in the environment as the participant moved the Moon. The simulation was created by our team using the Unity game engine for use with the Oculus Rift VR headset.

\subsubsection{Desktop Simulation}
The desktop simulation was designed to mimic the VR activity as closely as possible. In the desktop simulation, participants were shown a realistic Sun-Earth-Moon system on a laptop with controls over the camera position and time (Fig. \ref{fig:flowchart}, middle). The environment was created using the same system as the VR simulation (Fig. \ref{pic:earthmoonpic}). Participants could control the Moon phases by going forwards and backwards in time through arrow key presses. They could also navigate around the environment using the mouse and scroll to zoom. The participants were given the guiding questions using a tablet device outside of the simulated environment. The names of the Moon phases were displayed in the environment as the participant moved the Moon. This simulation was created by our team using the Unity game engine. 

\subsubsection{Hands-on activity}
The hands-on activity was based on traditional classroom activities about Moon phases~\citep{newbury}. In this activity, the participant's head represented the Earth; the Moon was represented by a small ball held at arm's length; and the Sun was represented by a stationary light (Fig. \ref{fig:flowchart}, right). Participants were asked to rotate counter-clockwise to observe changes in the Moon's phase, as observed in the shadowed portion of the ball's surface. Through this action, real-world complexities arise where the phenomenon may be inaccurately presented. For example, due to the relative distances and sizes between the ball and the participant's head, participants may see eclipses every month. Subtleties such as inclination and procession of the orbits are uncontrollable and missed in the hands-on activity. The participants were given the guiding questions using a tablet as well as the names of the Moon phases.

\subsection{Measures}
To evaluate the effects of the different conditions, we assessed student understanding of Moon phases using existing pre-post assessment instruments, their attitudes towards the activities and the environment, and tracked their movements in the simulations.

\subsubsection{Moon phases assessment}
The pre- and post-tests each consisted of 14 multiple choice questions about Moon phases and the Moon's motion relative to the Earth, sourced from existing research-validated assessments ~\citep{2002AEdRv...1a..47H,lindell2002,lindell2005,slater2014}. The selected items included ones where the activities were both likely (orbit direction and period) and unlikely (rise and set times) to impact learning. Because the time between the pre- and post-tests were so short, the questions on each test were isomorphic and matched on content, but not identical. The delayed post-test questions were similar to the ones found on the post-test. On all tests, each question had only one correct answer, and the participant's score was the sum of the number of correct answers, with all questions weighted equally. The item-test correlation and item difficulty turned out to be 0.37 and 0.34 for the pre-test and 0.43 and 0.59 for the post-test respectively based on all participant scores.

We also examined learning across conditions on each sub-topic on the assessments by comparing performance on the isomorphic questions. One of the deciding factors in choosing a certain technology as an educational tool is the affordances it offers to teach different aspects of the lesson. For learning about Moon phases, for example, the orbit and rotation periods are not controlled in the hands-on condition, but are constrained to be realistic in the desktop and VR simulations. Keeping the Moon's orbit fixed is a task the participants must preform in the hands-on condition that does not require conscious effort in the other conditions. Differences between the technologies to control or not control certain aspects of the activities, therefore, may lead to differences in participant performance on knowledge questions in different topic areas despite showing the same performance between conditions on the whole exam. Our test contained a question pair that, upon investigation, was not truly the same from pre-test to post-test. Though they referred to the same general topic of why phases occur, they were not isomorphic and the question pair was removed from our analysis. The removal of this question did not significantly alter the results of our analysis. 

\subsubsection{Demographic survey}
The demographic survey was provided at the end of the post-test and asked about a variety of participants' characteristics.

            \textbf{Gender:}
Participants were asked their gender with the choices of Male, Female, or Other. For the analysis involving gender as a variable we removed the three participants who chose Other, due to low sample size.

            \textbf{Video game Experience:}
Participants were asked ``On average, how frequently have you played video games over the past three years?" and the choices were daily, weekly, 1-2 times a month, 1-2 times a year, or never. We grouped the choices of daily, weekly, and 1-2 times a month into the category of `having significant video game experience' and the participants who chose 1-2 times a year or never as `not having significant video game experience.'

            \textbf{VR experience:}
Participants were asked ``How much virtual reality experience did you have before you participated in the experiment today?" and the choices were very minimum, moderate, a lot, or none. We grouped the participants who specified moderate or a lot as `having significant VR experience' and the other participants as `not having significant VR experience'.
Only one participant indicated having a lot of VR experience. We maintained three categories of VR experience: none, minimal VR experience, and moderate to high VR experience. 

            \textbf{Academic Major:}
Participants were asked to pick their academic major from a list or write their own. For our analysis we grouped participants according to whether their major was science-focused or non-science-focused. Non-science majors included arts, humanities, economics, social science, communication, and business. Science majors included physics, astronomy, engineering, biology, chemistry, computer science, information science, and math.

\subsubsection{Environmental attitudes survey}
We also added an exploratory set of questions to probe participants' socio-scientific attitudes through measurements of their individual differences in environmental attitudes using 15-items selected from the Environmental Attitude Inventory (EAI)~\citep{milfont2010environmental,Dunlap2000}. We suspected that participants in our VR activity might come out with different socio-scientific attitudes, based on similar outcomes found in previous work~\citep{Chang2016, ahn2016experiencing}. In our simulation, we wanted to test the possibility that viewing the planet from the unique vantage point of space (an astronaut-like perspective of the Earth) might have an impact on environmental consciousness~\citep{Riley2008,Poole2010,stepanova2019space}. This effect is attributed to a recognition of the the planet's limited resources~\citep{Dunlap1978}. Specifically, we selected five questions corresponding to each of three of the EAI’s subscales of particular interest to test whether the intervention influenced participants’ environmental movement activism, environmental threat, and human utilization of nature. Sample items were ``I would not want to donate money to support an environmentalist cause,” ``When humans interfere with nature it often produces disastrous consequences (R)” and ``In order to protect the environment, we need economic growth” (1 = Strongly agree to 7 = Strongly disagree), and the modified scale showed sufficiently reliability (Cronbach’s $\alpha=.83$). 

\subsubsection{Activity preference}
After completing the post-test, demographic questions, and EAI, participants were invited to try the other two conditions. They then completed a short survey that asked, ``Today, you have experienced three different ways of learning moon phases. Which of the three ways to simulate the moon phases is your favorite method?" Participants could select either ``Demonstration in virtual reality,'' ``Hands-on demonstration,'' or ``On-screen demonstration." They were then asked, ``Please briefly explain, why did you prefer this learning method?" with an open text box answer.

\subsubsection{Virtual Reality Only Analyses}
Analyses on presence and movement measures will be included in a subsequent publication, as only those in the virtual reality condition answered these questions or had head and hand movements, specifically, tracked.

            \textbf{Presence:}
Fourteen questions from two presence questionnaires \citep{aymerich2012effects,witmer1998measuring} measured participants' sense of spatial presence in the virtual reality activity. Questions drawn from these surveys involved asking participants how much they agreed with the following statements: ``I was really in outer space,” ``I felt surrounded by outer space,” ``I really visited outer space,” and ``The outer space seemed real.”  
The participants assigned to the VR condition were also asked if they had experienced any simulator sickness during the activity and if it was distracting. 

    \textbf{Movement tracking and controller use:}
The X, Y, and Z position and the pitch, yaw, and roll rotation of participants’ head and hands were recorded for the entire session in the VR condition. Movement from timepoint to timepoint may be calculated as the Euclidean distance (mm) between the positions of a tracker at time one and time two. Because the current paper focuses on the comparisons between the three conditions, analysis of the VR specific data, which includes movement, button presses, and presence measures, will be presented in a subsequent publication focused only on participants' experiences in the VR simulation.

    \subsection{Data analysis}
After all of the data had been collected we cleaned the data set by merging the survey results and verifying participant responses. The full data set was checked for inaccuracies, duplicates, and was translated into the proper form for analysis. Mathematica and R~\citep{R} packages were used for the analyses. All data and code are available through the CISER data archive (\url{ciser.cornell.edu/data/data-archive/})

First, performance on the pre-test, post-test, and delayed post-test were compared between conditions using Analysis of Variance. Effects of condition on environmental attitudes were compared between conditions using ANOVA. Linear regression analyses were used to evaluate effects of other variables on student performance on the post-test, controlling for pre-test score. Variables were selected based on prior literature and included condition, gender, video game experience, and VR experience. Variables were checked for correlation using the ggpairs function from the GGally package in R, recording each variable level as numeric. The only significant correlation was between gender and video game experience, with a correlation coefficient of 0.47. Gender and video game experience were, therefore, analyzed separately in all regression analyses. Variance Inflation Factors (VIF) were also calculated in all regression analyses to measure possible collinearity. Main effects were tested first without interactions, and then regressions with interactions between condition and video game experience, gender, and VR experience were tested next. For the regression analyses, the base variables were a female, non-science major in the hands-on condition, with no VR experience, and not having significant video game experience.

\section{Results}
We describe our results grouped by overall differences in learning and attitudes based on condition and then by interactions between other variables. 

    \subsection{Overall learning between conditions}
\begin{figure}[!ht]
\centering
\includegraphics[width=0.85\textwidth]{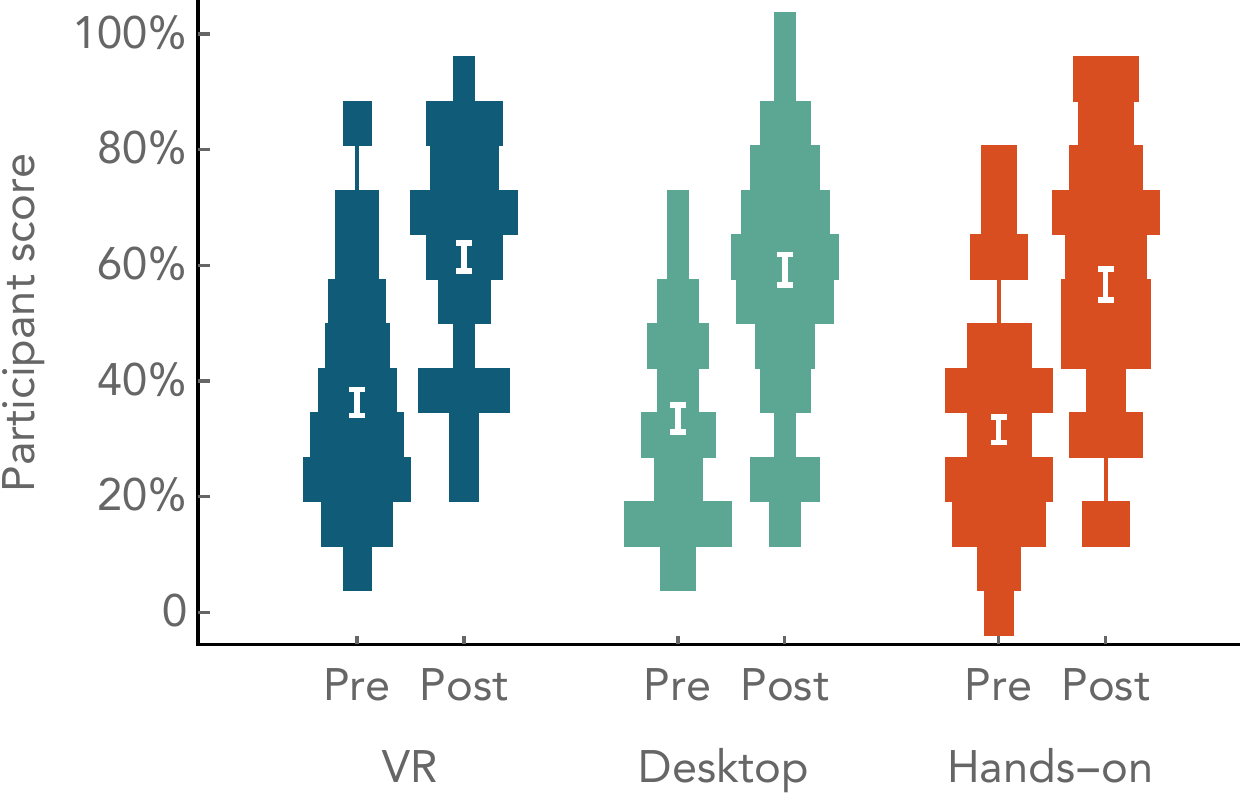}
\caption{\singlespace {\bf Pre and Post scores by condition}
Figure updated and adapted from~\cite{Madden2018VR}. An overall view of pre- to post-test performance. Violin plots show how scores were distributed across conditions and between pre- to post-test. Bins are 1 point wide. Average scores and standard error are indicated in white.}
\label{fig:histogram}
\vspace{-30pt}
\end{figure}

The average score on the pre-test was 33.7\% with no significant differences between the three conditions: \textit{F}(2,169)=1.04, $p=0.356$. Student performance significantly increased from pre- to post-test by 25.3\% on average (average score of 59.1\%), with no significant differences between the post-test scores across the three conditions: \textit{F}(2,169)=0.815, $p=0.444$ (Fig. \ref{fig:histogram}). An ANCOVA controlling for pre-score also showed no differences between the post-score between conditions: \textit{F}(2,168)=1.07, $p=0.344$. The average score on the delayed post-test(completed approximately four months after participation in the activity) was 39.0\%, again with no differences between conditions: \textit{F}(2,51)=0.571, $p=.568$. Our overall normalized gain of 0.38 is similar to that reported in an experiment involving pre-post moon phase assessments for a 20 minute inquiry-based tutorial of 0.54 \citep{lindell2005}.  

Our pre and post tests were not designed to be unidimensional and therefore did not reach a high Cronbach's alpha ($\alpha=0.52$ for pre test, $\alpha=0.66$ for post test). This is expected for multidimensional multiple choice tests \citep{Wilcox2014,Cortina1993}. Both of our tests gave scores across a sufficient range according to Ferguson's delta ($\delta=0.93$ for pre test, $\delta=0.96$ for post test).

\subsection{Learning by question}
\begin{figure}[!ht]
\centering
\includegraphics[width=0.9\textwidth]{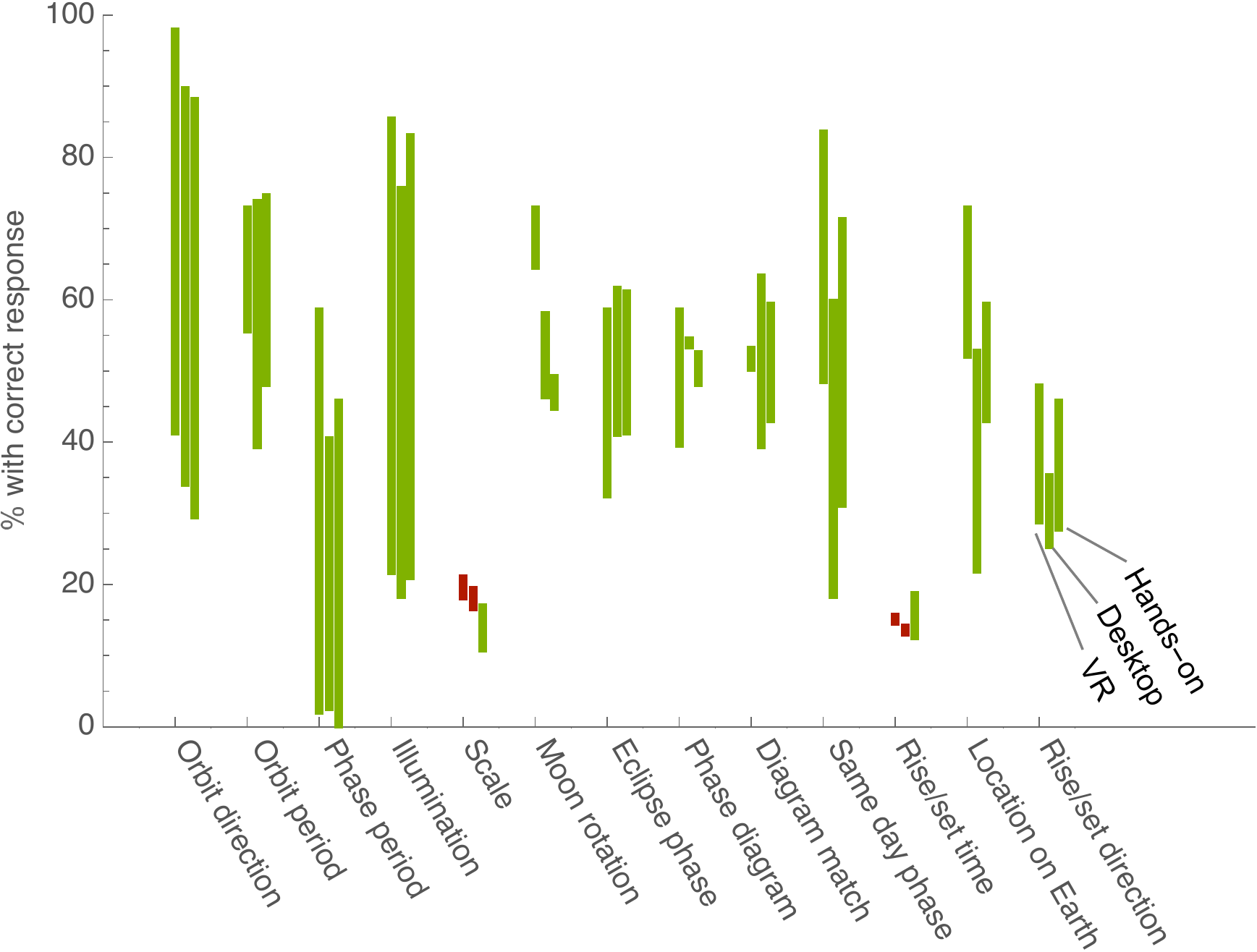}
\caption{\singlespace {\bf Learning by condition and topic}
Figure updated and adapted from~\cite{Madden2018VR}. The difference in participant responses from the pre-test to the post-test broken down by question topic. The percent of correct responses on the pre-test and post test for that topic are connected by a colored bar. A green bar signifies improvement, with the higher number representing the post-test score. A red bar means there were fewer correct responses on the post-test, with the higher number representing pre-test score.}
\label{fig:prePostbyQ}
\end{figure}

Breaking down the pre- and post-test scores by question show consistent gains between conditions, with no statistically significant differences (Fig. \ref{fig:prePostbyQ}). Across the conditions, some topic areas showed large gains over 40\% (orbit period, phase period, and illumination), while others showed small gains less than 10\%  (scale, Moon rotation, phase diagram, rise/set time), consistent with previous research on learning about Moon phases ~\citep{lindell2005,wilhelm2018}. This demonstrates that learning did not differ between conditions on sub-topics that may have held learning benefits within the different conditions.

    \subsection{Attitudes towards the conditions}
\begin{wrapfigure}{r}{.45\textwidth}
\centering
\includegraphics[width=0.43\textwidth]{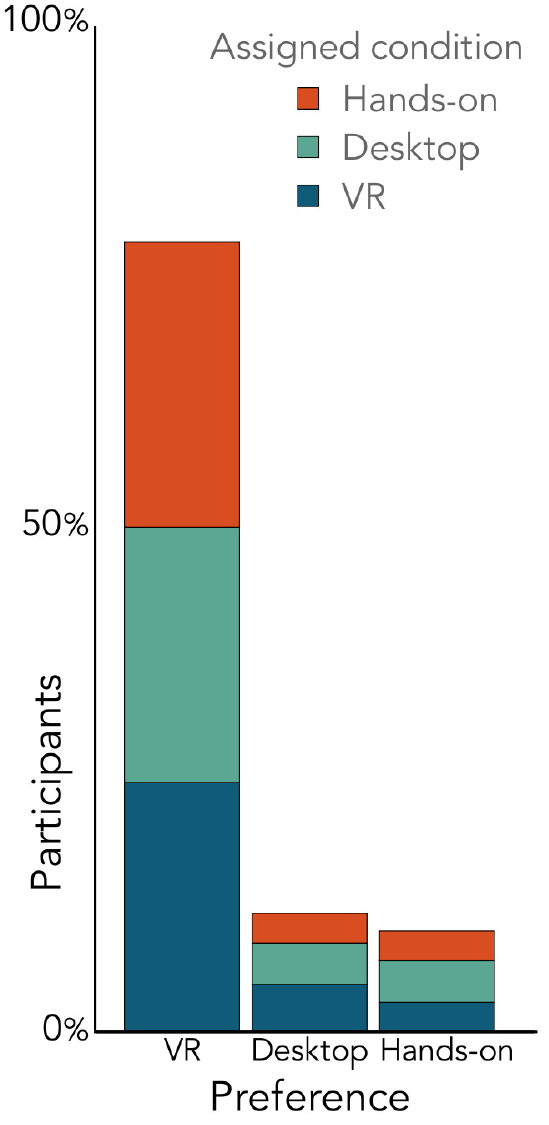}
\caption{\singlespace {\bf Participant preference}
Preferred choice after viewing each activity. }
\label{fig:preference}
\vspace{-40pt}
\end{wrapfigure}

Consistent with previous work~\citep{dede2009immersive} participants overwhelmingly ($Chi^{2}=152$, $p<0.00001$) preferred the VR condition as their favorite learning method independent of their study condition (Fig. \ref{fig:preference}). 

78\% preferred the VR condition and when describing their reason used phrases such as: the VR condition was ``easier to visualize", ``more realistic", ``more immersive", ``more fun", ``more interesting", and ``the most accurate". 
The participants who preferred VR generally said that seeing the full picture in a realistic way helped with their learning. The main contributors to this feeling were the ease of viewing different perspectives and having easier control over the system. For example, one participant wrote: 
\begin{quote}
``Having a overall space to see where everything is helps a lot. Even in class I still had a hard time understanding what they are talking about in concept. But I think I learned a lot in VR and being able to manipulate the environment on my own accord. It seems more engaging than the 2 other methods."
\end{quote}

For the few students who did not prefer VR, participants said that they  ``did not notice everything that was going on", found it ``a little too complex", it ``made me dizzy and confused," or generally indicated feeling uncomfortable or overwhelmed. They preferred either the hands-on or desktop conditions because they were more familiar. For example, a participant who preferred the desktop condition wrote:
\begin{quote}
``The VR was cool but since I'm very new to it I spent most of my time just trying to figure out how it worked--it was also tough to find where the Sun and Moon were at times because of how `large' the environment was. The desktop game was more familiar and easy-to-use for me. Personally."
\end{quote}

The 12\% of participants who preferred the desktop condition also used phrases such as: the desktop condition was ``less overwhelming", ``easier to control", and ``very easy to follow". When participants did not like the desktop condition they said: the desktop condition ``gave a limited field of view", and ``[was] a lot harder for me to navigate".

The 10\% of participants who preferred the hands-on condition used phrases such as ``easiest and fastest", ``I was able to more clearly focus", and ``I got distracted by the other methods". A participant who preferred the hands-on condition wrote: 
\begin{quote}
``I really liked the virtual reality method. And it gave me more information than the other two methods, for instance, what time of day certain Moon phases would rise and set. Nevertheless, it was almost too overwhelming and it was as if I was too excited to be in space to actually commit to learning the Moon phases. With the hands-on demonstration. there was nothing to distract me. And, obviously, controlling the demonstration felt about as natural as possible." 
\end{quote}

    \subsection{Effects on Environmental Attitudes}
Our second hypothesis, that there would be a main effect of condition on environmental attitudes, such that participants in the VR conditions would evince greater environmental support, was also not supported. A one-way ANOVA found no significant differences in environmental attitudes among treatment groups (means $\pm$ standard error were VR = 5.24 $\pm$ 0.09,  Desktop = 5.25 $\pm$ 0.1, Hands-on = 5.35 $\pm$ 0.08; \textit{F}(2, 169) 0.41, p = .66), suggesting that the methods of learning Moon phases did not affect participants’ environmental attitudes. Although we expected that brief exposure in the VR condition to the unique vantage point of the Earth could possibly increase participants’ concerns for environmental issues (by priming the concept of limited resources), the instrument's typical use as a trait-level instrument may make it unsurprising that we did not observe such an effect.

    \subsection{Effects due to participant demographics}
We conducted exploratory analyses to identify whether other variables were interacting with participants' experiences in the three conditions. Based on prior literature, we focused on gender, major, video game experience, and quantity of VR experience. We first explored main effects alone and then explored interactions between the variables and condition. When examining the role of the demographic variables on performance, we found a correlation between video game experience and gender (column 1 of Fig. \ref{fig:VGGenderCorr}), but no correlation between any of the other variables. Therefore, video game experience and gender were analyzed in separate regression models.

    \subsubsection{Main effects between condition and demographic variables}
As shown in Table \ref{table:model12}, students' pre-score was a significant predictor of their post-scores across conditions and, again, there was no significant differences in post-score between conditions. We interpret the regression coefficients as the simple effect size, $\beta$ \citep{baguley2009standardized}. The main effects model also found that there is a significant effect for major, with science majors slightly outperforming non-science majors, even when controlling for pre-score. Neither participants' gender, video game experience, or VR experience had significant main effects.

\subsubsection{Interactions between condition and demographic variables}
Experience in VR did not have a significant interaction with condition (Table \ref{table:model34}). Notably, this indicates that having experience with VR did not improve students' learning in the VR condition, even though some students claimed that not having experience in VR impeded their learning.

\begin{figure}[th!]
\centering
\includegraphics[width=0.85\textwidth]{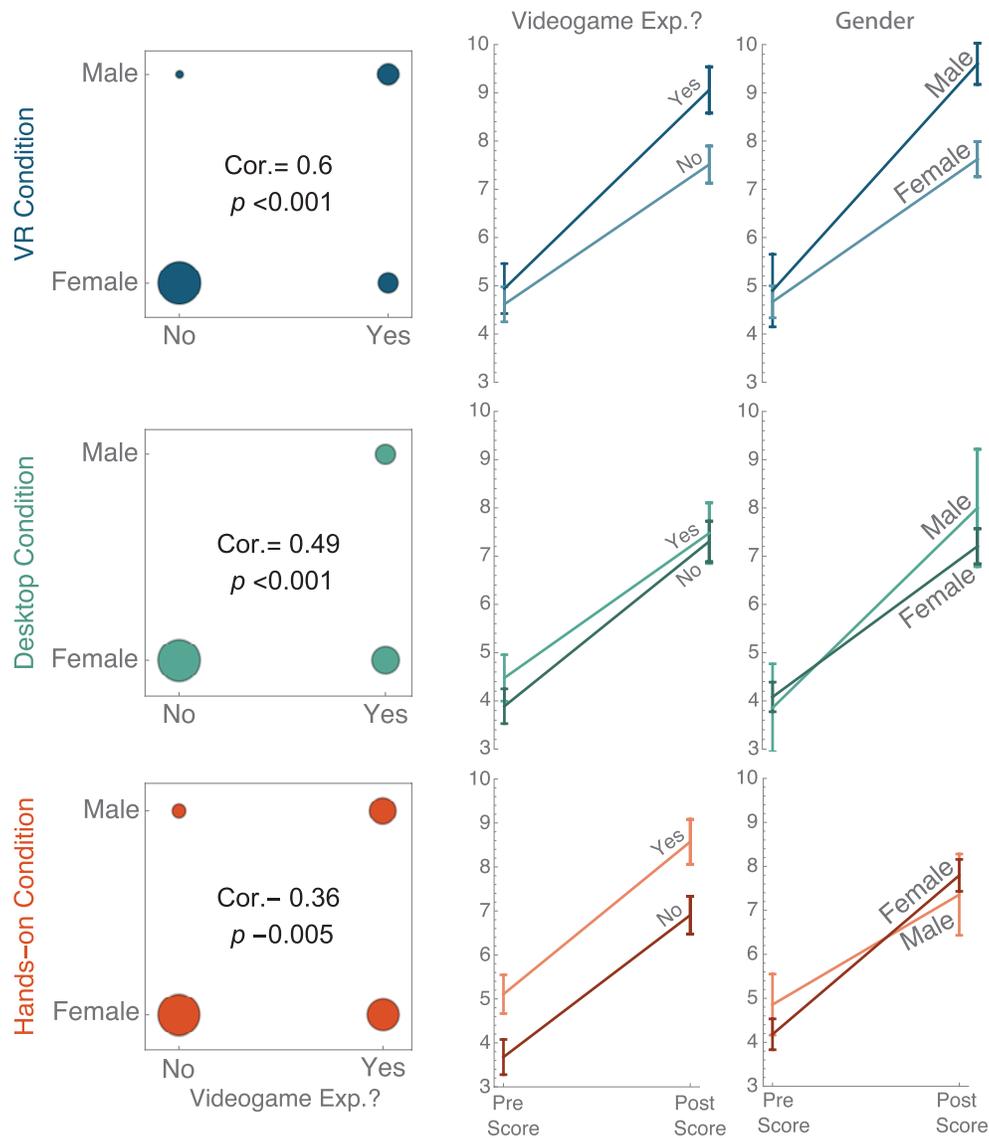}
\caption{\singlespace {\bf Relationship between gender, video game experience, and scores across conditions}
The first column shows the correlations between gender and video game experience. The second column shows the average scores and standard error at pre- and post-test for students with and without video game experience. The third column shows the average scores and standard error at pre- and post-test for students identifying as male and female. See Table S1 for line details.}
\label{fig:VGGenderCorr}
\end{figure}

The interactions with gender and video game experience are more complicated. As seen in Table \ref{table:model34}, there is an interaction between gender and condition in Model 3, with men outperforming in the VR condition. However, Model 4 indicates that there is a main effect with video game experience (seen as marginally significant in Model 2), and a potential interaction between condition and video game experience. From Fig. \ref{fig:VGGenderCorr}, this interaction is likely understood through the pre-test scores. Only in the hands-on condition do we see differences in pre-test scores between participants with and without video game experience. We believe these results can be understood such that men and participants with video game experience learn more from the VR condition, followed by the desktop and then hands-on conditions. Given the correlation between video game experience and gender, it is unclear which variable is responsible for the differences. This correlation between gender and video game experience may be more clearly understood by elaborating on the type and detailed quantity of video game experience in future studies. 
\begin{table}[!ht]
\centering
\caption{\singlespace 
{\bf Linear regression models for gender and video game experience including main effects only.} Regression analysis models for students’ post-score with main effects for pre-score, condition, gender, video game experience, VR experience, and students’ major. Gender and video game experience are analyzed in separate models as the variables were highly correlated.}
\resizebox{1\textwidth}{!}{
\begin{tabular}{|l|c|c|c|c|c|c|c|c|c|c|}
\hline
                     & \multicolumn{5}{c|}{Model 1}                           & \multicolumn{5}{c|}{Model 2}                          \\ \hline
Term                 & B     & SE   & $t$   & $p$          & VIF              & B     & SE   & $t$   & $p$          & VIF             \\ \hline
$Intercept$          & 4.52  & 0.53 & 8.45  & \textless{}.001$^{***}$ & 0.0  & 4.42  & 0.54 & 8.26  & \textless{}.001$^{***}$ & 0.0 \\ \hline
$Pre\mhyphen Score$  & 0.50  & 0.08 & 6.65  & \textless{}.001$^{***}$ & 1.1  & 0.50  & 0.08 & 6.62  & \textless{}.001$^{***}$ & 1.1 \\ \hline
$Condition:Desktop$  & -0.05 & 0.41 & -0.13 & 0.898        & 1.4              & 0.00  & 0.40 & 0.00  & 0.998        & 1.4             \\ \hline
$Condition:VR$       & 0.23  & 0.41 & 0.56  & 0.578        & 1.4              & 0.32  & 0.41 & 0.78  & 0.44         & 1.4             \\ \hline
$Gender:Male$        & 0.52  & 0.43 & 1.20  & 0.233        & 1.0              & \multicolumn{5}{c|}{}                                 \\ \hline
$VGexp:Significant$  & \multicolumn{5}{c|}{}                                  & 0.63  & 0.35 & 1.79  & 0.075        & 1.1             \\ \hline
$VRexp:Minimal$      & 0.70  & 0.37 & 1.89  & 0.060        & 1.3              & 0.70  & 0.37 & 1.91  & 0.058        & 1.3             \\ \hline
$VRexp:Moderate$     & -0.17 & 0.51 & -0.33 & 0.741        & 1.4              & -0.19 & 0.50 & -0.39 & 0.699        & 1.4             \\ \hline
$Major:Science$      & 1.14  & 0.36 & 3.21  & 0.002$^{**}$ & 1.1              & 1.00  & 0.35 & 2.83  & 0.005$^{**}$ & 1.2             \\ \hline
$R^{2}$              & \multicolumn{5}{c|}{0.325}                      & \multicolumn{5}{c|}{0.330}                                   \\ \hline
$R^{2}_{Adj}$        & \multicolumn{5}{c|}{0.296}                      & \multicolumn{5}{c|}{0.302}                                   \\ \hline
$AIC$                & \multicolumn{5}{c|}{748.41}                     & \multicolumn{5}{c|}{758.67}                                  \\ \hline
\end{tabular}
}
\begin{flushleft} 
$^* p<.05$, $^{**} p<.01$, $^{***} p<.001$ 
\end{flushleft}
\label{table:model12}
\end{table}

\begin{table}[!ht]
\centering
\caption{\singlespace  
{\bf Linear regression models including main effects with gender and video game interaction.} Regression analysis models for students’ post-score with main effects for pre-score, condition, gender, video game experience, VR experience, and students’ major and interactions between condition and gender, VR experience, and video game experience. Gender and video game experience are analyzed in separate models as the variables were highly correlated.}
\resizebox{\textwidth}{!}{
\begin{tabular}{|l|c|c|c|c|c|c|c|c|c|c|}
\hline
                                         & \multicolumn{5}{c|}{Model 3}                       & \multicolumn{5}{c|}{Model 4}                        \\ \hline
Term                                     & B    & SE   & $t$  & $p$ & VIF                     & B    & SE   & $t$  & $p$ & VIF                      \\ \hline
$Intercept$                              & 4.76 & 0.61 & 7.74 & \textless{}.001$^{***}$& 0.0 & 4.15 & 0.65 & 6.38 & \textless{}.001$^{***}$ & 0.0 \\ \hline
$Pre\mhyphen Score$                      & 0.50 & 0.08 & 6.71 & \textless{}.001$^{***}$& 1.1 & 0.48 & 0.08 & 6.28 & \textless{}.001$^{***}$ & 1.1 \\ \hline
$Condition:Desktop$                      & -0.55& 0.72 & -0.77& 0.443 & 4.4                   & 0.83 & 0.79 & 1.05 & 0.296 & 5.7                    \\ \hline
$Condition:VR$                           & -0.25& 0.68 & -0.36& 0.719 & 4.0                   & 0.55 & 0.75 & 0.73 & 0.465 & 5.1                    \\ \hline
$Gender:Male$                            & -0.76& 0.67 & -1.14& 0.258 & 2.6                   & \multicolumn{5}{c|}{}                               \\ \hline
$VGexp:Significant$                      & \multicolumn{5}{c|}{}                              & 1.42 & 0.58 & 2.43 & 0.016$^{*}$ & 3.3              \\ \hline
$VRexp:Minimal$                          & 0.88 & 0.62 & 1.41 & 0.160 & 3.7                   & 0.92 & 0.61 & 1.50 & 0.136 & 3.8                    \\ \hline
$VRexp:Moderate$                         & -0.42& 0.88 & -0.48& 0.634 & 4.3                   & -1.11& 0.87 & -1.27& 0.205 & 4.2                    \\ \hline
$Major:Science$                          & 1.18 & 0.35 & 3.37 & 0.001$^{**}$ & 1.1            & 1.06 & 0.36 & 2.99 & 0.003$^{**}$ & 1.2             \\ \hline
$Gender:Male \times Condition:Desktop$   & 1.77 & 1.12 & 1.59 & 0.115 & 1.8                   & \multicolumn{5}{c|}{}                               \\ \hline
$Gender:Male \times Condition:VR$        & 2.56 & 1.00 & 2.57 & 0.011$^{*}$ & 2.2             & \multicolumn{5}{c|}{}                               \\ \hline
$VRexp:Minimal \times Condition:Desktop$ & 0.28 & 0.89 & 0.32 & 0.752 & 4.3                   & -0.17& 0.87 & -0.20& 0.841 & 4.4                    \\ \hline
$VRexp:Minimal \times Condition:VR$      &-0.66 & 0.90 & -0.73& 0.467 & 3.4                   & -0.74& 0.90 & -0.83& 0.407 & 3.5                    \\ \hline
$VRexp:Moderate \times Condition:Desktop$&-0.18 & 1.29 & 0.14 & 0.888 & 2.5                   & -0.09& 1.28 & -0.07& 0.943 & 2.5                    \\ \hline
$VRexp:Moderate \times Condition:VR$     & 0.95 & 1.15 & 0.83 & 0.409 & 3.6                   & 1.78 & 1.14 & 1.56 & 0.120 &  3.5                   \\ \hline
$VG exp:Has \times Condition:Desktop$    & \multicolumn{5}{c|}{}                              & -1.82& 0.84 & -2.17& 0.03$^{*}$ & 3.0               \\ \hline
$VG exp:Has \times Condition:VR$         & \multicolumn{5}{c|}{}                              & -0.40& 0.86 & -0.47& 0.642 & 2.7                    \\ \hline
$R^{2}$                                  & \multicolumn{5}{c|}{0.373}                         & \multicolumn{5}{c|}{0.372}                          \\ \hline
$R^{2}_{Adj}$                            & \multicolumn{5}{c|}{0.321}                         & \multicolumn{5}{c|}{0.321}                          \\ \hline
$AIC$                                    & \multicolumn{5}{c|}{747.89}                        & \multicolumn{5}{c|}{759.59}                         \\ \hline
\end{tabular}
}
\begin{flushleft} 
$^* p<.05$, $^{**} p<.01$, $^{***} p<.001$ 
\end{flushleft}
\label{table:model34}
\end{table}

\clearpage
\section{Discussion}
\vspace{-5pt}
In this study, we performed a controlled experiment of student learning about \begin{wrapfigure}{r}{0.52\textwidth}
\vspace{-15pt}
\includegraphics[trim=100 0 100 0, clip, width=0.5\textwidth]{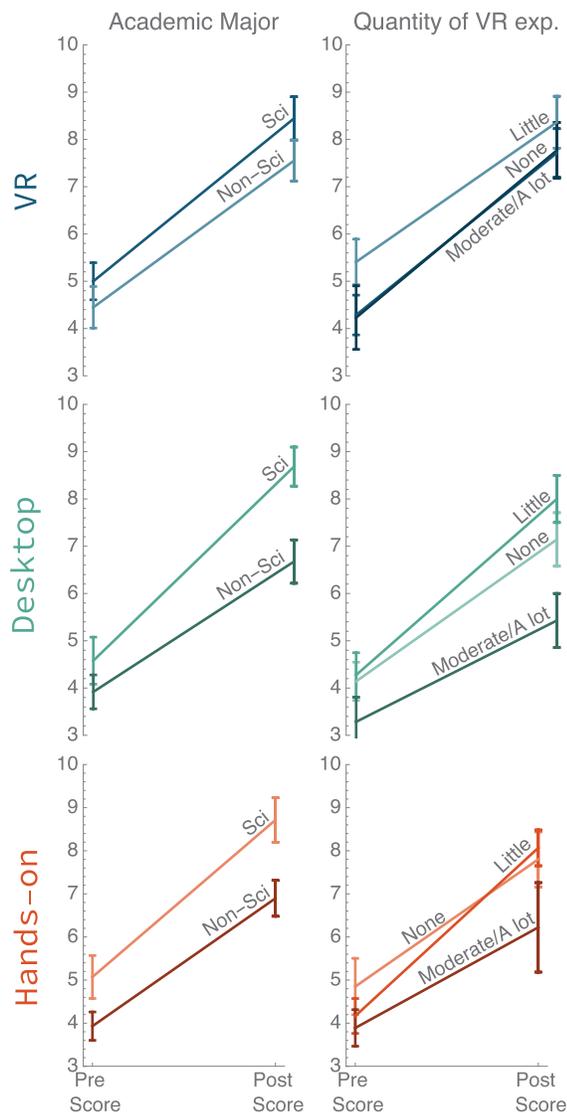}
\caption{\singlespace {\bf Interactions plots for academic major, and quantity of VR experience}
Each column shows the relationship present between one of these measures and pre to post score by condition. Mean score and standard error are shown, see Table S2 for line details.}
\label{fig:scitypeexp}
\end{wrapfigure}Moon phases through three different modalities: a hands-on activity, a desktop simulation, and a VR simulation.We found no overall effect for condition on students' learning on an immediate post-test, nor on a delayed post-test four months after the intervention. We did find that students with declared science majors outperformed non-science majors in all conditions. Student learning in the VR condition was not improved with VR experience. Despite these results, students overwhelmingly preferred learning in the VR condition. 

There are several possible interpretations of these results. First, our hypothesis had been that the VR simulation would improve performance by reducing the real-world complexities of a hands-on activity and providing a realistic, embodied learning experience. The results suggest that these affordances did not impact learning about Moon phases. On the other hand, one may interpret that the VR condition provided equivalent learning to the other two modalities, while dramatically improving students' attitudes towards the learning experience. 

The study also explored interactions between conditions and students' video game experience and gender. Gender and video game experience were significantly correlated in our study, and men performed better in the VR condition. This means that either video game experience, being male, or a combination of the two provides an advantage. 

There are several studies that suggest video game experience would provide this advantage. One study found that video games can have a beneficial effect on completing complex spatial tasks, visuomotor coordination, and multiple object tracking\citep{Spence2010}. Other studies suggest that the benefits of video games on such abilities can be gained after a short time playing an action game and is not dependent on gender~\citep{Feng2007}.  Furthermore, research has shown that men are more drawn to the type of video games suggested to provide these benefits~\citep{Phan2012}. Combining these studies paints a picture that these advantages are not inherently male but may be caused by a particular type of video game experience that so happens to be common for men. 

Alternatively, since the 1970's, several studies have suggested that men have better spatial reasoning than women. However, comprehensive studies have shown recently that, while differences in spatial reasoning between men and women may be present in certain cases, we may not fully understand their causes. Newcombe and Stieff claimed, ``from a practical educational standpoint, the most relevant fact is that the relevant skills can be improved in both men and women"~\cite[p. 962]{Newcombe2012}. Research has also previously found that men outperformed women on post-test conceptual assessments of Moon phase understanding, but only on items involving spatial reasoning~\citep{Wilhelm2009}. However, it has also been found that men's and women's scores improve similarly from pre- to post-test with appropriate instruction~\citep{Wilhelm2009, Jackson2015}, again suggesting that performance can be improved in both men and women. 

While our study was unable to prove video game experience was the true contributor to performance increases in the VR condition over gender, literature on the subject suggests that video game experience and not gender may be the affecting variable. While further experimental work should examine this proposition, this points the way to improving learning experiences in virtual reality such that they benefit all learners. If all participants did as well in the virtual reality condition as did men/people with video game experience, then virtual reality could be an overall more effective teaching tool than the other two tested modes of teaching moon phases. 

    \subsection{Limitations}
There are several limitations to this study that should be considered when interpreting our results. This study was focused on learning in an activity that was based on physics and astronomy concepts. Thus, we should be careful when applying our findings to learning in other subjects or even other concepts within physics and astronomy \citep{Winn2002}. Furthermore, the participant pool for this study was not perfectly representative of either a typical college classroom or typical physics/astronomy classroom. For example, our study was comprised of 80\% women while the American Physical Society reports that only around 20\% of undergraduate physics degrees are awarded to women ~\citep{APSGender}. In addition, the significant interaction between gender and condition was based on a small sample of male participants and future work should evaluate this result with more equally distributed samples.

The contribution that we believe this paper makes to the discussion on gender/video game effects on learning in virtual reality is based on exploratory analyses. Thus, future work should explicitly test these hypotheses in a pre-registered study and include more detailed questions related to the type and quantity of video game experience.  

While VR technology has advanced rapidly, it is still not ideal. Control responsiveness, motion sickness, limited resolution and field of view are all technological obstacles that can still break immersion and distract from learning using today's equipment. In contrast, the technologies we used for the hand-on condition and the desktop condition have essentially plateaued compared to VR, meaning the results of this study may change as VR technology progresses. 

Compared to VR, the participants were very familiar with the technology used in the hands-on condition and desktop condition. Many participants had never experienced VR before our study or had very limited experience with using a VR headset. This suggests that many of our participants were managing high cognitive load as they attempt to become comfortable with VR, understand the activity, and learn about Moon phases. Participants in the other two conditions did not have such a high barrier to getting comfortable with the technology handed to them. For participants who preferred the non-VR conditions, a common praise was familiarity with the equipment. This suggests that as a college population becomes more familiarly and comfortable using VR, the results of this study may change. Future research should measure students' cognitive load during the activity explicitly, such as through eye-tracking~\cite{Zu2018}, self-report surveys~\cite{Zu2018, Paas1992, Leppink2013, Paas2003, Paas1994}, physiological indicators~\cite{Paas1994, Paas2003}, or electroencephalography~\cite{Antonenko2010}. We note, however, that many of the common methods have several limitations~\cite{DeJong2010}, such as that different methods may or may not be able to distinguish different types of cognitive load~\cite{Leppink2013, DeLeeuw2008, DeJong2010, Paas2003}. It is unclear which may be at play in these activities.

\subsection{Future Work}

Our study joins others in suggesting that virtual reality is a promising technology as an educational tool but does not, in itself, guarantee a learning advantage over traditional hands-on activities or desktop simulations. There are several areas for future work, based on this analysis.

Future work should focus on further investigating the potential relationship between video game experience and learning gains. This strategy has three components. First, researchers should attempt to sample both male and female participants with equivalent video game experience. Video game experience should be more precisely characterized, focusing on game types that require more visuospatial navigation skills. Video game experiences tend to build procedural or motor skills \citep{rosser2007impact} which do not typically go away after long-term disuse. Measures should thus include lifetime experience, rather than frequency of use alone. If video game experience is confirmed to provide users with skills that allow them to learn better in VR, then researchers can use this knowledge to provide participants with tools to learn to interact with the environment in order to provide a level playing field to all participants. 

The VR educational experience can also be improved through enhancing the user interface to maximize VR’s full potential. This does not require waiting for the technology to improve or to become more widely used; instead, designers can take cues from participant's responses. One strategy would be to enhance users' sense of embodiment. In this study, users were not embodied in avatars, which may have affected their feeling of presence or ‘being there’ within an environment \citep{slater2017implicit}. This in turn could have affected how willing participants were to interact with the environment, thereby reducing their learning gains. Examining existing movement data in the VR condition could provide hints for how to design such a simulation. Similarly, our study used guiding questions that must be answered before proceeding into the environment which may have prevented exploration. Thus, future studies should evaluate how avatar embodiment and interactive text affect movement and learning within an environment as well as how visuospatial ability relates to movement in VR. 

Finally, participants worked alone in our study, but each of our conditions could also be used to collaborate. Two students could help each other learn concepts through answering questions together, or an instructor could emphasize key components that are easily missed~\citep{Chi2009}. Indeed in virtual reality in particular, the anonymity can  make it a safer space for learning \citep{yu2009creating}, thereby encouraging students to make mistakes and learn from them without fear of judgement. If participants experience gains from social learning, such gains may be more noticeable in virtual environments.

However, future work must also remain open to the possibility that the excitement and engagement produced by virtual reality experiences may not translate into learning gains in all domains.

\section{Conclusion}
This study has several takeaway findings. First, participants' learning gains from pre- to post-test were not significantly different, on average, between the VR, desktop, and hands-on conditions. Participants preformed similarly well on each question topic across the three conditions. We found no strong evidence that participants' retention differed between conditions after four months. Our hypothesis, that VR would improve learning by simplifying real-world complexities and providing an embodied learning experience, was not supported. Nonetheless, participants strongly favored learning in the VR activity. 

Guided by the literature on virtual reality and learning, we also collected data on demographic measures (gender, video game experience, virtual reality experience, and major) to explore predicted interactions, all of which are reported here. We did find a positive effect from gender within the VR condition. However, video game experience and gender were significantly correlated in our study, and the literature suggests that video game experience may be the main reason for the performance increase. 

This study has allowed us to determine new experiment designs that will help explore the reasons we saw similar learning gains across conditions. What remains promising about VR is that, relative to a ball on a stick and 2D computer games, it is a technology that is rapidly advancing. Given that participants' unfamiliarity with VR and the technical roughness of the simulation, the fact that participants were able to learn as much as those in the other conditions may bode well to support VR as a better educational tool when the majority of students are comfortable learning in a virtual environment.

Given that the learning was the same regardless of condition what remains is the fact that participants widely favored the VR experience. As a method of engaging students, using VR was successful in our study and was not achieved at the cost of learning gains. Novelty does diminish however, so an advantage based on novelty alone will cease to be an advantage as exposure rises. 

All together, there are many avenues to explore with educational VR. Future work should encourage a more comprehensive look into VR’s ability as an educational tool, such that a participant’s experience is considered from multiple perspectives. 

\clearpage

\section*{Data Access}
DOI for accompanying data: ~\href{https://doi.org/10.6077/f12c-m973}{10.6077/f12c-m973}

\section*{Acknowledgments}
This work was supported by Oculus Education with special thanks to Cindy Ball and Cindi McCurdy. We would like to thank all the graduate and undergraduate students who helped develop the simulations and run participants: Tristan Stone, Yilu Sun, Akhil Gopu, Kristi Lin, Jason Wu, Anirudh Maddula, Frank Rodriguez, Alice Nam, Phil Barrett, Dwyer Tschantz, Connor Lapresi, Giulia Reversi, Keun Youk, Albert Tsao, and Annie Hughey. We would also like to thank Kimberly Williams from the Cornell Center for the Integration of Research, Teaching, and Learning, Stephen Parry of the Cornell Statistical Consulting Unit, Daniel Alexander and Florio Arguillas from the Cornell Institute of Social and Economic Research, and the entire Cornell Physics Education Research Lab for support, comments, and assistance.

\newpage
\section{Supporting information}
\paragraph*{S1 Table}
{\bf Modeling interactions between video game experience and gender with condition} Supporting data for Fig. \ref{fig:VGGenderCorr}
\begin{table}[!ht]
\centering
\begin{tabular}{|c|c|c|c|c|c|c|c|}
\hline
\multicolumn{8}{|c|}{Video game experience}                               \\ \hline
Condition & Exp.   & Intercept & SE   & p       & Slope & SE   & p       \\ \hline
VR        & Yes    & 4.94      & 0.50 & 3.0E-11 & \textbf{4.12}  & 0.71 & 1.8E-06 \\ \hline
VR        & No     & 4.62      & 0.37 & 7.1E-20 & \textbf{2.90}  & 0.53 & 5.1E-07 \\ \hline
Desktop   & Yes    & 4.48      & 0.56 & 7.7E-10 & 3.00  & 0.79 & 4.9E-04 \\ \hline
Desktop   & No     & 3.89      & 0.39 & 5.5E-15 & 3.42  & 0.55 & 4.0E-08 \\ \hline
Hands-on  & Yes    & \textbf{5.11}      & 0.48 & 5.6E-15 & 3.46  & 0.67 & 3.9E-06 \\ \hline
Hands-on  & No     & \textbf{3.68}      & 0.41 & 1.4E-12 & 3.23  & 0.58 & 7.6E-07 \\ \hline
\multicolumn{8}{|c|}{Gender}                                             \\ \hline
Condition & Gender & Intercept & SE   & p       & Slope & SE   & p       \\ \hline
VR        & Male   & 4.90      & 0.61 & 2.4E-07 & \textbf{4.70}  & 0.86 & 3.7E-05 \\ \hline
VR        & Female & 4.67      & 0.35 & 5.2E-23 & \textbf{2.96}  & 0.49 & 4.1E-08 \\ \hline
Desktop   & Male   & 3.86      & 1.07 & 3.7E-03 & 4.14  & 1.52 & 1.8E-02 \\ \hline
Desktop   & Female & 4.08      & 0.34 & 4.2E-21 & 3.12  & 0.48 & 2.6E-09 \\ \hline
Hands-on  & Male   & 4.86      & 0.82 & 2.8E-06 & 2.50  & 1.16 & 4.0E-02 \\ \hline
Hands-on  & Female & 4.18      & 0.36 & 1.6E-19 & 3.61  & 0.50 & 2.5E-10 \\ \hline
\end{tabular}
\begin{flushleft} 
\end{flushleft}
\end{table}
\clearpage
\paragraph*{S2 Table}
{\bf Modeling interactions between academic major, and VR experience quantity with condition} Supporting data for Fig. \ref{fig:scitypeexp}
\begin{table}[!ht]
\centering
\begin{tabular}{|c|c|c|c|c|c|c|c|}
\hline
\multicolumn{8}{|c|}{Academic Major}                                             \\ \hline
Condition & Major          & Intercept & SE   & p       & Slope & SE   & p       \\ \hline
VR        & Science        & 5.40      & 0.62 & 1.6E-09 & 3.27  & 0.87 & 8.1E-04 \\ \hline
VR        & Non-Science    & 5.43      & 0.99 & 1.5E-04 & 2.29  & 1.41 & 1.3E-01 \\ \hline
Desktop   & Science        & 4.38      & 0.37 & 1.2E-08 & \textbf{3.88}  & 0.52 & 3.4E-06 \\ \hline
Desktop   & Non-Science    & 4.17      & 0.79 & 2.7E-05 & \textbf{2.17}  & 1.12 & 6.5E-02 \\ \hline
Hands-on  & Science        & \textbf{5.75}      & 0.89 & 1.7E-06 & 2.83  & 1.26 & 3.5E-02 \\ \hline
Hands-on  & Non-Science    & \textbf{4.00}      & 0.72 & 1.3E-04 & 3.00  & 1.02 & 1.3E-02 \\ \hline
\multicolumn{8}{|c|}{Quantity of VR experience}                                  \\ \hline
Condition & Quantity       & Intercept & SE   & p       & Slope & SE   & p       \\ \hline
VR        & None           & \textbf{5.41}      & 0.52 & 2.8E-13 & 2.95  & 0.73 & 2.2E-04 \\ \hline
VR        & Little         & 4.29      & 0.47 & 2.4E-11 & 3.43  & 0.66 & 6.8E-06 \\ \hline
VR        & Moderate/A lot & 4.23      & 0.63 & 6.3E-07 & 3.54  & 0.89 & 5.8E-04 \\ \hline
Desktop   & None           & 4.14      & 0.49 & 2.1E-10 & 3.00  & 0.70 & 1.0E-04 \\ \hline
Desktop   & Little         & 4.28      & 0.48 & 3.5E-12 & 3.72  & 0.69 & 1.3E-06 \\ \hline
Desktop   & Moderate/A lot & 3.29      & 0.55 & 6.2E-05 & \textbf{2.14}  & 0.77 & \textbf{1.7E-02} \\ \hline
Hands-on  & None           & 4.85      & 0.65 & 5.2E-09 & 2.95  & 0.91 & 2.6E-03 \\ \hline
Hands-on  & Little         & 4.17      & 0.41 & 1.7E-14 & 3.90  & 0.58 & 8.3E-09 \\ \hline
Hands-on  & Moderate/A lot & 3.89      & 0.79 & 1.6E-04 & 2.33  & 1.12 & 5.4E-02 \\ \hline
\end{tabular}
\begin{flushleft} 
\end{flushleft}
\end{table}

%% file: 5_observe/ms.tex
\section*{Abstract} 
The search for life in the universe is currently focused on Earth-analog planets. However, we should be prepared to find a diversity of terrestrial exoplanets not only in terms of host star but also in terms of surface environment. Simulated high-resolution spectra of habitable planets covering a wide parameter space are essential in training retrieval tools, optimizing observing strategies, and interpreting upcoming observations. Ground-based extremely large telescopes like ELT, GMT, and TMT; and future space-based mission concepts like Origins, HabEx, and LUVOIR are designed to have the capability of characterizing a variety of potentially habitable worlds. Some of these telescopes will use high precision radial velocity techniques to obtain the required high-resolution spectra ($R\approx100,000$) needed to characterize potentially habitable exoplanets.

Here we present a database of high-resolution (0.01 cm$^{-1}$) reflection and emission spectra for simulated exoplanets with a wide range of surfaces, receiving similar irradiation as Earth around 12 different host stars from F0 to K7.

Depending on surface type and host star, we show differences in spectral feature strength as well as overall reflectance, emission, and star to planet contrast ratio of terrestrial planets in the Habitable zone of their host stars. Accounting for the wavelength-dependent interaction of the stellar flux and the surface will help identify the best targets for upcoming spectral observations in the visible and infrared. 
 
All of our spectra and model profiles are available online. 

\section{Introduction}
Currently, about 4000 extrasolar planets have been detected orbiting Main Sequence stars with dozens of terrestrial planets orbiting in their habitable zone (HZ) \citep{Kane2016, Johns2018, Berger2018}. The detected rocky exoplanets in the HZ show a wide variety of sizes and stellar hosts. For now, we are unable to characterize their atmospheres.

If our Solar System is any indication \cite[e.g.][]{Madden2018Abio,Krissansen2016,Cahoy2010,Lundock2009,Traub2003}, exoplanets should show a large diversity in composition and surface type. Therefore, it is important to model a wide range of surfaces and stellar hosts for rocky planets to expand the spectral database we will use to characterize planets and search for signs of life in their atmospheres.  

Direct observations that provide reflection and emission spectra of habitable zone exoplanets are critical to identifying signs of life on exoplanets \cite[e.g.][]{Kaltenegger2017, Scheieterman2018, Fujii2018} and should be within the capabilities of the next generation of ground-based telescopes like the extremely large telescopes (ELTs) and mission concepts such as Origins, HabEx, and LUVOIR \cite[e.g.][]{Arney2018,Snellen2017}. Spectrographs on the Extremely Large Telescope (ELT) like HIRES (0.3-2.5$\mu m$) and METIS (3-19$\mu m$) are designed for a resolution of $R=100,000$ \citep{Ramsay2020}. Our database provides spectra modelled at 0.01$cm^{-1}$, which translates into a minimum resolution of $R=100,000$ from 0.4 to 10$\mu m$ and a minimum of $R=50,000$ from 10 to 20$\mu m$

Here we present a high-resolution database of 360 reflection and emission spectra of Earth-like planets with diverse surfaces, which evolved in the HZ of a wide range of Sun-like host stars. Our spectra are based on the atmosphere models described in detail in \cite{Madden2020mnras}. 

This database enables us to explore which of these planets provide the strongest atmospheric features for overall characterization as well as signs of life. Biosignatures in this work represent disequilibrium atmospheric chemistry suggesting biotic sources, namely the biosignature pairs of O$_2$ and CH$_4$, and CH$_4$ and O$_3$ \citep{Lovelock1965,Lederberg1965}.  

Our high-resolution spectra show the effects surfaces and host stars can have on the detectability of atmospheric features of habitable-zone planets and is a tool to prioritize promising targets in upcoming observations.

Our spectra provide an important step in expanding the references used for optimizing upcoming observations, training retrieval algorithms as well as providing comparison model datasets to analyze future observations. In addition, studies show that high-resolution ($R\approx100,000$) exoplanet spectrum can be isolated from the combined star-planet spectrum, using the radial velocity difference between the two objects \citep{Snellen2015,Rodler2014,Brogi2014,Fischer2016,LopezMorales2019}. high-resolution spectra models of habitable atmospheres are important in refining this technique and may allow characterization of planets even if they can't be resolved.

Section 2 describes our models, section 3 presents our results, and section 4 discusses and summarizes our paper.

Our high-resolution spectra are available online at \href{https://doi.org/10.5281/zenodo.3912065}{DOI: 10.5281/zenodo.3912065}.

\section{Methods}
\subsection{Planetary and Atmospheric Model}
The atmospheric composition of Earth-like planets depends on the outgassing rates, the irradiation from its host star, subsequent photochemistry, surface type, and cloud coverage. Here, we define `Earth-like' to refer to an Earth-radius and Earth-mass planet with similar outgassing rates to the modern Earth. Our spectra use planetary models generated using a coupled 1D climate and photochemistry model with wavelength-dependent albedo, described in detail in \cite{Madden2020mnras}.  

By incorporating wavelength-dependent reflection of surfaces and decoupling clouds from the surface reflection \cite{Madden2020mnras} explored the relationship between surface type and stellar type in the context of habitability. \cite{Madden2020mnras} found that surfaces with high variability across the visible and near-IR displayed a wide range of surface temperatures across star type. Surfaces like vegetation and sand showed the biggest change in surface temperature between cool K and hot F-stars while flatter overall albedo such as basalt, granite, coast, and seawater showed less change in surface temperature between star type. The surface temperature ranges for the different planet models are shown in Table \ref{tab:albedos}. 

\subsection{Generating reflection and emission spectra}
We use EXO-Prime2 to generate the high-resolution reflection and emission spectra for each simulated exoplanet from 0.4 to 20$\mu m$ at a resolution of $0.01 cm^{-1}$. The radiative transfer model used was originally developed for stratospheric measurements in Earth's atmosphere \citep{TraubStier1976, TraubJucks2002} and has been updated for use with exoplanets \citep[e.g.][]{DesMarais2002,TraubJucks2002,Kaltenegger2007,KalteneggerTraub2009, OMalleyJames2019}. For our calculations, we used 38 plane-parallel layers for an 80km atmosphere with an observation zenith angle of 60 degrees giving an approximation of quadrature viewing.

\begin{center}
\begin{table*}[p]
\begin{center}
\resizebox{\textwidth}{!}{
\singlespace
\begin{tabular}{@{}cccc@{}}
\toprule
\multirow{3}{*}{\textbf{Surface}}  & \multirow{3}{*}{\textbf{Source}}   & \textbf{Temp. Range} & \textbf{$\Delta$Temp.}    \\
                          &                                             & \textbf{(K)}         & \textbf{(K)}               \\
                          &                                             & \textbf{(F0V-K7V)}   &              \\ \midrule
Basalt                    & ASTER Basalt: Solid: Basalt.H5      & 315.5-296.4                  & 19.1                  \\
Granite                   & ASTER Alkalic: Solid: Granite.H1    & 314.4-295.2                  & 19.2                  \\
Sand                      & ASTER Brown loamy fine: 87P3468     & 311.8-280.2                  & 31.6                  \\
Grass                     & ASTER Grass: Unknown             & 314.7-280.8                  & 33.9                  \\
Trees                     & ASTER Deciduous: Unknown              & 312.4-278.8                  & 33.6                  \\ \cmidrule(lr){2-2}
\multirow{2}{*}{Seawater} & USGS Open Ocean SW2 (0.2-2.4$\mu m$) & \multirow{2}{*}{326.4-304.7} & \multirow{2}{*}{21.7} \\
                          & ASTER Seawater: Liquid (2.4+$\mu m$) &                              &                       \\ \cmidrule(lr){2-2}
\multirow{2}{*}{Coast}    & USGS Coast SW1 (0.2-2.4$\mu m$)      & \multirow{2}{*}{326.6-303.9} & \multirow{2}{*}{22.7} \\
                          & ASTER Seawater: Liquid (2.4+$\mu m$) &                              &                       \\ \cmidrule(lr){2-2}
Cloud                     & Modis $20\mu m$ Cloud Model          & 249.9-260.0                  & -10.1                 \\ \midrule
Basalt+Cloud              & 56.3\% Basalt, 43.7\% Cloud                 & 286.7-281.9                  & 4.8                   \\
Granite+Cloud             & 56.3\% Granite, 43.7\% Cloud                & 286.1-280.8                  & 5.3                   \\
Sand+Cloud                & 56.3\% Sand, 43.7\% Cloud                   & 284.0-271.9                  & 12.1                  \\
Grass+Cloud               & 56.3\% Grass, 43.7\% Cloud                  & 285.0-272.7                  & 13.8                  \\
Trees+Cloud               & 56.3\% Trees, 43.7\% Cloud                  & 283.9-270.1                  & 12.3                  \\
Seawater+Cloud            & 56.3\% Seawater, 43.7\% Cloud               & 297.0-287.8                  & 9.2                   \\
Coast+Cloud               & 56.3\% Coast, 43.7\% Cloud                  & 297.1-287.9                  & 9.2                   \\ \midrule
Basalt+Seawater           & 30\% Basalt, 70\% Seawater                  & 323.1-302.7                  & 20.4                  \\
Granite+Seawater          & 30\% Granite, 70\% Seawater                 & 322.9-302.1                  & 20.8                  \\
Sand+Seawater             & 30\% Sand, 70\% Seawater                    & 322.5-299.2                  & 23.3                  \\
Grass+Seawater            & 30\% Grass, 70\% Seawater                   & 323.2-299.0                  & 24.1                  \\
Trees+Seawater            & 30\% Trees, 70\% Seawater                   & 322.7-298.6                  & 24.2                  \\
Snow+Seawater             & 30\% Snow, 70\% Seawater                    & 290.6-288.5                  & 2.1                   \\ \midrule
Basalt+Seawater+Cloud     & 56.3\% (Basalt+Seawater), 43.7\% Cloud      & 293.7-286.1                  & 7.6                   \\
Granite+Seawater+Cloud    & 56.3\% (Granite+Seawater), 43.7\% Cloud     & 293.5-285.9                  & 7.6                   \\
Sand+Seawater+Cloud       & 56.3\% (Sand+Seawater), 43.7\% Cloud        & 292.6-283.1                  & 9.5                   \\
Grass+Seawater+Cloud      & 56.3\% (Grass+Seawater), 43.7\% Cloud       & 293.1-283.4                  & 9.5                   \\
Trees+Seawater+Cloud      & 56.3\% (Trees+Seawater), 43.7\% Cloud       & 292.6-283.1                  & 9.7                   \\
Snow+Seawater+Cloud       & 56.3\% (Snow+Seawater), 43.7\% Cloud        & 277.9-277.2                  & 0.7                   \\ \midrule
\multirow{3}{*}{Earth}    & 70\% Seawater, 2\% Coast, 2.52\% Basalt,    & \multirow{3}{*}{319.6-298.6} & \multirow{3}{*}{21.0} \\
                          & 2.52\% Granite, 1.96\% Sand, 8.4\% Grass,   &                              &                       \\
                          & 8.4\% Trees, 4.2\% Snow                     &                              &                       \\ \cmidrule(lr){2-2}
Earth+Cloud               & 56.3\% Earth, 43.7\% Cloud                  & 290.4-283.1                  & 7.3                   \\ \midrule
Flat                      & Flat reflectence of 0.31                    & 291.3-280.8                  & 10.5                  \\ \bottomrule
\end{tabular}
}
\end{center}
\caption{\singlespace The 30 simulated surface types with source and surface temperature range across star types. USGS: \cite{Kokaly2017usgs} (https://crustal.usgs.gov/speclab/), ASTER: \cite{Baldridge2009aster} (https://speclib.jpl.nasa.gov/library), Modis: \cite{King1997}}
\label{tab:albedos}
\end{table*}
\end{center}
\clearpage

We include the molecular species with prominent absorption features expected in the atmospheres of Earth-like planets orbiting F to K stars as modeled in \cite{Madden2020mnras}. We use the 2016 HITRAN database for our opacities for H$_2$O, CO$_2$, CH$_4$, N$_2$O, O$_3$, O$_2$, H$_2$CO, OH, C$_2$H$_6$, HO$_2$, CO, NO, NO$_2$, H$_2$O$_2$, H$_2$S, and SO$_2$ \citep{Gordon2017hitran}. We include CO$_2$ line mixing \citep{Niro2005}. For CO$_2$, H$_2$O, and N$_2$, we use measured continua data instead of line-by-line calculations in the far wings \citep{TraubJucks2002}.

With no clear answer on how cloud-feedback should affect clouds on exoplanets orbiting different host stars, we use Earth's clouds as a first approximation for all our models. We include 3 cloud layers in our models (following \cite{Kaltenegger2007}) at 1km (40\%), 6km (40\%), and 12km (20\%) and an overall cloud coverage of 44\% \citep{Madden2020mnras}. This simulates an observation of a cloudy exoplanet by having the spectrum represent the sum of different layers in the atmosphere.

\subsection{Stellar Spectra \& Surface Albedos}
The effects of wavelength-dependent surface and cloud albedo on habitability are most apparent when comparing the planetary models across star type. We used the same ATLAS model \citep{CastelliKurucz2004} F, G, and K star as in \cite{Madden2020mnras} for our calculations of planetary spectra and star to planet contrast. In total, we simulated planets around 12 star types spaced roughly 250K in temperature between an F0V (7,400K) and a K7V (3,900K). Note that the simulations in \cite{Madden2020mnras} used lower total stellar incident flux on the planet for cooler host stars to achieve temperatures similar to modern Earth across star type ($288 K \pm 2$ percent). 

The albedos used here and in \cite{Madden2020mnras} focus on the dominant surfaces on Earth: seawater, coastal water, basalt, granite, sand, trees, grass, snow, and clouds. A modern Earth albedo can be made by combining these surfaces with weights based on their modern Earth surface coverage \citep{Kaltenegger2007}. Surface albedos were taken from the USGS and ASTER spectral libraries ~\citep{Baldridge2009aster,Kokaly2017usgs,Clark2007usgs}. For all three cloud layers, we use the $20\mu m$ Modis cloud albedo ~\citep{King1997,RossowSchiffer1999} (Table \ref{tab:albedos}). 

In this paper we show four planetary scenarios for each surface: i) a single planetary surface to show the maximum effect of a specific surface on the spectra, ii) a 30\% single surface and 70\% seawater combination, and iii) and iv) two more scenarios where these cases have the 44\% cloud coverage, derived to simulate the modern Earth model in \cite{Madden2020mnras}. 

For 30 different surfaces around 12 host star types, we simulated in total, 360 terrestrial planetary spectra from 0.4 to 20$\mu m$.
\section{Results}
Our high-resolution spectra database contains the combined reflection and emission spectra for 360 Earth-like planets with 30 different surfaces orbiting 12 different Sun-like host stars. All spectra shown are a combination of planet reflection and emission. Emission begins to dominate the flux between 3 and 4$\mu m$ depending on the star and surface temperature.

We highlight a subset of these spectra in our figures to show a balance of variety and specific effects while keeping figures uncrowded. We do not show the simulated spectra for granite, grass, coast, or cloud surface only. The surface reflectivity of granite is similar to first-order with basalt, grass with trees, and coast with seawater. However, these spectra can all be downloaded from our database. 

\subsection{Reflection Spectra}
The star-surface interaction leads to drastic differences in a planet's appearance, which are most apparent in the exoplanets' reflectance spectra in the visible. 

\begin{figure*}[ht!]
\centering
\includegraphics[width=\textwidth]{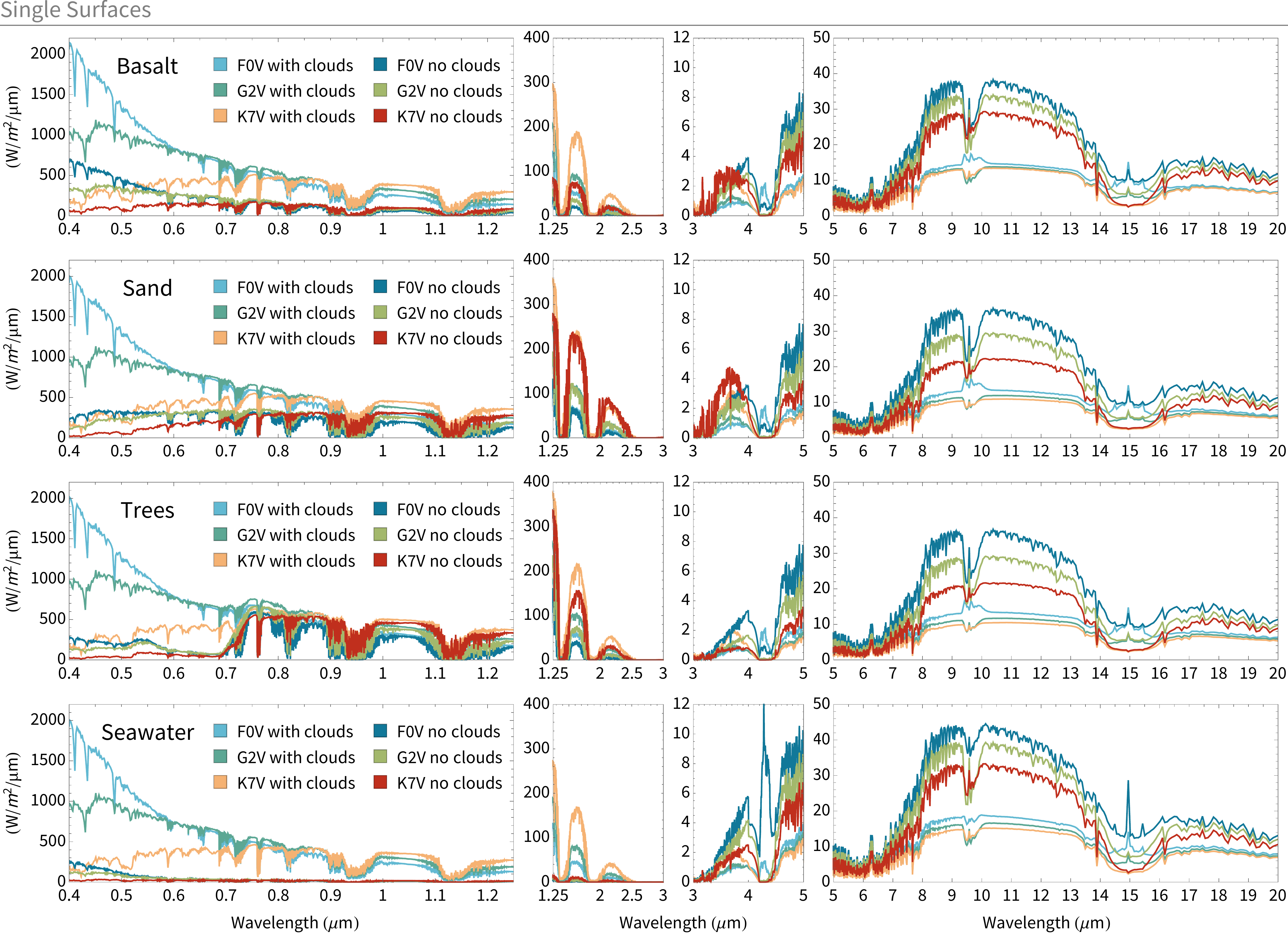}
\caption{\singlespace A sample of the combined reflection and emission spectra from the simulated exoplanets with 100\% of a single surface type both with and without clouds added. \label{fig:single}}
\end{figure*}

\begin{figure*}[ht!]
\centering
\includegraphics[width=\textwidth]{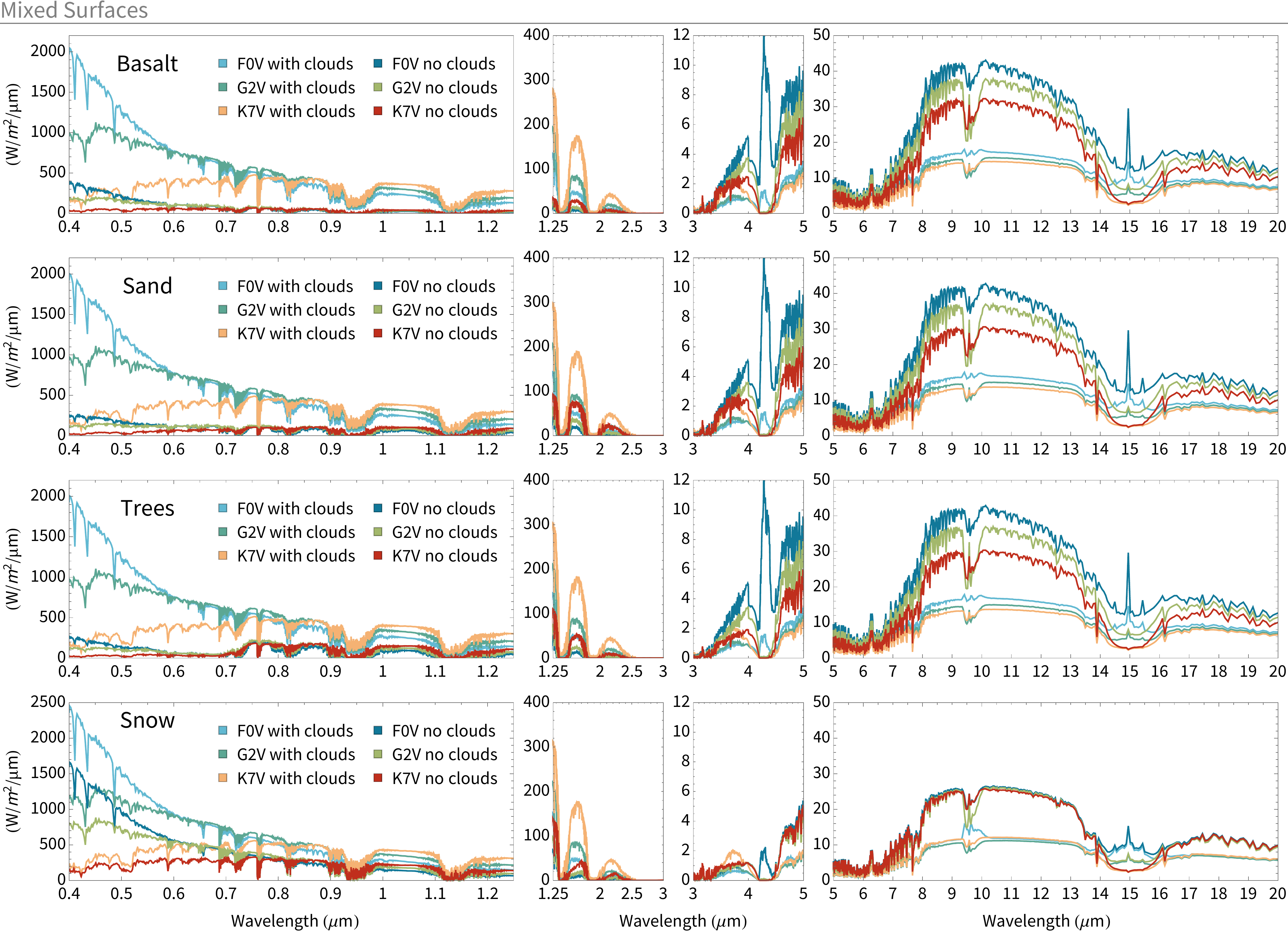} 
\caption{\singlespace A sample of the combined reflection and emission spectra from the simulated exoplanets with mixed surfaces of 30\% of one type and 70\% of seawater both with and without clouds added. \label{fig:mixed}}
\end{figure*}

Even though the incident stellar flux decreases for cooler star types to provide similar surface temperatures in our models, the reflected flux of a planet can vary by more than an order of magnitude at specific wavelengths depending on a planet's surface reflectivity, as shown in Fig. \ref{fig:single}, and Fig. \ref{fig:mixed}. It can result in planets with highly reflective surfaces orbiting cooler stars reflecting more starlight than planets with low surface reflectivity orbiting hotter stars: For example, at $0.5 \mu m$, a desert planet (sand surface) orbiting a K7V star is twice as bright as an ocean planet (seawater surface) orbiting an F0V host star, despite the higher incident flux of an F0V star at that wavelength.

Surfaces with high reflectivity generally lead to more prominent spectral features at visible wavelengths. The deepest absorption features can be seen for planetary models with high reflective surfaces like vegetation, sand, and snow orbiting the hottest grid host stars, which provide the highest incident flux (Fig. \ref{fig:single} and \ref{fig:mixed}).

The shape of the surface albedo also modulates the flux of the visible exoplanet spectra models. For example, the vegetation 'red-edge' near $0.7 \mu m$ shows as a strong increase in reflectivity in the spectra of tree-covered planets (Fig. \ref{fig:single} and Fig. \ref{fig:mixed}).  

\subsection{Emission}
Planetary surface albedos can have a large effect on the planetary surface temperature as well as atmospheric temperature structure \citep{Madden2020mnras}. Planetary models with highly reflective surfaces generally lead to lower surface temperatures and therefore lower infrared emission, while models with less reflective surfaces lead to higher surface temperatures and thus higher infrared emission of the planet for a specific host star (Table \ref{tab:albedos}, Fig. \ref{fig:single} and \ref{fig:mixed}).

The models in \cite{Madden2020mnras} used a reduced incident flux for cooler stars to achieve similar modern Earth temperatures for a constant surface albedo of 0.31. Therefore, while the surface reflectivity of a planet changes the surface temperature for similar incident flux, that difference has been compensated for in the modeling for a specific wavelength-independent surface albedo case, resulting in a slight increase in surface temperature and overall emission for planets orbiting hotter host stars (see Fig. \ref{fig:single} and \ref{fig:mixed}).

Infrared spectral feature depth depends on both the abundance as well as the difference in temperature of the overall emitting and absorption layer. Thus the deepest absorption features are not seen for the hottest planetary models with low reflective surfaces like oceans (Fig. \ref{fig:single} and \ref{fig:mixed}), because of the similarity in temperature of the two layers compared to planets with different surfaces.

\subsection{Planet-to-star contrast}
\begin{figure*}[ht!]
\centering
\includegraphics[width=\textwidth]{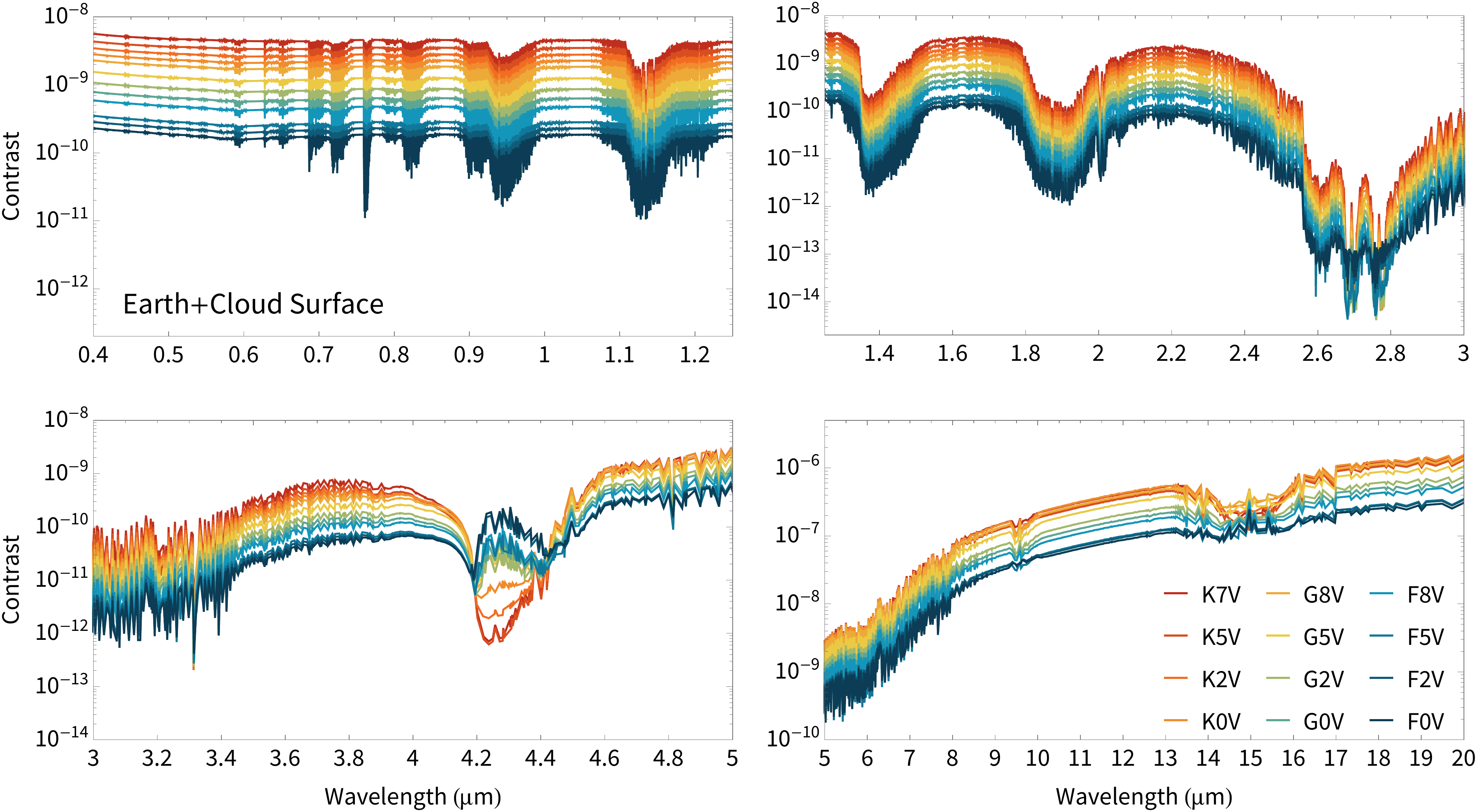}
\caption{\singlespace The model contrast spectra for a modern Earth surface including clouds across all star types. \label{fig:Earth}}
\end{figure*}

\begin{figure*}[ht!]
\centering
\includegraphics[width=\textwidth]{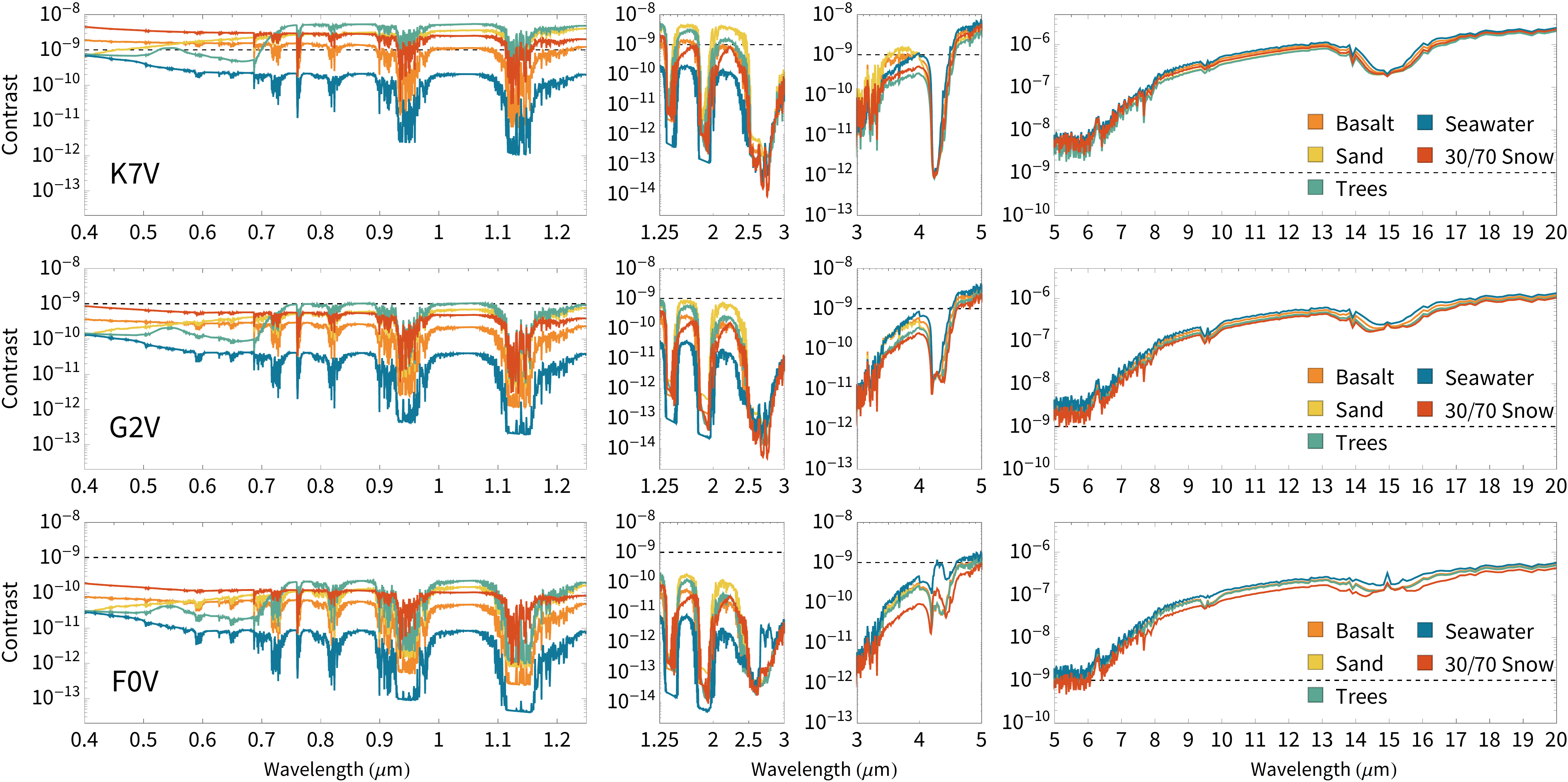}
\caption{\singlespace Contrast for planets modeled with basalt, sand, tree, seawater, and snow surfaces around K7V (top), G2V (middle), and F0V stars (bottom). A line at $10^{-9}$ is shown for reference between panels.  \label{fig:contrast}}
\end{figure*}

Planet-star contrast is generally higher for similar planets around cooler stars versus hotter stars. Fig. \ref{fig:Earth} shows that Earth-like planets orbiting our coolest grid stars have the highest contrast across the spectrum compared to hotter host stars. Planets with the same surface also show this based on our spectra (Fig. \ref{fig:contrast}). 

However, an ocean planet covered with dark seawater orbiting a K7V-star shows a similar contrast ratio in the visible and near-infrared ($0.7-4\mu m$) as a planet covered in vegetation around an F0V host star (Fig. \ref{fig:contrast}). Planets covered by highly reflective surfaces such as grass, trees, snow, and sand around G-stars will be consistently as high or higher in contrast at visible wavelengths than planets covered by darker surfaces such as coast, seawater, basalt, or granite around K-stars. Therefore, a planet’s surface can influence the contrast ratio significantly in the visible and near-IR. When comparing surfaces with extreme differences in reflectivity, cooler stars may not always provide the highest contrast habitable zone targets in the visible and near-IR, depending on their surface composition.

\clearpage

\subsection{Atmospheric composition change with host star}
The stellar energy distribution (SED) of a star influences the atmospheric composition of a planet \citep{Kasting1993,Rugheimer2013,Rugheimer2015Spectra,Segura2003,Segura2005,Madden2020mnras,Rauer2011}. Fig. \ref{fig:Earth}, Fig. \ref{fig:single}, and Fig. \ref{fig:mixed} show the varying depth of the atmospheric spectral features such as O$_3$, CO$_2$, and CH$_4$ for different stellar hosts.

In the visible, the depth of an absorption feature is proportional to the abundance of a molecule, the amount of incident stellar radiation, and the reflectivity of the planet. For similar reflectivity, the change in the absorption features depth reflects the change in abundance of the chemicals due to the stellar SED and subsequent photochemistry in the planet's atmosphere as well as incident irradiation (Fig. \ref{fig:Earth}). Hotter stars in our grid emit higher UV flux, thus altering the profiles of these molecules and subsequent reactions in a planet's atmosphere

In the infrared, the depth of the absorption features depends on the abundance of a chemical as well as the temperature difference between the emitting/absorbing layer and the continuum.

Surfaces can modify the atmosphere composition based on how the surface albedo alters the surface temperature of the planet as well as the temperature profiles of the atmosphere, for example how much water is evaporated. 

Fig. \ref{fig:Earth} shows the change in planet-to-star contrast ratio for a planet model with a modern Earth-analog surface for our grid stars to isolate the effect of the host star on the planet's spectra.

The most notable spectral features between 0.4 and 20$\mu m$ relevant to biosignature detection include oxygen and ozone at 0.69, 0.76, and 9.6$\mu m$; in combination with methane at 0.88, 1.04, 2.3, 3.3, and 7.66$\mu m$; ; N$_2$O shows features at 7.75, 8.52, 10.65, and 16.89$\mu m$. H$_2$O has features at 0.6, 0.65. 0.73, 0.82, 0.95, 1.14, 1.4, 1.85, 2.5-3.5, 3.7, and 5-8$\mu m$. Another greenhouse gas seen in the spectra is CO$_2$ at 2.7, 4.25, and 15$\mu m$.

\begin{figure*}[ht!]
\centering
\includegraphics[width=\textwidth]{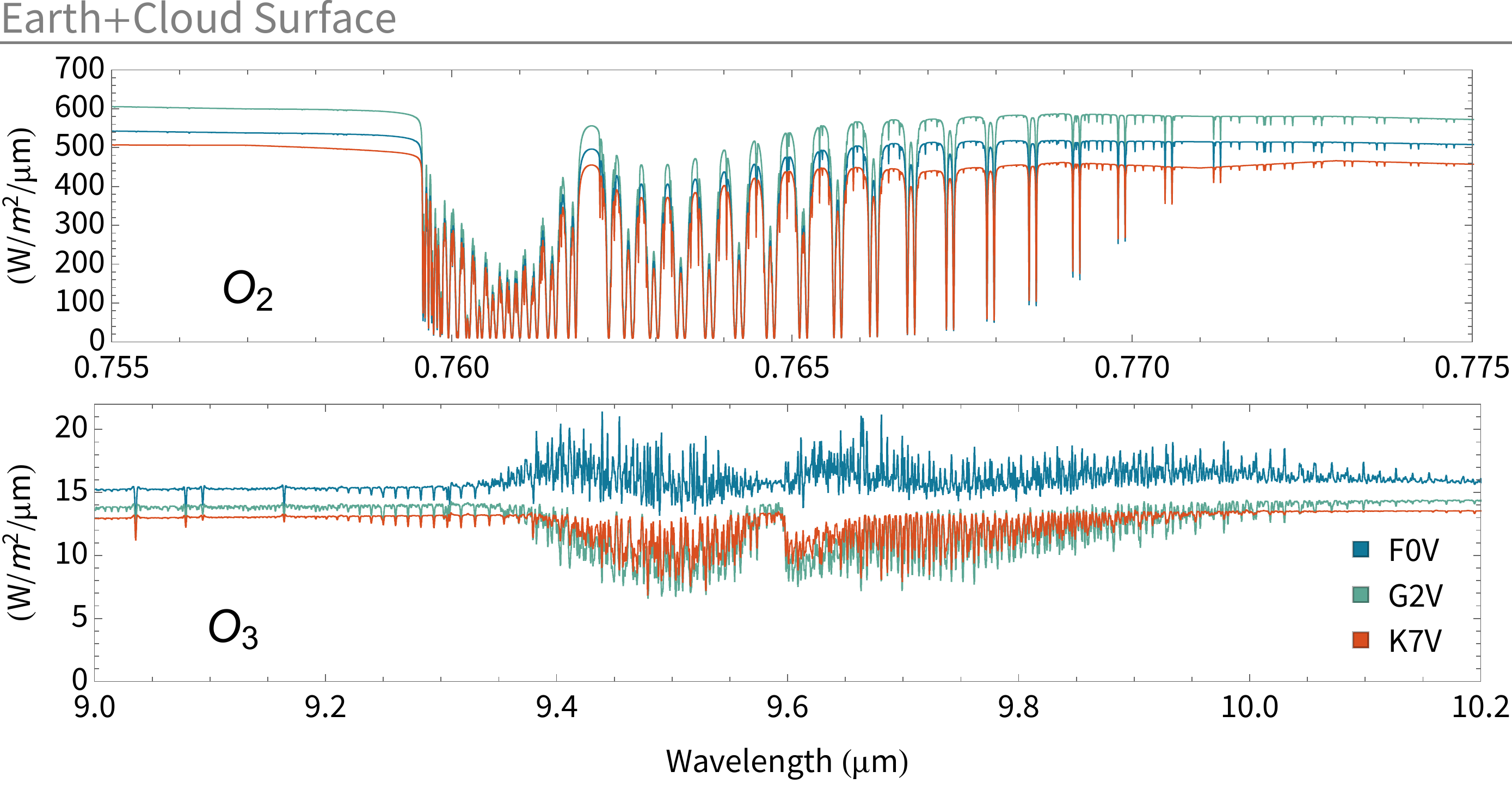} 
\caption{\singlespace Two oxygen features shown at high-resolution ($0.01cm^{-1}$, $R>100,000$) for an exoplanet with an Earth-like albedo with clouds around an F0V, G2V, and K7V host star.  \label{fig:highres}}
\end{figure*}

As an example, for the 0.76$\mu m$ O$_2$ and the 9.6$\mu m$ O$_3$ feature, Fig. \ref{fig:highres} shows that at a high-resolution of $0.01cm^{-1}$, all spectral features have a distinct series of lines to identify such Doppler-shifted lines uniquely on exoplanets which move predictably around their host star \citep{Rodler2014,Snellen2015,Brogi2014}. For specific absorption features of interest, our high-resolution spectra database can be used to optimize observation strategies for specific features and specific wavelength regions.

\section{Discussion \& Conclusion}
We present 360 emission and reflection spectra at a high-resolution of $0.01cm^{-1}$ for habitable zone planets orbiting 12 F, G, and K star types. These model spectra show the interaction of the host star's stellar energy distribution and a planet's wavelength-dependent albedo. We used the 9 dominant surfaces on modern Earth and isolated the effects of rock (basalt and granite), vegetation, sand, snow, clouds, water (ocean and coast) as well as clouds on a planet's reflection and emission spectra from the visible to the infrared (0.4-20$\mu m$). 

To show the variety in this database, we include models with single surface planets to show the maximum effect of each surface on the spectra as well as the planet star to planet contrast ratio (Fig. \ref{fig:contrast} and \ref{fig:single}), mixtures of seawater and cloud coverage (Fig. \ref{fig:single} and \ref{fig:mixed}) as well as modern Earth surface coverage models (Fig. \ref{fig:Earth}) for the 12 host stars from F0V to K7V.

Other surface albedos and fractional combinations of surfaces are of course possible for exoplanets. As a first-order estimate our single surface spectra can be combined to create spectra for new surface mixtures. In this way, our spectral database provides a toolkit to generate estimated spectra of Earth-like planets with different surface combinations with and without clouds. One aspect that won't be captured by such a combination is the potential difference in surface temperature introduced by the different surface reflectivity \cite[see][]{Madden2020mnras}. Exoplanets may also have many different surface types not addressed in this study including mineral surfaces\citep{Shields2018}, a wide range of different biota \citep{Hegde2013} or biofluorescent organisms\citep{OMalleyJames2018fluor}.

We've shown that surface albedos can deepen spectral features. In the visible, where the spectra show reflected starlight, highly reflective surfaces generally increase the depth of absorption features. In the infrared, on the other hand, low reflective surfaces increase the surface temperature and thus a planet's overall emission. 

Our high-resolution spectra database provides a critical tool in the planning and analysis of observations with upcoming ground-based telescopes like ELT, GMT, and TMT and future space mission concepts Origins, HabEx, and LUVOIR. Ground-based telescopes plan to employ high precision radial velocity techniques that require high-resolution ($R>100,000$) in order to characterize potentially habitable exoplanets.

Studying the wide range of changes caused by different surfaces and host stars improves our understanding of biosignatures and their remote observability. Obtaining high-resolution spectra of terrestrial planets in the habitable zone is an essential milestone in discovering life beyond our Solar System. Our database is intended to support this objective by presenting a wide range of cases for further study, planning, training, and comparison.
\clearpage
\section*{Data Access}
DOI for accompanying data: ~\href{https://doi.org/10.5281/zenodo.3912065}{10.5281/zenodo.3912065}

\section*{Acknowledgments}
This work was supported by the Carl Sagan Institute and the Brinson Foundation.